\documentclass{article}
\usepackage{latexsym}
\usepackage{graphics}
\usepackage{color}
\usepackage{amsmath, amsthm, amssymb}
\newtheorem{theorem}{Theorem}[section]
\newtheorem{proposition}{Proposition}[section]
\newtheorem{lemma}{Lemma}[proposition]
\newtheorem{corollary}{Corollary}[theorem]
\newtheorem{conjecture}{Conjecture}[section]
\newcommand{\tl}{\frac{\theta}{\lambda}}
\newcommand{\zn}{\frac{\zeta}{\nu}}
\begin{document}
\author{\sc Mihalis Dafermos\footnote{supported in part by
the NSF grant DMS-0302748} and Igor Rodnianski\footnote{a Clay Prize
Fellow supported in part by the NSF grant DMS-01007791}}
\title{A proof of Price's law for the collapse
of a self-gravitating scalar field}
\maketitle
\begin{abstract}
A well-known open problem in general relativity, dating back to 1972, 
has been to prove \emph{Price's law} for an appropriate model 
of gravitational collapse. This law postulates 
inverse-power decay rates
for the gravitational radiation flux on the event horizon
and null infinity with respect to appropriately normalized
advanced and retarded time coordinates. It is intimately
related both to astrophysical observations of black holes
and to the fate of observers who dare cross the event horizon.
In this paper, we prove a well-defined (upper bound) formulation
of Price's law for the collapse of a self-gravitating scalar field
with spherically symmetric initial data. We also allow
the presence of an additional gravitationally coupled Maxwell field. 
Our results are obtained by a new mathematical technique
for understanding the long-time behavior of large data solutions to the resulting coupled non-linear
hyperbolic system of p.d.e.'s in $2$ independent variables.
The technique is based on the interaction of the conformal geometry, the celebrated red-shift
effect, and local energy conservation; we feel
it may be relevant for the problem of non-linear stability of
the Kerr solution. When combined with previous work of the first author concerning
the internal structure of charged black holes, which \emph{assumed}
the validity of Price's law, our results can be applied to the strong cosmic censorship
conjecture for the Einstein-Maxwell-real scalar field system
with complete spacelike asymptotically flat spherically symmetric 
initial data. Under Christodoulou's $C^0$-formulation, 
the conjecture is proven to be false. 
\end{abstract}
\tableofcontents
\section{Introduction}
A central problem in general relativity is to understand
the final state of the collapse of isolated self-gravitating systems. 
One of the most
fascinating predictions of this theory is the formation of 
so-called \emph{black holes}. The fundamental issue is then 
to investigate their structure, both from the point of view
of far away observers, and of those
who dare cross the ``event horizon''.

The problem of gravitational collapse
can be formulated mathematically 
as the study of the initial value problem for
the Einstein equations
\[
R_{\mu\nu}-\frac12Rg_{\mu\nu}=2T_{\mu\nu}
\]
coupled to appriopriate matter equations,
where the initial data are assumed to be ``asymptotically flat''.
The unknown in the Einstein equations is a time-oriented $4$-dimensional
Lorentzian manifold $(\mathcal{M},g)$ known as \emph{spacetime},
the points of which are commonly called \emph{events}. 
We employ in what follows standard language and concepts of global Lorentzian 
geometry.\footnote{In the effort of
making this paper as self-contained as possible, 
all notions referred to in what follows are reviewed in a series
of Appendices. Standard references in the subject are
given at the end of this Introduction.}
In these terms, the notion of a black hole can be understood roughly as follows:
Asymptotic flatness and the causal geometry of $g$ allow 
one to distinguish a certain subset of the spacetime $\mathcal{M}$,
namely the set of those events in the causal past of events 
arbitrarily ``far away''.
The black hole region of spacetime is precisely the above
subset's complement in $\mathcal{M}$.\footnote{In the context of this paper, we will be able
to make this rigorous by defining first the concept of \emph{null infinity}.
One way to think about this is as a subset $\mathcal{I}^+$
of an appropriately defined ``boundary'' 
of $\mathcal{M}$, for which causal relations can still be applied.
The black hole is then $\mathcal{M}\setminus J^{-}(\mathcal{I}^+)$.
We call $J^-(\mathcal{I^+})$ the \emph{domain of outer communications} or
more simply the \emph{exterior}.}  The black hole's past boundary, a null
hypersurface in $\mathcal{M}$, is what is known as the \emph{event horizon}.

Current physical understanding of the basic nature of 
black holes--in particular, considerations regarding their stability,
as measured by outside observers--rests 
more on folklore than on rigorous results. Central to this folklore
is a heuristic study of the problem of linearized stability\footnote{In
the linearized context, one can
see already the most basic phenomena by considering the scalar wave equation
$\Box_g\phi=0$ on a fixed background $g$.} around
the Schwarzschild solution, 
carried out by Price~\cite{rpr:ns} in 1972. 
Price's heuristics suggested that solutions of the wave equation on this background
decay polynomially with respect to the static coordinate along timelike surfaces 
of constant area-radius. Later, these heuristics 
were refined~\cite{gpp:de1} to suggest 
polynomial decay on the event horizon\footnote{Here, decay is measured
with respect to an appropriately defined advanced time.} itself,
as well as along null infinity. In the context of the non-linear theory,
Price's heuristics were interpreted to suggest
that black holes with regular event horizons are stable from the point
of view of far away observers and
can form dynamically in collapse; moreover, that all ``parameters'' of 
the exterior spacetime 
other than mass, charge, and angular 
momentum--so-called ``hair''--should
decay polynomially when measured along the event horizon or null infinity. 
The conjectured generic decay rates are now widely known as \emph{Price's law}.

In this paper, we shall show that Price's law, as an upper bound for
the rate of decay, indeed holds in the context of a suitable mathematical
theory of gravitational collapse.
The theory will be described by the Einstein-Maxwell-real scalar field equations  
\begin{equation}
\label{Einstein-xs}
R_{\mu\nu}-\frac12Rg_{\mu\nu}=2T_{\mu\nu},
\end{equation}
\begin{equation}
\label{Maxwell-xs}
F^{,\mu}_{\mu\nu}=0, F_{[\mu\nu,\lambda]}=0,
\end{equation}
\begin{equation}
\label{sf-xs}
g^{\mu\nu}\phi_{;\mu\nu}=0,
\end{equation}
\begin{equation}
\label{emtensor-xs}
T_{\mu\nu}=\phi_{;\mu}\phi_{;\nu}
-\frac12g_{\mu\nu}\phi^{;\alpha}\phi_{;\alpha}
+F_{\mu\alpha}F^{\alpha}_{\ \nu}-\frac14g_{\mu\nu}F^{\alpha}_{\ \lambda}
F^{\lambda}_{\ \alpha}.
\end{equation}
We will assume our initial data to be spherically symmetric\footnote{See Appendix~\ref{sphsymsec} 
for a geometric definition. This assumption 
will imply that the system $(\ref{Einstein-xs})$--$(\ref{emtensor-xs})$
reduces to a system in $2$ spacetime dimensions.}.
Physically, the above equations correspond to a self-gravitating system comprising
of a neutral scalar field and an electromagnetic field.
The motivation for this system has been discussed in previous work of 
the first author~\cite{md:si, md:cbh}. 
Basically, $(\ref{Einstein-xs})$--$(\ref{emtensor-xs})$ 
is the simplest self-gravitating system
that radiates to infinity under spherical symmetry\footnote{Recall, that in view
of Birkhoff's theorem, the vacuum equations $R_{\mu\nu}=0$ have no dynamical degrees of freedom in
spherical symmetry.}, and in addition has a repulsive mechanism (provided by the charge) 
competing with the attractive force of gravity. It can thus be thought of as a good 
first attempt at a spherically symmetrical
model of the collapse of the vacuum equations \emph{without} symmetry, 
where gravitational radiation and the centrifugal effects of angular momentum play
an important role.\footnote{See, however, the remarks at the end of Section~\ref{mapl}.} 

We shall give a precise statement of the main theorem of this paper in Section~\ref{K9e} below.
We then  turn in Section~\ref{mapl} to a discussion 
of the \emph{mass inflation scenario},
first proposed in 1990 by Israel and Poisson. According to this scenario, based
on heuristic calculations, Price's law
has implications for the internal structure of black holes, and the
validity of the strong cosmic censorship conjecture.
In view of previous results of the first author~\cite{md:cbh}, we shall deduce from
Theorem~\ref{int-the} that a suitable formulation of
strong cosmic censorship is \emph{false} for $(\ref{Einstein-xs})$--$(\ref{emtensor-xs})$
under spherical symmetry (Corollary~\ref{cor-1}).  In Section~\ref{grpro}, we shall 
give a precise statement (Theorem~\ref{unth}) of Price's law in the linearized 
setting, for which our results also apply.
Finally, we provide an outline for the body of the paper in Section~\ref{outl}, and in addition,
point out some useful references.

\subsection{The main theorem}
\label{K9e}
Price's law is a statement about precise decay rates in the exterior of a black hole
in the Cauchy development of an appropriate initial data set. 
Because our system is self-gravitating, i.e.~the spacetime
$(\mathcal{M}, g)$ is itself unknown \emph{a priori}, 
it follows that before one can talk about decay, one must show that indeed 
a black hole exists in the spacetime, 
identify its event horizon, and define time parameters
in the exterior
with respect to which decay is to be measured. All this must thus be contained in
the statement of our theorem. 
To describe concisely the global geometry of spacetime, it will 
be convenient to employ so-called \emph{Penrose diagrams}. 
This notation and related conventions are described in detail in Appendix~\ref{pdveo}.
It is an easy exercise to translate geometric statements exhibited by 
Penrose diagrams
to statements about the existence and properties of certain null coordinate 
systems.\footnote{For the convenience of the reader uncomfortable with 
diagrams, starting from Section~\ref{assumsec}, we will 
always supplement information inferred from diagrams with explicit formulas in terms of a related null 
coordinate system. In particular,
the Penrose diagram of Theorem~\ref{int-the} is ``translated'' in 
${\bf A}'$ of Section~\ref{assumsec}.} 

The main result of this paper is:
\begin{theorem} 
\label{int-the}
Consider spacelike spherically symmetric initial 
data for the Einstein-Maxwell-scalar field equations, with at least one
asymptotically flat end, such that
the scalar field and its gradient have compact support on the initial
hypersurface. 
Let $\mathcal{Q}$ denote the $2$-dimensional Lorentzian quotient
of the future Cauchy development.
Assume either
\begin{enumerate}
\item 
$\mathcal{Q}$ contains a trapped surface, \emph{or}
\item 
The initial hypersurface is complete, and the Maxwell field does not identically
vanish, \emph{or}
\item
The initial hypersurface is complete and has two asymptotically flat 
ends.\footnote{It turns out that assumption 2 implies 3.}
\end{enumerate}
Then, $\mathcal{Q}$ contains a subset $\mathcal{D}$
with Penrose diagram:
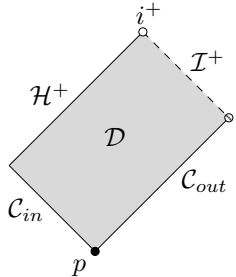
\begin{figure}[h]
\centering
\input{arxiko.pstex_t}
\label{ff}
\end{figure}

\noindent
Let $r$ denote the area-radius function, $m$ the Hawking mass function
and $e$ the charge\footnote{Under spherical symmetry,
the Maxwell part decouples, and its contribution
to the evolution of the metric and the scalar field depends only on
a constant--denoted $e$ and called \emph{charge}--which can be 
computed from initial data. We have $e=0$ iff the 
Maxwell tensor $F_{\mu\nu}$ vanishes identically. See Appendix~\ref{sphsymsec}.}.
One can define a regular null coordinate 
system $(u,v)$ covering $\mathcal{D}\cap J^{-}(\mathcal{I}^+)$ 
by setting $p=(1,1)$, and
defining the $v$-coordinate such that $v\sim r$ along the outgoing null
ray
$\mathcal{C}_{out}$, while
defining the $u$-coordinate such that $\partial_ur=-1$
(in a suitable limiting sense) along null infinity $\mathcal{I}^+$. 
With this normalization, $u\to\infty$ along $\mathcal{I}^+$ as $i^+$ is approached,
and, moreover, null infinity is complete in the sense of 
Christodoulou~\cite{chr:mg}. The points of the event horizon
$\mathcal{H}^+$ can be parametrized by $(\infty,v)$, and 
$\mathcal{I}^+$ can be parametrized by $(u,\infty)$. 
On $\mathcal{H}^+$, the quantities $m(\infty,v)$, $r(\infty,v)$, $\phi(\infty,v)$,
and $\partial_v\phi(\infty,v)$ are defined, and
$r\phi(u,\infty)$ and $\partial_u(r\phi)(u,\infty)$ can be defined
on $\mathcal{I}^+$ by a limiting procedure. 
It follows that either ``the black hole is extremal in the limit'', i.e.
\begin{equation}
\label{kritiknp}
m(\infty,v)+\frac{e^2}{2r(\infty,v)}-|e|\to0\hbox{\rm\ as\ }v\to\infty,
\end{equation}
along $\mathcal{H}^+$,
or the following hold:
\begin{equation}
\label{kavovikopoinsn}
|\partial_v r|(\infty,v)\ge C\left(1-\frac{2m(\infty,v)}{r(\infty,v)}\right),
\end{equation}
\begin{equation}
\label{eisa-1}
|\phi|(\infty,v)+|\partial_v\phi|(\infty,v)\le C_\epsilon v^{-3+\epsilon},
\end{equation}
along $\mathcal{H}^+$,
and
\begin{equation}
\label{eisa-2}
|r\phi|(u,\infty)\le C_\epsilon u^{-2},
\end{equation}
\begin{equation}
\label{eisa-3}
|\partial_u{(r\phi)}|(u,\infty)\le C_\epsilon u^{-3+\epsilon}
\end{equation}
along $\mathcal{I}^+$, 
for any $\epsilon>0$, and for some positive constants $C$, $C_\epsilon$.
\end{theorem}

One should note that the
system $(\ref{Einstein-xs})$--$(\ref{emtensor-xs})$ 
has previously been studied
numerically \cite{bo:lte,gpp:de,mc:bhsf}, again under spherical symmetry; 
Theorem~\ref{int-the} is in complete agreement
with the findings of these studies.

In the case where $e=0$,
Christodoulou has shown~\cite{chr:ins} that
for generic\footnote{Here, the notion of genericity is 
defined in the context of a space of functions of bounded variation, 
which in turn is defined directly in terms of the spherically 
symmetric reduction. We refer to~\cite{chr:bv}
for details.} initial data, either the solution disperses
of a black hole forms.
In the latter case, it is easy to deduce the assumptions of
Theorem~\ref{int-the}. We thus have
\begin{corollary}
If $e=0$, generic complete asymptotically flat $BV$-data in the sense of Christodoulou 
either have a future causally geodesically complete maximal development, or
the result of Theorem~\ref{int-the} applies.\footnote{Previously, Christodoulou
had shown in this case that $\phi(\infty,v)\to0$ and $r\phi(u,\infty)\to0$, 
without, however, a rate.} 
\end{corollary}

The global features of $(\ref{Einstein-xs})$--$(\ref{emtensor-xs})$ 
with $e\ne0$ in the \emph{interior} of black hole regions turn out
to be completely different from the case were $e=0$. This is already clear,
for example, by comparing the Penrose diagrams of the Schwarzschild
and Reissner-Nordstr\"om solutions. (See Appendix~\ref{paradeigmata}.)
It is the existence
of a smooth Cauchy horizon in the latter solutions
 that raises the question of strong cosmic censorship. 
 We will turn in the next section
to a discussion of this issue. 
The story turns out to be much more interesting 
than what one could conjecture from the examination of special solutions; 
moreover, the validity of Price's law plays a central role.

\subsection{Mass inflation}
\label{mapl}
The most basic mathematical questions regarding the internal structure
of black holes concern the presence and nature of singularities.
The old story goes that ``deep inside'' the black hole, spacetime
should terminate in a singularity, strong enough to eventually destroy
any observer who crosses the event horizon. The classic example is
Schwarzschild (see Appendix~\ref{paradeigmata}). 
The set labelled $\mathcal{B}$ on the boundary
indeed represents a ``singularity''; the
spacetime is \emph{inextendible} as a $C^0$ metric. 
The statement of strong cosmic censorship (see Appendix~\ref{eikasies}) 
thus holds for this particular solution.

The Schwarzschild family of solutions can be considered, however,
as a very special case of the Reissner-Nordstr\"om family of solutions
of the Einstein-Maxwell system,
for which the charge $e$ happens to vanish. 
In the case $e\ne 0$, the situation is entirely different
(see Appendix~\ref{paradeigmata}).
The spacetime is future causally geodesically incomplete, but there is no
``singularity'';
the Cauchy development $(\mathcal{M},g)$ can be extended to a 
larger $C^\infty$  spacetime $(\tilde{\mathcal{M}},\tilde{g})$,
still satisfying $(\ref{Einstein-xs})$--$(\ref{emtensor-xs})$, such that
all inextendible causal curves in $\mathcal{M}$
enter $\tilde{\mathcal{M}}\setminus\mathcal{M}$. 
The statement of strong cosmic censorship fails thus
for this solution.
The boundary of $\mathcal{M}$ in $\tilde{\mathcal{M}}$ is known as  
the \emph{Cauchy horizon}.\footnote{In the Penrose diagram of the quotient
of such an extension, this boundary would correspond to the set
labelled $\mathcal{CH}^+$.}

The basis for the strong cosmic censorship conjecture were heuristic arguments,
originally due to Penrose, which indicated that the
Reissner-Nordstr\"om Cauchy horizon should be unstable.
These arguments were put forth in the context of the linear theory 
and made use of the so-called \emph{blue-shift effect}.\footnote{This effect
is in some sense dual to the \emph{red-shift effect} which will play such 
an important role in the analysis in this paper.} (See
Hawking and Ellis~\cite{he:lssst}, p.~161, for a discussion.)
In 1990, Israel and Poisson took a closer look at this
effect in a non-linear context motivated by the system
$(\ref{Einstein-xs})$--$(\ref{emtensor-xs})$. 
The picture they obtained was more complex than Penrose's original 
expectations,
and, in particular, its implications for cosmic censorship turned 
out to be mixed. 

To understand the strength of the blue-shift effect in the non-linear setting, 
one needs a quantitative estimate on the 
strength of infalling radiation into 
the black hole. But this is precisely the rate of decay of radiation tails
on the event horizon, i.e.~Price's law. 
On the one hand, Israel and Poisson calculated--in the context
of their heuristics--that  
Price's power-law decay was in fact sufficiently \emph{fast} so as 
to ensure Cauchy horizons still arise in the black hole interior, 
beyond which the spacetime metric can be \emph{continuously} extended.
On the other hand, they calculated, provided that the rate postulated by Price is
sharp, in the sense that it provides--generically--a \emph{lower} bound
on the rate of decay, that it is sufficiently
\emph{slow} to ensure that these Cauchy horizons would generically 
be weakly singular, in particular the Hawking mass would blow up identically,
and thus, the above-mentioned $C^0$ metric extensions could in fact not be 
$C^1$. In view of the blow-up of the mass, Israel and Poisson dubbed this
scenario~\emph{mass inflation}.

In \cite{md:cbh}, the first author proved that under the assumption that
a slightly weaker version of Price's law holds, the mass-inflation 
scenario of Israel and Poisson is correct
for spherically-symmetric solutions of $(\ref{Einstein-xs})$--$(\ref{emtensor-xs})$
with $e\ne0$.
In particular, it is assumed in~\cite{md:cbh} that an event horizon
has formed, and that that the resulting black hole is non-extremal.
The precise polynomial decay assumption imposed in~\cite{md:cbh}
that guaranteed continuous extendibility
of the metric was that
\begin{equation}
\label{giavabgei}
|\partial_v\phi|\le Cv^{-1-\epsilon}
\end{equation}
along the event horizon\footnote{The even weaker assumption
$|\partial_v\phi|\le Cv^{-\frac12-\epsilon}$ was sufficient to show that the area
radius remained bounded away from $0$. The results of~\cite{md:cbh} improved
previous work \cite{md:si} of the first
author, where the mass-inflation scenario was shown to hold provided
that the scalar field is compactly supported on the event horizon.}, 
where $\epsilon$
is an arbitrarily small positive number and $v$ a suitably normalized
advanced time coordinate.

In fact, the condition $(\ref{kavovikopoinsn})$ ensures
that the $v$-coordinate described in Theorem~\ref{int-the}
is equivalent to the $v$-coordinate
employed in the formulation of~\cite{md:cbh}. Inequality
$(\ref{eisa-1})$ is then seen to be stronger than the necessary condition
$(\ref{giavabgei})$.
It follows that, under the assumptions of Theorem~\ref{int-the}, 
if $e\neq0$ and $(\ref{kritiknp})$ does not 
hold\footnote{
It is easy to identify large open classes of initial data such that
$(\ref{kritiknp})$ does not occur: For instance, if $r>e$ 
everwhere on the initial hypersurface.},
then $\mathcal{Q}$ contains 
a short outgoing null ray $\mathcal{C}'_{in}$ emanating
from a point $q$ sufficiently close to $i^+$ (see the Penrose
diagram below), so that
our solution restricted to
$(\mathcal{H}^+\cap J^+(q))\cup\mathcal{C}'_{in}$  satisfies the assumptions
on characteristic initial data of Theorem 1.1 of~\cite{md:cbh}. 
Thus, we have
\begin{corollary}
\label{cor-1}
Under the assumptions of the previous theorem, if in addition $e\neq0$,
then either $(\ref{kritiknp})$ holds on $\mathcal{H}^+$,
or the Cauchy development $(\mathcal{M},g)$ is extendible 
to a larger manifold with $C^0$
metric. In particular, the Lorentzian
quotient $\mathcal{Q}$ contains
a subset with Penrose diagram depicted
below as the union $\mathcal{D}\cup\mathcal{G}$:
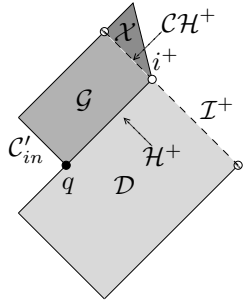
\[
\input{arxikoedw.pstex_t}
\]
The extension $\tilde{\mathcal{M}}$ can be taken spherically symmetric
with quotient manifold $\tilde{\mathcal{Q}}$  
extending globally ``across'' $\mathcal{CH}^+$, i.e.~such that
$\tilde{Q}$ contains a subset $\mathcal{D}\cup\mathcal{G}\cup
\mathcal{CH}^+\cup\mathcal{X}$ depicted 
in the Penrose diagram above.
It follows that
the strong cosmic censorship conjecture, under the
formulation\footnote{See Appendix \ref{cccon}. In this formulation,
inextendibility is required in $C^0$.} of Christodoulou~\cite{chr:givp}, 
is \emph{false}
in the context of spherically symmetric solutions of this system.
\end{corollary}

To understand the significance of the above it is instructive to compare with
the case $e=0$. In~\cite{chr:fbh}, Christodoulou 
showed that if a black hole forms\footnote{Recall that, as discussed
above, in the case $e=0$ it is proven in~\cite{chr:ins}
that for generic initial data, either
a black hole forms or the solution is future causally
geodesically complete.}, then the
$2$-dimensional quotient $\mathcal{Q}$ has Penrose diagram:
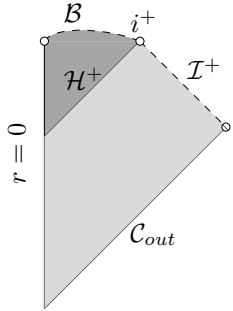
\[
\input{christodoulou.pstex_t}
\]
The future boundary $\mathcal{B}$ of the black hole region 
$\mathcal{Q}\setminus
J^-(\mathcal{I}^+)$ is such that the
spacetime is locally inextendible across it as a $C^0$ metric.
In particular, strong cosmic censorship is
\emph{true} in this case. 

Finally, in connection to the system $(\ref{Einstein-xs})$--$(\ref{emtensor-xs})$,
it is important, to point out the following:
While modelling the important physical effect of the centrifugal
force of angular momentum, our system introduces another 
feature, which is entirely foreign to the physical problem.
If $e\neq0$, the equations do not admit complete spherically
symmetric asymptotically flat initial data
with one end.\footnote{This arises from the fact that $(\ref{Maxwell-xs})$
and $(\ref{sf-xs})$ are not coupled directly. The source for Maxwell tensor
must thus be topological.} 
In particular, the black holes to which our theorem applies are
generated by the topology of the initial data. A model that would
allow for \emph{both} realistic initial data \emph{and} the possibility of
the formation of regular or partially regular Cauchy horizons is provided
by the charged scalar field, discussed in Section 3.3 of Hawking and 
Ellis~\cite{he:lssst}. Unfortunately, at present there is neither
a theory of the formation of black holes for this system of equations,
in the spirit of~\cite{chr:ins}, nor 
the internal causal structure of these black holes, in the spirit of~\cite{md:cbh}.
The local existence theory for this system and small data results, however, have been
proven (under spherical symmetry) by Chae~\cite{doc:ge}.

\subsection{The linearized problem}
\label{grpro}
The method of this paper can in fact be specialized to
provide a proof of decay in the uncoupled  problem $\Box_g\phi=0$,
where $g$ is Schwarzschild or Reissner-Nordstr\"om, heuristically
studied in~\cite{rpr:ns,gpp:de1, jb:gcc}. We obtain
\begin{theorem}
\label{unth}
Fix real numbers $0\le |e|< M$, and consider the standard maximally
analytic
Reissner-Nordstr\"om solution $(\mathcal{M},g_{RN})$, 
with parameters $e$, $M$. (See Hawking and Ellis~\cite{he:lssst}, p.~158.)
Let $\mathcal{Q}$ denote the Lorentzian quotient $\mathcal{M}/SO(3)$
and $\pi_1$ the projection $\pi_1:\mathcal{M}\to\mathcal{Q}$.
Consider a subset $\mathcal{D}\subset\mathcal{Q}$ with 
Penrose diagram\footnote{See Appendix \ref{pdiag}. In particular,
note that according to the conventions
described there, $\mathcal{H}^+\cup\mathcal{H}^-\subset\mathcal{D}$}:
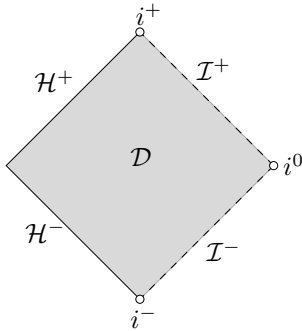
\[
\input{Re-No-ext.pstex_t}
\]
Let $\phi$ denote a $C^2$ solution to the 
wave equation on $\pi_1^{-1}(\mathcal{D})$, and 
define $\phi_0:\mathcal{Q}\to{\bf R}$
by $\phi_0(p)=\frac1{4\pi r^2}\int_{\pi_1^{-1}(p)}\phi$.\footnote{This
is the $0$'th spherical harmonic. Recall that $p\in\mathcal{Q}$ means
that $\pi_1^{-1}(p)$ 
represents a spacelike sphere in $\mathcal{M}$. The integral
is with respect to the area element of $\pi_1^{-1}(p)$.}
Let $u$ and $v$ denote standard Eddington-Finkelstein coordinates on
$J^-(\mathcal{I}^+)\cap J^+(\mathcal{I}^-)\cap\mathcal{D}$,
i.e.~such that 
\[
g_{RN}=-\left(1-\frac{2M}r+\frac{e^2}{r^2}\right)dudv+r^2 \gamma,
\]
where $\gamma$ is the standard metric on the $2$-sphere.
If $u$ is such that constant-$u$ curves are outgoing, then
the future event horizon $\mathcal{H}^+$
thus corresponds to points $(\infty, v)$
and the past event horizon $\mathcal{H}^-$ to $(u, -\infty)$. 
Assume $\phi$ and $\nabla\phi$ satisfy $|r\phi|\le \bar{C}r^{-2}$, 
$|\partial_u(r\phi)|+|\partial_v(r\phi)|\le \bar{C}r^{-3}$ on a
Cauchy surface $\mathcal{S}\subset\mathcal{D}$. 
Since the points $(\infty,v)$ and $(u,-\infty)$ are contained in $\mathcal{D}$, 
$\phi_0(\infty,v)$ and $\phi_0(u,-\infty)$ are clearly defined,
and one can see that $\partial_v\phi_0(\infty,v)$, $\partial_u\phi_0(u,-\infty)$
are likewise defined. Moreover,
in this coordinate system, points $(u,\infty)$, $(-\infty,v)$ can be 
identified with points of $\mathcal{I}^+$, $\mathcal{I}^-$, respectively,
and $r\phi_0(u,\infty)$, $r\phi_0(-\infty,v)$, $\partial_u(r\phi_0)(u,\infty)$,
$\partial_v(r\phi_0)(-\infty,v)$ can be defined. 
The following decay
rates hold
\[
|\partial_v\phi_0(\infty,v)|+|\phi_0(\infty,v)|
\le C_\epsilon (\max\{v,1\})^{-3+\epsilon},
\]
\[
|\partial_u\phi_0(u,-\infty)|+|\phi_0(u, -\infty)|
\le C_\epsilon (\max\{-u, 1\})^{-3+\epsilon},
\]
\[
|r\phi_0|(u,\infty)\le C|u|^{-2},
\]
\[
|r\phi_0|(-\infty,v)\le C|v|^{-2},
\]
\[
|\partial_u(r\phi_0)|(u,\infty)\le C|u|^{-3+\epsilon},
\]
\[
|\partial_v(r\phi_0)|(-\infty,v)\le C|v|^{-3+\epsilon}.
\]
In terms of $(r,t)$ coordinates where $t$ is the static time
defined by $t=\frac12(v+u)$,
we have
\[
|\partial_t\phi|(r_0,t)+|\partial_r\phi|(r_0,t)+|\phi|(r_0,t)
\le C_{r_0,\epsilon}|t|^{-3+\epsilon}
\]
for each fixed $r_0>r_+=M+\sqrt{M^2-e^2}$.
\end{theorem}

What connects the linear and the non-linear problem,
in such a way so as to allow the above ``specialization'',
is the existence, in both cases, of an analogous energy estimate for $\phi$. 
In the static case, this estimate
is deduced by contracting the energy-momentum tensor
of $\phi$ with the $\frac\partial{\partial t}$ Killing vector field, 
while in the coupled case, the estimate arises through the coupling,
via the remarkable properties of the (renormalized) Hawking mass.

A proof of a weaker\footnote{Decay rates are derived
in the Schwarzschild $t$ coordinate but do not extend
to decay rates on the event horizon with respect to
advanced time. Also, it is assumed that the solution
vanishes in a neighborhood of
$\mathcal{H}^+\cap\mathcal{H}^-$ and $i^0$.} 
version of this linearized Price's law has 
recently been given by Machedon
and Stalker~\cite{st:pc}. Other studies of this linear
problem have appeared in the physics literature, 
e.g.~\cite{clsy:wpgs}.
The boundedness of the solution (without the assumption
of spherical symmetry) was proven 
by Kay and Wald~\cite{kw:lss}, 
while decay in $t$, without however a rate, was proven by Twainy~\cite{ft:td}.
 
The standard approach for the linearized problem
has been to adopt the special Regge-Wheeler coordinates, which transform the
wave equation on a static backround to the Minkowski-space wave 
equation with time-independent potential. Fourier analytic and/or spectral theoretic 
methods can then be employed. It would be interesting to investigate whether 
these methods would also be applicable to the non-linear 
problem studied here. The method of the present paper, however, is in a 
completely different, new direction. Here it is the global characteristic geometry of the 
event horizon, together with the celebrated red-shift effect,
that will play the central role. Both these aspects seem to
be somewhat obscured by the Regge-Wheeler coordinate approach.

\subsection{Outline of the paper}
\label{outl}
A brief outline of the paper is as follows: In Section~\ref{assumsec},
we shall formulate the most general assumptions for which we can prove our results.
These will refer to a $2$-dimensional Lorentzian manifold
$(\mathcal{Q},\bar{g})$ on which two functions $r$, $\phi$ are
defined, satisfying various properties postulated \emph{a priori}.
In Section~\ref{reduc}, we retrieve the assumptions of Section~\ref{assumsec} 
from various more recognizable assumptions, including (but not limited
to) those of Theorem~\ref{int-the}. In Section~\ref{ecsec},
we shall introduce the most basic analytic features of the system of equations
formulated in Section~\ref{assumsec}. We shall then proceed
to prove a uniform boundedness (and uniform decay in $r$) result in
Section~\ref{largersec}. In Sections \ref{rssection}--\ref{akoma2sec},
we prove certain localized estimates in characteristic rectangles. 
These estimates relate behavior on various boundary segments to behavior 
in the interior. Decay is then proven in Section~\ref{partI},
where the estimates of Sections \ref{rssection}--\ref{akoma2sec} are applied to sequences
of rectangles constructed with the help of the global causal geometry, energy conservation,
and the pigeonhole principle. The final results are given by Theorems~\ref{final1} and~\ref{final2} of 
Section~\ref{finalsec}. We formulate some open problems in 
Section~\ref{sharpsec}, and finally, we outline and the modifications 
of our argument necessary to prove Theorem~\ref{unth}, in
Section~\ref{uncoupledsec}.

As noted in the very beginning, we have included,
for the benefit of the reader not so familiar with the standard language of
the global study of the Einstein equations, several appendices. 
In Appendix~\ref{ivpinl}, we discuss the terminology necessary to understand
the global initial value problem for the Einstein equations:
basic Lorentzian geometry, the notion of the maximal Cauchy development,
Penrose's singularity theorems, asymptotically flat initial data,
and finally, the statement of the cosmic censorship conjectures.
Whenever convenient, we will restrict consideration to
the system $(\ref{Einstein-xs})$--$(\ref{emtensor-xs})$.  
In Appendix~\ref{sphsymsec}, we review the definition and basic facts about 
spherical symmetry, in particular the definition of quotient manifold $\mathcal{Q}$,
area radius $r$, and Hawking mass $m$. Finally, in Appendix~\ref{pdveo}, we introduce 
the notational convention of Penrose diagrams.

For more, the reader should consult
standard textbooks, like~Hawking and Ellis~\cite{he:lssst}, or~\cite{bee:glg}, as well as 
the nice survey article~\cite{chr:givp}. 
References~\cite{md:bb, md:ouggaria} might also be helpful.

\vskip1pc
\noindent{\bf Acknowledgement.} 
The authors thank Demetrios Christodoulou
for his very useful comments on a preliminary version of this paper. 
The authors also thank the organizers of the
General Relativity Summer School at Carg\`ese, Corsica,
where this collaboration began in the Summer of 2002, 
as well as the Newton Institute of Cambridge University for its
hospitality in June 2003, while this work was being completed.

\section{Assumptions}
\label{assumsec}
In this section, we will formulate the most general assumptions 
under which the results of our paper apply. These 
assumptions (${\bf A}'$--${\bf \Sigma T}'$ below)
will refer to a $2$-dimensional spacetime $(\mathcal{Q},\bar{g})$,
on which two functions $r$ and $\phi$
are defined, and a constant $e$. 
As we shall see, only weak regularity assumptions
need be imposed. 

\begin{enumerate}
\item[${\bf A}'$.]
We assume
we have a $C^3$ $2$-dimensional manifold $\mathcal{Q}$ on which a
$C^1$ Lorentzian metric $\bar{g}$ and
a $C^2$ nonnegative function $r$ are defined, such 
that $\mathcal{Q}$
contains a subset $\mathcal{D}$ on which $r$ is strictly positive,
and such that $\mathcal{D}$ has Penrose 
diagram\footnote{See Appendix~\ref{pdveo} 
for an explanation of this notation, or else, skip to the
next paragraph!} 
depicted below:
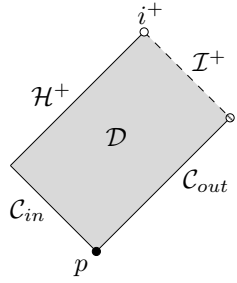
\[
\input{arxiko.pstex_t}
\]
Note
that in view of our conventions of the Appendices,
the above diagram encodes in particular the
inclusions $\mathcal{H}^+\subset\mathcal{D}$, 
$\mathcal{C}_{out}\subset\mathcal{D}$, $\mathcal{C}_{in}\subset\mathcal{D}$,
while, 
again by our convention,
$i^+\not\in \mathcal{I}^+$, and thus, by the definition of $\mathcal{I}^+$,
$\sup_{\mathcal{H}^+}r=r_+<\infty$.

For the reader uncomfortable with diagrams, we translate:
There exists 
a global null coordinate chart $(u,v)$ covering $\mathcal{D}$
such that
\[
\bar{g}=-\frac{\Omega^2}2(du\otimes dv+dv\otimes du)
\]
with $\Omega$ a strictly positive $C^1$ function.
We assume that $\mathcal{D}$ corresponds precisely to the
coordinate range $[1,U_0]\times[1,V_0)$ for some $1<U_0<\infty$,
$1<V_0\le\infty$, and moreover, that $u$ and $v$ increase towards the future,
i.e.~$\nabla u$, and $\nabla v$ are future pointing.
We define  
$\mathcal{C}_{out}\subset\mathcal{D}$ to be the constant-$u$
curve $\{1\}\times[1,V_0)$, we define
$\mathcal{C}_{in}$ to be the constant-$v$ curve
$[1,U_0]\times\{1\}$, we define $p=(1,1)$,
and we define $\mathcal{H}^+$ to be $\{U_0\}\times [1,V_0)$. 
We assume that for all $u\in[1,U_0)$, $\sup_{v} r(u,v)=\infty$,
whereas we assume $\sup_{v}r(U_0,v)=r_+<\infty$.

\item[${\bf B}'$.] We assume a $C^1$ real-valued
function $\phi$ to be defined on $\mathcal{D}$.

\item[${\bf \Gamma}'$.]
Let $(u,v)$ be a null coordinate system as in
${\bf A}'$,
With respect to such coordinates
define the functions $\nu$, $\lambda$ by
\begin{equation}
\label{ruqu}
\partial_ur=\nu,
\end{equation}
\begin{equation}
\label{rvqu}
\partial_vr=\lambda.
\end{equation}
We assume
\begin{equation}
\label{lamov}
\lambda\ge0
\end{equation}
and 
\begin{equation}
\label{vumov}
\nu<0. 
\end{equation}
On $J^-(\mathcal{I}^+)\cap\mathcal{D}=\mathcal{D}\setminus\mathcal{H}^+$ we assume
\begin{equation}
\label{lamov2}
\lambda>0.
\end{equation}

\item[${\bf \Delta}'$.]
Define the functions $\theta$, $\zeta$ by
\begin{equation}
\label{zeta-orismos}
r\partial_u\phi=\zeta
\end{equation}
\begin{equation}
\label{theta-orismos}
r\partial_v\phi=\theta.
\end{equation}
Fix some real number $\omega$ such that $1<\omega\le 3$.
On $\mathcal{C}_{out}$, assume\footnote{Note that $\tl$ is invariant with respect
to reparametrization of the $v$ coordinate.}
\begin{equation}
\label{byil}
r\left|\frac\theta\lambda\right|\le C,
\end{equation}
\begin{equation}
\label{pioduvato}
r^\omega\left|\phi+\frac\theta\lambda\right|\le C,
\end{equation}
for some constant $C$.

\item[${\bf E}'$.]
Define the function $m$
by
\begin{equation}
\label{m-orismos}
m=\frac{r}2(1-\bar{g}(\nabla r,\nabla r)).
\end{equation}
We will call $m$ the \emph{Hawking mass}.
Define $\mu$, the so-called \emph{mass-aspect function}, by
\begin{equation}
\label{ma-orismos}
\mu=\frac{2m}r.
\end{equation}
Assume that
\begin{equation}
\label{pepmaza}
0<\sup_{\mathcal{C}_{out}}m<\infty.
\end{equation}
Fix a constant $e$, to be called the \emph{charge}, and
define the 
\emph{renormalized Hawking mass} $\varpi$ by
\begin{equation}
\label{pi-orismos}
\varpi=m+\frac{e^2}{2r}=\frac{r}2(1-\bar{g}(\nabla r,\nabla r))+\frac{e^2}{2r}.
\end{equation}
In view of $(\ref{vumov})$, a
function $\kappa$ can be defined everywhere on $\mathcal{D}$
by
\begin{equation}
\label{kappaori}
\kappa=-\frac14\Omega^{2}\nu^{-1}.
\end{equation}
By $(\ref{m-orismos})$, we have the identity
\begin{equation}
\label{star}
\lambda=\kappa(1-\mu).
\end{equation}
Assume that $\varpi$, $\kappa$, $\theta$,
$\zeta$ satisfy pointwise throughout $\mathcal{D}$ the equations
\begin{equation}
\label{puqu}
\partial_u\varpi=\frac{1}{2}(1-\mu)\left(\zn\right)^2\nu,
\end{equation}
\begin{equation}
\label{pvqu}
\partial_v\varpi=
\frac{1}{2}\kappa^{-1}\theta^2,
\end{equation}
\begin{equation}
\label{fdb1}
\partial_u\kappa=\frac{1}r\left(\frac{\zeta}{\nu}\right)^2\nu\kappa
\end{equation}
\begin{equation}
\label{sign1}
\partial_u\theta=-\frac{\zeta\lambda}r,
\end{equation}
\begin{equation}
\label{sign2}
\partial_v\zeta=-\frac{\theta\nu}r.
\end{equation}
(In particular, assume $\theta$, $\zeta$ are differentiable
in the $u$ and $v$ directions, respectively.)
\item[${\bf \Sigma T}'$.]
Assume
$r_+\ne |e|$.
\end{enumerate}

The above assumptions are clearly independent of the choice of
null coordinate system in ${\bf B}'$. That is to say, given a null
coordinate system $(u,v)$ covering $\mathcal{D}$ 
and functions $\nu,\lambda,\theta,\zeta,\kappa$ satisfying Assumptions
${\bf \Gamma}'$--${\bf \Sigma T}'$, then if $(u^*,v^*)$ is a new null 
coordinate system
covering $\mathcal{D}$
with $\frac{\partial u}{\partial u^*}$,
$\frac{\partial v}{\partial v^*}$ positive $C^1$ functions,
defining $\nu^*=\frac{\partial r}{\partial u^*}=\frac{\partial u}{\partial u^*}\nu$, 
$\lambda^*=\frac{\partial r}{\partial v^*}=
\frac{\partial u}{\partial u^*}\lambda$, etc., these
quantities again satisfy Assumptions ${\bf \Gamma}'$--${\bf \Sigma T}'$.
Indeed, we have formulated the assumptions in the above manner precisely
in order to emphasize their invariant geometric content.
We defer till the next section the task of identifying a particular
convenient null coordinate system.

From $(\ref{star})$, $(\ref{puqu})$, and $(\ref{ruqu})$ we obtain
\begin{equation}
\label{lqu}
\partial_u\lambda=\nu\frac{2\kappa}{r^2}
\left(\varpi-\frac{e^2}{r}\right).
\end{equation}
By equality of mixed partials, we then have
\begin{equation}
\label{nqu}
\partial_v\nu=
\nu\frac{2\kappa}{r^2}
\left(\varpi-\frac{e^2}{r}\right).
\end{equation}
In addition, note that where $1-\mu\ne0$, $\lambda\ne0$, we
have,
in analogy to $(\ref{fdb1})$,
\begin{equation}
\label{fdb2}
\partial_v\left(\frac\nu{1-\mu}\right)
=\frac{1}r\left(\frac{\theta}{\lambda}\right)^2\lambda\left(\frac\nu{1-\mu}\right).
\end{equation}

Finally, we note the following:
When denoting causal relations in $\mathcal{D}$, 
\emph{e.g.}~$p\in J^+(q)$, it goes without saying
that $J^+$, etc.~will always
refer to the causal structure of $\bar{g}$ on $\mathcal{D}$. It is clear, however, that,
in any future pointing null coordinate system on $\mathcal{D}$, 
as in ${\bf A}'$, the causal structure of $\bar{g}$ is identical to the causal 
structure of the domain of the coordinates as a subset of
$2$-dimensional Minkowski space, i.e., the set $J^+(u,v)$, $J^-(u,v)$, etc., are 
given in null coordinates as in the formulas of Appendix~\ref{2mink}.

\section{Retrieving the assumptions}
\label{reduc}
In this section we give three alternative propositions
all of which retrieve the
assumptions ${\bf A}'$--${\bf \Sigma T}'$ 
of the previous section.

In the first proposition, we retrieve ${\bf A}'$--${\bf \Sigma T}'$
from the assumptions
of Theorem \ref{int-the}:
\begin{proposition}
\label{sxesntwvduo}
Let $\{\Sigma, \stackrel\circ{g}, K_{ij}, E_i, B_i, 
\stackrel{\circ}\phi, \phi'\}$ be an asymptotically flat spherically symmetric
initial data set\footnote{See Appendix \ref{cccon}.}, such 
that $\stackrel{\circ}\phi$, $\phi'$ are
of compact support. Let 
\begin{equation}
\label{lampei}
\{\mathcal{M},g,F_{\mu\nu},\phi\}
\end{equation}
denote its future Cauchy development, and let $(\mathcal{Q},\bar{g})$ be the quotient
manifold of $(\ref{lampei})$
with its induced Lorentzian metric.
Suppose either
\begin{enumerate}
\item
$\mathcal{Q}$ contains a trapped surface, \emph{or}
\item
$(\Sigma,\stackrel\circ{g})$ is complete
and $E_i$ and $B_i$ do not both identically vanish, \emph{or}
\item
$(\Sigma,\stackrel\circ{g})$ is complete with two ends. 
\end{enumerate}
Then $\mathcal{Q}$ contains a subset $\mathcal{D}$,
such that if 
$r$ is defined by
$(\ref{r-orismos})$,
 $\phi:\mathcal{Q}\to{\bf R}$ denotes
the induced map of $\phi:\mathcal{M}\to{\bf R}$ on group orbits, 
and $e$ is defined by $(\ref{edw})$,
then
$\{\mathcal{Q}, \bar{g}, r, \phi, e\}$ satisfy
Assumptions ${\bf A}'$--${\bf E}'$
with $\omega=3$. 
If, in addition, $e=0$ or
$r>e$ identically along $\mathcal{S}=\Sigma/SO(3)$, 
then Assumption ${\bf \Sigma T}'$ is also satisfied.
\end{proposition}

\noindent\emph{Proof.} The equations $(\ref{ruqu})$--$(\ref{sign2})$
were derived from $(\ref{Einstein-xs})$--$(\ref{emtensor-xs})$ 
in~\cite{md:si}. The global structure
implicit in the Penrose diagram of ${\bf A}'$ follows
from the results of~\cite{md:sssts}.
$\Box$ 

In the spherically symmetric $e=0$ case considered by Christodoulou,
we can formulate a different proposition by
appealing to the results of~\cite{chr:mt}:
\begin{theorem}
\label{touxristodoulou}
Consider the initial
value problem of~\cite{chr:bv}, where it is assumed
that $(\ref{byil})$ and $(\ref{pioduvato})$ hold initially,
and let $\{\mathcal{Q}, \bar{g}, r, \phi\}$ denote the $BV$
future Cauchy development. Assume $\mathcal{Q}$ possesses a
complete\footnote{Complete can be taken in the sense of~\cite{chr:givp},
or merely as the statement that the $u$-coordinate defined as in 
Section~\ref{veosxrovos} satisfies
$u\to\infty$.} null infinity $\mathcal{I}^+$, and that the final Bondi mass
is strictly positive. Assume, moreover, that $\frac\zeta\nu$ is uniformly 
bounded on some ingoing null ray intersecting the event horizon
$\mathcal{H}^+$.\footnote{See Appendix~\ref{pdiag} for
an explanation of this notation.}
Then
$\{\mathcal{Q}, \bar{g},r,\phi\}$ satisfy
Assumptions ${\bf A}'$--${\bf \Sigma T}'$ with $e=0$.
\end{theorem}
We note that the existence of a trapped surface in the $BV$-Cauchy development
implies that null infinity is complete and that the final Bondi mass is
strictly positive (see for instance~\cite{md:sssts}). Thus, in the context
of Christodoulou's initial value problem, Theorem~\ref{touxristodoulou} contains
Proposition~\ref{sxesntwvduo}.
It would be nice if the results of this paper could also
be proven for the more weak general class of solutions considered
in~\cite{chr:mt}. This remains, however, an open problem.

Finally, we note that in 
the original heuristic analysis of Price, besides
data of compact support, data which have an ``initially static'' tail
are also considered. For this
we have the following:
\begin{proposition}
In Proposition~\ref{sxesntwvduo}, 
if the assumption that $\stackrel{\circ}\phi$, $\phi'$ are of
compact support is replaced by the assumption that $K_{ij}$, $\phi'$ are of compact
support, then Assumptions ${\bf A}'$--${\bf \Sigma T}'$ hold as before, 
with $\omega=2$. 
\end{proposition}
Our final conclusions (Theorems~\ref{final1} 
and~\ref{final2}) will depend on $\omega$
in complete agreement with the heuristics of Price.

\section{Basic properties of the system}
\label{ecsec}
In the rest of the paper, we assume always ${\bf A}'$--${\bf \Sigma T}'$.
Before proceeding into the details of our argument, we make some introductory
comments about the structure of the system $(\ref{ruqu})$--$(\ref{sign2})$.

\subsection{Coordinates}
To get started, let us settle for the time being on a definite coordinate system.
(We will change it soon enough, but no matter!) 
First note that 
by $(\ref{lamov2})$ and $(\ref{star})$
it follows that
\[
1-\mu>0
\]
on $\mathcal{C}_{out}$. 
Since $r$ is strictly increasing to the future on $\mathcal{C}_{out}$, and
since\footnote{If the reader wishes to interpret directly the Penrose diagram, note
that by our conventions, $\mathcal{C}_{out}$, as depicted, has a limit point
on $\mathcal{I}^+$!}  
$\sup_{\mathcal{C}_{out}}r=\infty$, it follows that
the set $\mathcal{C}_{out}\cap\{r\le 3\sup_{\mathcal{C}_{out}}m\}$ is compact.
Thus $1-\mu$ is bounded above and
below by positive constants on this set, while
$\frac53\ge 1-\mu\ge \frac13$ on 
$\mathcal{C}_{out}\cap\{r\ge 3\sup_{\mathcal{C}_{out}}m\}$.
It follows that
\[
\tilde{d}>1-\mu>d>0
\]
on $\mathcal{C}_{out}$ for some constants $d$, $\tilde{d}$.

We now set $p$ by $p=(1,1)$, and then require 
\begin{equation}
\label{INITv}
\frac{\partial_vr}{1-\mu}=\kappa=1
\end{equation}
along $\mathcal{C}_{out}$,
and 
\[
\nu=\partial_vr=-1
\]
along $\mathcal{C}_{in}$. This clearly completely determines a null
regular coordinate
system in $\mathcal{D}$. 
To examine its range, note first that
since $1-\mu<\tilde{d}$ on $\mathcal{C}_{out}$, we
have, for $q\in\mathcal{C}_{out}$
\begin{eqnarray*}
v(q)	&=&	\int_1^v{1d\tilde{v}}+1\\
	&=&	\int_1^v\frac{\partial_vr}{1-\mu}d\tilde{v}+1\\
	&\ge&	\tilde{d}^{-1}(r(1,v)-r(1,1))+1,
\end{eqnarray*}
Thus, as $q\to\mathcal{I}^+$, it follows that, since $r(q)\to\infty$,
we must have $v(q)\to\infty$.
On the other hand, 
\begin{eqnarray*}
u(\mathcal{C}_{out}\cap\mathcal{H}^+)	&=&	\int_1^ud\tilde{u}+1\\
					&=&	\int_1^u(-\nu)(u,1)d\tilde{u}+1\\
					&=&	r(p)-
						r(\mathcal{C}_{out}\cap\mathcal{H}^+)
							+1\\
					&\equiv&	U.
\end{eqnarray*}
Thus, with respect to these coordinates, we have
\[
\mathcal{D}=[1,U]\times[1,\infty)
\]
and 
\[
\mathcal{H}^+=\{U\}\times[1,\infty).
\]

\subsection{Monotonicity}
Define the constants
\[
\varpi_1=\varpi(U,1)
\]
\begin{equation}
\label{r_1def}
r_1=r(U,1)
\end{equation}
and 
\begin{equation}
\label{MDEF}
M=\sup_{C_{out}}{\varpi}.
\end{equation}
By Assumptions ${\bf A}'$ and ${\bf E}'$, it follows that both $r_1>0$
and $M>0$.  
We have the following
\begin{proposition}
In $\mathcal{D}$, we have
\begin{equation}
\label{kappa9etiko}
1\ge\kappa>0
\end{equation}
\begin{equation}
\label{xwrisovoma1}
1-\mu\ge0
\end{equation}
\begin{equation}
\label{pumov}
\partial_u\varpi\le0,
\end{equation}
\begin{equation}
\label{pvmov}
\partial_v\varpi\ge0.
\end{equation}
\begin{equation}
\label{II}
\partial_u\kappa\le0,
\end{equation}
\begin{equation}
\label{rlower}
r\ge r_1,
\end{equation}
\begin{equation}
\label{pi-pavwkatw}
\varpi_1\le\varpi\le M.
\end{equation}
In $J^{-}(\mathcal{I}^+)\cap\mathcal{D}$, we have in addition
\begin{equation}
\label{xwrisovoma}
1-\mu>0,
\end{equation}
\begin{equation}
\label{etaprosnmo}
\frac{\nu}{1-\mu}<0
\end{equation}
\begin{equation}
\label{IIb}
\partial_v\left(\frac\nu{1-\mu}\right)\le0.
\end{equation}
\end{proposition}

\noindent\emph{Proof.} 
Inequality $(\ref{rlower})$ is immediate from $(\ref{vumov})$, $(\ref{lamov})$.
In view of $(\ref{fdb1})$, 
inequality $(\ref{vumov})$ immediately implies $(\ref{II})$ and the left hand
side of $(\ref{kappa9etiko})$.
On any segment $[1,u]\times\{v\}$, where $u\le U$,
$r^{-1}\frac{\zeta^2}{\nu}$ is necessarily uniformly bounded by
continuity and compactness.
Thus, integrating $(\ref{fdb1})$,
\[
\kappa(u,v)=\kappa(1,v)
e^{\int_1^u{r^{-1}\nu^{-1}\zeta^2(u',v)du'}}>0,
\]
i.e., we have obtained the right hand side of $(\ref{kappa9etiko})$. 
Together with $(\ref{lamov})$, this yields $(\ref{xwrisovoma1})$,
with $(\ref{lamov2})$, this yields $(\ref{xwrisovoma})$,
and with $(\ref{vumov})$, this yields $(\ref{etaprosnmo})$, $(\ref{IIb})$. 
The remaining inequalities follow immediately from
$(\ref{puqu})$, $(\ref{pvqu})$, $(\ref{fdb2})$. $\Box$

We note that all inequalities in the above proposition are in fact
independent of our normalization of coordinates $(u,v)$, except
for the left hand side of $(\ref{kappa9etiko})$, which depends
on our normalization of $v$.

\subsection{The energy estimate}
In view of $(\ref{pumov})$, $(\ref{pvmov})$, we can view equations
$(\ref{puqu})$ and $(\ref{pvqu})$ as yielding an energy \emph{estimate}
for the scalar field: For any null
segment $\{u\}\times[v_1,v_2]$,
integrating $(\ref{pvqu})$, 
we obtain
\begin{eqnarray}
\label{EE-v}
\nonumber
\int_{v_1}^{v_2}{\frac12\theta^2\kappa^{-1}(u,v)dv}	& = &	
\varpi(u,v_2)-\varpi(u,v_1)\\
					&\le&	M-\varpi_1
\end{eqnarray}
and similarly, integrating $(\ref{pvqu})$ on any
$[u_1,u_2]\times \{v\}$, 
\begin{equation}
\label{EE-u}
-\int_{u_1}^{u_2}{\frac12\zeta^2\frac{1-\mu}{\nu}(u,v)du}	\le	M-\varpi_1.
\end{equation}
In fact, as
will be discussed in Section \ref{uncoupledsec}, the above estimates are intimately
related to the energy estimate arising from Noether's theorem applied
to a scalar field
on a fixed spherically symmetric \emph{static} background.

Although the energy estimates $(\ref{EE-v})$, $(\ref{EE-u})$ will certainly 
play an important role at all levels in this paper,
central to our proof will be sharper estimates, taylored to the geometry
of various subregions of $\mathcal{D}$.

\subsection{An ``almost'' Riemann invariant}
Where $r$ is large, we shall be able to exploit the quantity
$\partial_v(r\phi)=\phi\lambda+\theta$. Recall that this quantity
is a ``Riemann invariant''
in the case of the spherically symmetric wave equation on a fixed
Minkowski background, i.e.~it satisfies
$\partial_u(\phi\lambda+\theta)=0$.
From $(\ref{sign1})$, $(\ref{lqu})$, we obtain here
\begin{equation}
\label{toallo}
\partial_u(\phi\lambda+\theta)=\frac{\phi}{r^2}2\nu\kappa
				\left(\varpi-\frac{e^2}r\right).
\end{equation}
While the right hand side of $(\ref{toallo})$ does not vanish,
the $r^{-2}$ factor,
together with our assumption $(\ref{pioduvato})$,
allows us to show that the quantity $\phi\lambda+\theta$
has better decay in $r$ than either $\phi$ or $\theta$ alone. We shall
see already the important role
of $(\ref{toallo})$ in the estimates of Section \ref{largersec}. The equation
will be further exploited in Section \ref{maxsec}.

\subsection{The red-shift effect}
In a neighborhood of $\mathcal{H}^+$, where $r\sim r_+$, very different
estimates will apply, estimates which have no analogue in Minkowski space.
These estimates are intimately related to the celebrated ``red-shift effect'',
the shift to the lower end of the spectrum for radiation emitted by
an observer crossing $\mathcal{H}^+$, as observed by an observer travelling
to $i^+$, or crossing $\mathcal{H}^+$ at a later advanced time.
To understand the analytic content of this effect in the context of
our system,
consider the equation satisfied by the quantity $\zn$ appearing
in $(\ref{puqu})$:
\begin{equation}
\label{zex}
\partial_v\left(\zn\right)=
-\frac{\theta}r-\left(\zn\right)\left(\frac{2\kappa}{r^2}\left(
\varpi-\frac{e^2}r\right)\right).
\end{equation}
(This equation is easily obtained from $(\ref{nqu})$ and $(\ref{sign2})$.)
Let us fix some $r_0>r_+$, and some sufficiently large $V$,
and restrict attention
to the region defined by 
\[
\mathcal{R}=\{r\le r_0\}\cap\{v\ge V\}\cap\mathcal{D},
\]
depicted as the darker-shaded
region in the Penrose diagram below\footnote{Note that $\nu<0$, $\lambda>0$
implies that constant-$r$ curves in $J^-(\mathcal{I}^+)$ are timelike!}:
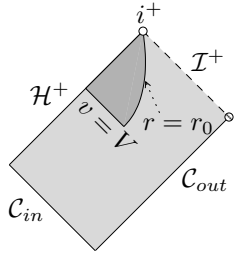
\[
\input{redshift.pstex_t}
\]
We will show in Section \ref{UB}  that 
\[
\kappa>\hat{c}>0,
\] 
and (using Assumption ${\bf E}'$!)
that
\[
\varpi-\frac{e^2}r>\tilde{c}'>0
\]
throughout this region, for constants $\hat{c}$, $\tilde{c}'$.
Thus, since clearly $r_1\le r\le r_0$ in $\mathcal{R}$,
we have 
\[
2r_0^{-2}M\ge
\frac{2\kappa}{r^2}\left(
\varpi-\frac{e^2}r\right)\ge2\hat{c}\tilde{c}'r_1^{-2},
\]
and consequently, integrating $(\ref{zex})$,
\begin{equation}
\label{prwtnmatia}
\left|\zn(u,v)\right|
\le
\int_V^{v}\frac{|\theta|}{r}e^{-2\hat{c}\tilde{c}'r_1^{-2}(v-v^*)}dv^*
+\left|\zn(u,V)\right|e^{-2\hat{c}\tilde{c}'r_1^{-2}(v-V)}.
\end{equation}

In Section \ref{UB}, we shall see that $(\ref{prwtnmatia})$ immediately
gives uniform bounds for $\left|\zn\right|$ in $\mathcal{R}$, since the first
term on the right can be bounded by
$(\ref{EE-v})$ and the second term is clearly bounded by $\left|\zn(u,V)\right|$.
Later on, however, we will want to exploit the fact that if
$v\gg V$, then the second term
on the right hand side acquires an exponentially decaying factor.
Of course, this information is only useful if the first term also can be shown
to be suitably small. In Section \ref{rssection}, 
we shall integrate $(\ref{prwtnmatia})$
in characteristic
rectangles $\mathcal{W}$ chosen so that we can indeed make use of the exponential
decaying factor in a quantitative way. 
The significance of this estimate will only become apparent
in the context of our induction scheme in Section \ref{partI}.

\section{Uniform boundedness and decay for large $r$}
\label{largersec}
Before turning to decay in $u$ and $v$, which is the ultimate goal
of this paper, we shall begin by showing a series of 
global bounds yielding decay in $r$ for $\phi$,
$\theta$, and $\lambda\phi+\theta$. 

\subsection{$r\ge R$}
We consider first the region $r\ge R$, for sufficiently large $R$. Our tools
will be energy estimates, as expressed by $(\ref{EE-u})$
and $(\ref{EE-v})$, and the additional estimate provided by the ``good'' right
hand side of 
$(\ref{toallo})$.

Note that Assumption ${\bf \Gamma}'$ implies in particular that
constant-$r$ curves not crossing $\mathcal{H}^+$ are 
timelike.
It is clear from $(\ref{pi-pavwkatw})$ that for large enough $R$, 
\begin{equation}
\label{pekcokmuhim}
1-\mu(u,v)\ge c
\end{equation}
for all $(u,v)$ such that $r(u,v)\ge R$, 
where $c$ is some positive constant, depending on $R$, such that
$c\to1$ as $R\to\infty$.

\begin{proposition} 
\label{giapolumegaloR}
For $R$ sufficiently large,
in $\{(u,v):r(u,v)\ge R\}$, we have the bounds 
\label{mprop}
\begin{equation}
\label{makrua1}
|\phi|\le \frac{\hat{C}}r,
\end{equation}
\begin{equation}
\label{makrua2}
|\phi\lambda+\theta|\le \frac{\hat{C}}{r^{\min\{2,\omega\}}},
\end{equation}
\begin{equation}
\label{makrua3}
\hat{c}\le\kappa\le1,
\end{equation}
for some constants $\hat{c},\hat{C}>0$, where
$\hat{c}\to1$ as $R\to\infty$.
\end{proposition}

\noindent\emph{Proof.} 
We will prove the bounds above in several steps. 
Note first that for 
\begin{equation}
\label{blakeia}
R>\max\{r_+,r(p)\},
\end{equation}
the set
$\{r(u,v)\ge R\}$ has structure  depicted by the darker-shaded region of
the Penrose diagram below:
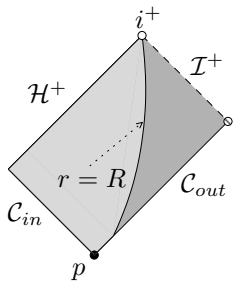
\begin{figure}[ht]
\centering
\input{rgeR.pstex_t}
\end{figure}

\noindent
In particular, if $(u,v)\in\{r\ge R\}$, then 
$[1,u]\times\{v\}\subset\{r\ge R\}$. Choose $R$ sufficiently large
so that $(\ref{pekcokmuhim})$ and $(\ref{blakeia})$ hold.
We estimate, using $(\ref{EE-u})$,
\begin{eqnarray*}
\kappa(u,v)	& = &	\kappa(1,v)e^
								{\int_1^u{\frac1r\zeta^2\frac{1-\mu}\nu
								\frac{1}{1-\mu}(u^*,v)du^*}}\\
					&\ge&	e^{c^{-1}R^{-1}
								\int_1^u{\zeta^2
								\frac{1-\mu}{\nu} du^*}}\\
					&\ge&	\hat{c},
\end{eqnarray*}	
where $\hat{c}=\exp\left(2c^{-1}R^{-1}(M-\varpi_1)\right)$.
Thus, in view also of $(\ref{INITv})$ and $(\ref{II})$, we 
have established $(\ref{makrua3})$.

Note that a bound
\[
|\phi(1,v)|\le\frac{2C}r
\]
holds on $\mathcal{C}_{out}$ 
by $(\ref{byil})$, $(\ref{pioduvato})$, and the triangle inequality.
Integrating the equation
\[
\partial_u\phi=\frac{\zeta}r,
\]
from $\mathcal{C}_{out}$
yields
\begin{eqnarray*}
|\phi(u,v)|	&\le&	\frac{2C}{r(1,v)}+
					\sqrt{\int_1^u\zeta^2\frac{1-\mu}{-\nu}
					(u^*,v)du^*}
					\sqrt{\int_1^u\frac{-\nu}
					{1-\mu}\frac{1}{r^2}(u^*,v)du^*}\\
		&\le&	\frac{\tilde{C}_1}{\sqrt{r}},
\end{eqnarray*}
for
\[
\tilde{C}_1=\frac{2C}{\sqrt{r_1}}+\sqrt{2(M-\varpi_1)}\sqrt{r_1^{-1}c^{-1}},
\]
where the last inequality follows in view of the bound $(\ref{pekcokmuhim})$
and the energy estimate $(\ref{EE-u})$.

Integrating now
$(\ref{toallo})$
from $\mathcal{C}_{out}$, in view of $(\ref{makrua3})$, the bound
\begin{equation}
\label{all1ast}
\left|\varpi-\frac{e^2}r\right|\le e^2r_1^{-1}+M+|\varpi_1|,
\end{equation}
and
$(\ref{pioduvato})$,
we obtain
\[
|\phi\lambda+\theta|(u,v)\le \tilde{C}_2r^{-\min\{\frac32,\omega\}},
\]
where
\[
\tilde{C}_2=C+\frac{2}3\tilde{C}_1(e^2r_1+M+|\varpi_1|).
\]

By $(\ref{blakeia})$, if $r(u,v)\ge R$, then
there exists a unique $v_*\le v$ such that $r(u,v_*)=R$. 
We can thus integrate the equation
\[
\partial_v(r\phi)=\phi\lambda+\theta
\]
on $\{u\}\times[v_*,v]$
 to obtain
\begin{eqnarray*}
|r\phi|(u,v)	&\le&	\frac{\tilde{C}_1R}{\sqrt R}+
\left|\int_{v_*}^v{{\tilde{C}_2}r^{-\min\left\{\frac32,\omega\right\}}dv}\right|\\
		&\le&	\tilde{C}_1\sqrt{R}+\max\left\{\frac{2\tilde{C}_2}{\sqrt{R}},
				\frac{\tilde{C}_2}{\omega-1}R^{1-\omega}\right\}
\end{eqnarray*}
and thus
\begin{equation}
\label{kaluterob}
|\phi|(u,v)\le\frac {\tilde{C}_3}r.
\end{equation}
Integrating again $(\ref{toallo})$
using the bound $(\ref{kaluterob})$ we obtain finally
\[
|\lambda\phi+\theta|(u,v)\le\frac{\tilde{C}_4}{r^{\min\{2,\omega\}}}.
\]
Thus $(\ref{makrua1})$ and $(\ref{makrua2})$ hold with 
$\hat{C}=\max\{\tilde{C}_3,\tilde{C}_4\}$.
$\Box$

\subsection{$r\le R$: Uniform boundedness near $\mathcal{H}^+$}
\label{UB}
We now turn to the region $r \le R$.
We will obtain uniform bounds on $\left|\zn\right|$, and then
extend our bounds to all of $\mathcal{D}$. 

\begin{proposition} 
\label{ZBD}
Let $R$ be as in Proposition \ref{giapolumegaloR}.
In the region $\{(u,v):r(u,v)\le R\}$, we have a bound
\begin{equation}
\label{2A}
\left|\zn\right|\le \hat{C},
\end{equation}
for some $\hat{C}>0$.
\end{proposition}

\noindent\emph{Proof.}
We note that by Assumptions ${\bf B}'$ and ${\bf \Gamma}'$ and compactness, 
it follows that $\sup_{1\le u\le U}\left|\zn(u,v)\right|\le C(v)$.
Thus, the difficulty involves obtaining uniform estimates in $v$.
We have already seen in Section \ref{ecsec} how we hope to proceed. The main difficulty
is to show that the bound $(\ref{3A})$ actually holds.

We first wish to exclude the case where there exists a $v'$ and
a \emph{positive} constant $c'$ such that
\begin{equation}
\label{excludeit}
1-\mu(U,v)\ge c',
\end{equation}
for all $v\ge v'$.

Recall from ${\bf A}'$ the constant $r_+$, and
define
\[
\varpi_+=\sup_{\mathcal{H}^+}\varpi.
\]
Note that by $(\ref{pi-pavwkatw})$, $\varpi_+\le M$. 
Assuming for the time being $(\ref{excludeit})$, 
we can choose 
sufficiently small intervals $[r',r'']$, $[\varpi',\varpi'']$,
containing in their interior $r_+$, $\varpi_+$, respectively,
so that for $r'\le r^*\le r''$, $\varpi'\le \varpi^*\le \varpi''$,
we have $1-\frac{2\varpi^*}{r^*}+\frac{e^2}{(r^*)^2}\ge \frac {c'}2$.

Choose $v'$ such that in addition to $(\ref{excludeit})$, we have
$r(U,v')\ge r'$, and $\varpi(U,v')\ge \varpi'$,
and consider the region 
\[
\mathcal{R}=\{v\ge v'\}\cap\{\varpi'\le\varpi(u,v)\le \varpi''\}
\cap\{r'\le r(u,v)\le r''\}.
\]
It is clear from inequalities $(\ref{vumov})$, $(\ref{lamov})$, 
$(\ref{pumov})$, and $(\ref{pvmov})$
that if $(u,v)\in\mathcal{R}$, then 
\[
\{u\}\times[v',v]\subset\mathcal{R}, [u,U]\times \{v\}\subset\mathcal{R};
\]
moreover, the inequality
\begin{equation}
\label{excludeit2}
1-\mu\ge\frac{c'}2
\end{equation}
holds in $\mathcal{R}$.
Integrating $(\ref{zex})$
from $v=v'$, we obtain:
\begin{eqnarray*}
\left|\zn(u,v)\right|	&\le&	\left(\sqrt{\int_{v'}^v{\frac{1-\mu}\lambda\theta^2
							(u,v^*)dv^*}}
							\sqrt{\int_{v'}^v
							{\frac{\lambda}{1-\mu}
							\frac1{r^2}(u,v^*)dv^*}}\right.\\
				&   & \hbox{\ \ }\left.+\left|\zn(u,v')\right|\right)
							e^{\int_{v'}^v{
							\frac\lambda{1-\mu}
							\frac2{r^2}\left|
							\frac{e^2}r-\varpi\right|
							(u,v^*)dv^*}}\\
				&\le&	\left(\sqrt{2(M-\varpi_1)}
							\sqrt{\int_{v'}^v{\frac{\lambda}
							{1-\mu}
							\frac1{r^2}(u,v^*)dv^*}}
							+\left|\zn(u,v')\right|
							\right)\\
				&   &\hbox{\ \ }\cdot e^{\int_{v'}^v{\frac\lambda{1-\mu}
							\frac2{r^2}\left|
							\frac{e^2}r-\varpi\right|
							(u,v^*)dv^*}}.
\end{eqnarray*}
For $(u,v)\in\mathcal{R}$, it follows 
from $(\ref{excludeit2})$ and the bounds on $\varpi$ and $r$ that
\begin{equation}
\label{stobfR}
\left|\zn\right|\le\tilde{C},
\end{equation}
where
\begin{eqnarray*}
\tilde{C}&=&\left(\sqrt{4c'^{-1}(M-\varpi_1)(r'^{-1}-r''^{-1})}+\sup_u\left|\zn(u,v')\right|\right)\cdot\\
	&&\hbox{\ \ }\cdot
		\exp\left(2c^{-1}(r'^{-1}-r''^{-1})(e^2r'^{-1}+M+|\varpi_1|)\right).
\end{eqnarray*}

Let $\partial\mathcal{R}$ denote the boundary of $\mathcal{R}$ in the topology of
$\mathcal{D}\cap\{v>v'\}$.
It is clear from the properties of $\mathcal{R}$ discussed above
that $\partial\mathcal{R}$ must be a non-empty connected causal curve
terminating 
at $i^+$.
Moreover, we can choose a sequence of points 
\[
p_i=(u_i,v_i)\in\partial\mathcal{R}
\]
such that $(u_i,v_i)\to(\infty,\infty)$,
and 
either
\begin{equation}
\label{3ava1}
r(u_i,v_i)=r''
\end{equation}
or
\begin{equation}
\label{3ava2}
\varpi(u_i,v_i)=\varpi''.
\end{equation}

Assume first $(\ref{3ava1})$. We will show this leads to a contradiction.
Refer to the Penrose diagram below:
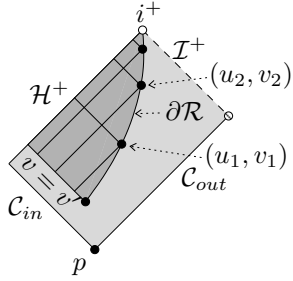
\[
\input{avtifasnN.pstex_t}
\]
Equation $(\ref{3ava1})$ immediately yields
\begin{equation}
\label{aptnmia0}
\int_{u_i}^U{\nu(u,v_i)du}\to r_+-r'',
\end{equation}
as $i\to\infty$.
Since $\{u_i\}\times[v',v_i]\subset\mathcal{R}$,
we have the bound
\[
\int_{v'}^{v_i}{\frac{\lambda}{1-\mu}\frac{2}{r^2}
\left(\frac{e^2}r-\varpi\right)(u_i,v^*)dv^*}<
2c^{-1}(r'^{-1}-r''^{-1})(e^2r'^{-1}+M+|\varpi_1|).
\]
Integrating $(\ref{nqu})$ in $v$, and then integrating in $u$,
we obtain that
\begin{eqnarray}
\label{aptnvalln0}
\nonumber
\int_{u_i}^U{(-\nu)(u,v_i) du}&\le& 
\exp\left(2c^{-1}(r'^{-1}-r''^{-1})(e^2r'^{-1}+M+|\varpi_1|)\right)\cdot\\
&&\hbox{}\cdot\int_{u_i}^U{(-\nu)(u,v') du}
\end{eqnarray}
As $i\to\infty$, the domain of integration $[u_i,U]\times \{v'\}$ of the integral
on the left shrinks to the point
$(U,v')$, and thus the right hand side of $(\ref{aptnvalln0})$ tends to $0$.
But
this contradicts $(\ref{aptnmia0})$. Thus, $(\ref{3ava1})$ is impossible.

We must have then $(\ref{3ava2})$.
But this gives
\begin{eqnarray*}
\varpi''-\varpi_+			&\le&	\varpi(u_i,v_i)-\varpi(\infty,v_i)\\
					&=&	\int_{u_i}^ U
							\frac12\left(\zn\right)^2
							(-\nu)(1-\mu)(u,v_i)du\\
					&\le&	\frac12\tilde{C}^2(1+2|\varpi_1|
							r'^{-1}+e^2r'^{-2})
							\int_{u_i}^U{(-\nu)(u,v_i)du}.
\end{eqnarray*}
Since, by $(\ref{aptnvalln0})$, the right hand side of the above tends to $0$, whereas
the left hand side is a strictly positive number independent of $i$, we immediately
obtain a contradiction.

We have thus excluded $(\ref{excludeit})$ and can assume
in what follows that
there is a sequence
of points $(U,\hat{v}_i)\in\mathcal{H}^+$ 
such that $1-\mu(U,\hat{v}_i)\to0$. But then 
by $(\ref{pvmov})$, it follows that in fact $1-\mu(U,v)\to0$.
In particular:
\[
\varpi_+=\frac{r_+}2+\frac{e^2}{2r_+}.
\]
This implies that $|e|\le\varpi_+$, and either
\begin{equation}
\label{duvato}
r_+=\varpi_++\sqrt{\varpi_+^2-e^2}
\end{equation}
or 
\begin{equation}
\label{aduvato}
r_+=\varpi_+-\sqrt{\varpi_+^2-e^2}.
\end{equation}

The possibilities $(\ref{duvato})$ and $(\ref{aduvato})$
coincide if and only if $e=\varpi_+$. This would give, however, $r_+=|e|$, 
which contradicts Assumption ${\bf \Sigma T}'$. Thus $(\ref{duvato})$ and $(\ref{aduvato})$ are 
necessarily distinct. Note that if $e=0$ we certainly have $(\ref{duvato})$. 
We show now that in general,
$(\ref{aduvato})$ leads to a contradiction.\footnote{If we modify our
initial assumptions by requiring in addition $m\ge0$ on $\mathcal{C}_{in}$, 
then $(\ref{duvato})$ follows immediately from
the inequality $\partial_v m\ge0$ in $\mathcal{D}$.}

In the case ($\ref{aduvato})$, 
we have by a simple computation that
\[
\varpi_+-\frac{e^2}{r_+}<0,
\]
and thus,
it is clear that we can choose $0<r_2<r_+$, $0<\varpi_2<\varpi_+$ close enough to 
$r_+$, $\varpi_+$, respectively, so that
\[
\varpi^*-\frac{e^2}{r^*}<-2\tilde{c},
\]
for some constant $\tilde{c}$ and
for all $r^*\in[r_2,r_+]$, $\varpi^*\in[\varpi_2,\varpi_+]$,
and we can then choose $v'$ sufficiently large so that
$r(U,v')\ge r_2$, $\varpi(U,v')\ge \varpi_2$.

Fix an $\epsilon>0$. By the inequalities
$1-\mu(u,v)\ge0$ and $r(u,v)\ge r_2$, for $v\ge v'$,
it follows that for $(u,v)\in\{r(u,v)\le r_++\epsilon\}\cap\{v\ge v'\}$, 
we have 
$\varpi(u,v)\le\varpi_++\tilde{\epsilon}$, 
where
\[
\tilde{\epsilon}=\epsilon\left(\varpi_++\frac{2e^2(r_++\epsilon)^2\epsilon}{r_+^2r_2^2}\right).
\]
If $\epsilon$ is chosen sufficiently small, we thus have
\begin{equation}
\label{ivo}
\varpi-\frac{e^2}r<-\tilde{c},
\end{equation}
in $\{r\le r_++\epsilon\}\cap\{v\ge v'\}$.

The timelike curve $r=r_++\epsilon$ cannot cross $\mathcal{H}^+$ so must
``terminate'' at $i^+$. Choose a sequence of points $(u_i,v_i)\to i^+$
on $r=r_++\epsilon$, with $v_i\ge v'$. Clearly, we must have
\begin{equation}
\label{aptnmia}
\int_{u_i}^U{(-\nu)(u,v_i) du}\to\epsilon.
\end{equation}
On the other hand, integrating equation
$(\ref{nqu})$,
noting that its right hand side is positive in view of $(\ref{ivo})$,
we obtain that 
\begin{equation}
\label{aptnvalln}
\int_{u_i}^U{(-\nu)(u,v_i) du}\le\int_{u_i}^U{(-\nu)(u,v') du}.
\end{equation}
Since the right hand side of $(\ref{aptnvalln})$ tends to $0$,
$(\ref{aptnvalln})$ contradicts $(\ref{aptnmia})$. Thus $(\ref{aduvato})$ is impossible.

We are thus in the case $(\ref{duvato})$. Again, choosing
$r_3<r_+$, $0<\varpi_3<\varpi_+$ sufficiently close to $r_+$, $\varpi_+$,
respectively,
we can arrange so that
\[
\varpi_3-\frac{e^2}{r_3}=\tilde{c}'>0,
\]
and $r(U,v')\ge r_3$, $\varpi(U,v')\ge\varpi_3$, for some $v'$.
In $[1,U]\times[v',\infty)$, it follows from
$(\ref{vumov})$, $(\ref{lamov})$, $(\ref{pumov})$, and $(\ref{pvmov})$
that
\begin{equation}
\label{3A}
\varpi-\frac{e^2}r\ge\varpi_3-\frac{e^2}{r_3}=\tilde{c}'.
\end{equation}
Thus, for $v\ge v'$, we have, integrating $(\ref{zex})$,
that
\begin{eqnarray*}
\left|\zn(u,v)\right|	&\le&	\left|\int_{v'}^v{\frac\theta{r}(u,v^*)e^{\int_{v^*}^v
							{-\frac{2\kappa}{r^2}
							\left(\varpi-\frac{e^2}r
							\right)(u,\tilde{v})
							d\tilde{v}}}dv^*}
							\right|\\
				&   &	\hbox{\ \ }
							+\left|\zn\right|(u,v')
							e^{\int_{v'}^v{-
							\frac{2\kappa}{r^2}\left(
							\varpi-\frac{e^2}r
							\right)(u,v^*)dv^*}}\\
				&\le&	r_1^{-1}\sqrt{\int_{v'}^v{\kappa^{-1}
							\theta^2}(u,v^*)dv^*}
							\sqrt{\int_{v'}^v{\kappa(u,v^*)
							e^{2\int_{v^*}^v{-
							\frac{2\kappa}{r^2}
							\left(\varpi-\frac{e^2}r
							\right)(u,\tilde{v})
							d\tilde{v}}}}}\\
				&	&	\hbox{\ \ }+\left|\zn\right|(u,v')
							e^{\int_{v'}^v{\kappa
							\frac{2}{r^2}
						\left(\frac{e^2}r-\varpi\right)
							(u,v^*)dv^*}}\\
				&\le& 	r_1^{-1}\sqrt{2(M-\varpi_3)}
						\sqrt{\int_{v'}^v{\kappa(u,v^*) 
							e^{-4\tilde{c}'R^{-2}\int_{v^*}^v
							{\kappa(u,\tilde{v})d\tilde{v}}}}}\\
				&&\hbox{}
							+\sup_{1\le u\le U}
							\left|\zn\right|(u,v').
\end{eqnarray*}
Setting
\[
\hat{C}=\sqrt{2(M-\varpi_3)}\left(2r_1R\sqrt{\tilde{c}}\right)^{-1}+
\sup_{1\le u\le U,1\le v\le v'}\left|\zn\right|(u,v),
\]
we obtain $(\ref{2A})$ throught $\{r\le R\}$.
This completes the proof. $\Box$		

Finally, we have the following
\begin{proposition}
\label{decaystor}
The inequalities  $(\ref{makrua1})$, $(\ref{makrua2})$,
$(\ref{makrua3})$, and $(\ref{2A})$ hold throughout $\mathcal{D}$,
for some constants $\hat{C},\hat{c}>0$.
\end{proposition}

\noindent\emph{Proof.} The bounds $(\ref{makrua1})$, $(\ref{makrua2})$,
and $(\ref{makrua3})$ applied to the region $\{r(u,v)\le R\}$ 
follow immediately by integrating
the equations:
\[
\partial_u\phi=\zn\frac{\nu}r,
\]
\[
\partial_u\theta=-\zn\frac\nu{r}\kappa(1-\mu),
\]
and $(\ref{fdb1})$
to the future 
from the curve $r=R$, in view of the
previous two propositions and the bounds 
$r\ge r_1$, $|1-\mu|\le1+2|\varpi_1|r_1^{-1}+e^2r_1^{-2}$.
The bound $(\ref{2A})$ applied to $r\ge R$ follows by integrating $(\ref{zex})$ to the right
from $r=R$, in view of the
bounds $(\ref{all1ast})$, $(\ref{pekcokmuhim})$.
$\Box$

\subsection{A new retarded time $u$}
\label{veosxrovos}
Before continuing, we will use the results of this section
to renormalize our retarded time coordinate $u$. This renormalization,
although not essential,
will be useful, because it emphasizes a certain symmetry between the $u$ and 
the $v$ directions which will play
an important role later on. It has the disadvantage, however, that it will put
$\mathcal{H}^+$ ``at'' $u=\infty$. As we shall see in this section,
we can easily get around this disadvantage by employing a simple convention.

Choose $R$ as in Proposition \ref{giapolumegaloR}.
In the region $\{r(u,v)\ge R\}$, 
by $(\ref{pekcokmuhim})$ and the bound
$(\ref{all1ast})$, we have a
uniform bound
\begin{equation}
\label{eb}
\left|\int_v^{v'}{\frac{2\kappa}{r^2}\left(\varpi-\frac {e^2}r\right)
(u,v^*)dv^*}\right|\le 2R^{-1}c^{-1}(e^2r_1^{-1}+M+|\varpi_1|),
\end{equation}
for any $v'>v$.
Thus, defining, for $u<U$, the function $\tilde{f}(u)$ by
\begin{eqnarray*}
\tilde{f}(u)&=&
\int_1^{\infty}{\frac{2\kappa}{r^2}\left(\varpi-\frac {e^2}r\right)
(u,v^*)dv^*}\\
&=&
\int_1^{\hat{v}(u)}{\frac{2\kappa}{r^2}\left(\varpi-\frac {e^2}r\right)
(u,v^*)dv^*}\\
&&\hbox{}+
\int_{\hat{v}}^{\infty}{\frac{2\kappa}{r^2}\left(\varpi-\frac {e^2}r\right)
(u,v^*)dv^*},
\end{eqnarray*}
where $\hat{v}(u)$ is defined by $r(u,\hat{v}(u))=R$,
it is clear that $\tilde{f}(u)$ is finite (on account of $(\ref{eb})$ and the 
fact that the domain of integration in the first term on the right is finite,
and so is the integrand) and (by an easy argument) 
continuously differentiable.
Define now a new null coordinate $u^*$ by
\[
u^*(u)=\int_1^u{e^{-\tilde{f}(u)}du}+1.
\]
This clearly defines a regular change of coordinates for $u<U$.
Moreover $\nu^*=\partial_{u^*}r\to-1$ as $v\to\infty$,
and this defines a continuous (in the topology of the Penrose diagram)
extension of $\nu^*$ to $\mathcal{I}^+$. 
We will
say that $u^*$ is retarded time normalized at null infinity $\mathcal{I}^+$.

In what follows, we shall drop the $*$ and refer to our new coordinate as $u$,
and $\nu^*$, etc., as $\nu$.
Note that the bound $(\ref{eb})$ 
implies for this new coordinate that in the region $r\ge R$ we have
\begin{equation}
\label{ebgtv}
|\nu|\le\exp\left(2R^{-1}c^{-1}(e^2r_1^{-1}+M+|\varpi_1|)\right).
\end{equation}

\begin{proposition} The new $u$-coordinate covers $J^{-}(\mathcal{I}^+)\cap\mathcal{D}$
with range $[1,\infty)$.
\end{proposition}

\noindent\emph{Proof.}
Integrating $(\ref{ruqu})$ with $(\ref{ebgtv})$ we have
\[
r(u,v)\ge r(1,v)-\exp\left(2R^{-1}c^{-1}(e^2r_1^{-1}+M+|\varpi_1|)\right)u,
\]
for $(u,v)$ in the region $r\ge R$.
By Assumptions ${\bf A}'$ and ${\bf \Gamma}'$, for $R>r_+$, $r(u,v)=R$ is
a timelike curve which terminates at $i^+$, in the topology of the Penrose
diagram. In particular, there exist
points on this curve with arbitrarily large $v$. But since 
$r(1,v)\to\infty$ as $v\to\infty$, by Assumption ${\bf A}'$, it follows
that for these points $u\to\infty$. $\Box$.

An elaboration of the
method of the proof of the above proposition immediately yields that
null infinity is \emph{complete} in the sense of Christodoulou~\cite{chr:givp}.
See~\cite{md:sssts}.

As remarked above, the disadvantage of this new coordinate system is that
the event horizon $\mathcal{H}^+$ is not covered. We can formally
parametrize it, however, by $(\infty,v)$.
Moreover, since $\zn$, $\lambda$, $\theta$ do not depend on the choice
of $u$-coordinate, one can make sense of 
expressions like $\zn(\infty,0)$,
$\int_0^\infty\lambda(\infty,v)dv$. These can be formally defined by
changing back to the old $u$-coordinate, but we shall save ourselves
this inconvenience. 

With respect to our new coordinates, we have the following bounds:
\begin{proposition}
\label{taG}
Fix $r_0>r_+$. There exist constants $c$ and $G$, depending on $r_0$, such that
the inequalities 
\begin{equation}
\label{1-mPK}
c\le1-\mu\le1,
\end{equation}
\begin{equation}
\label{upo1}
0<G\le\lambda\le 1,
\end{equation}
\begin{equation}
\label{upo2}
-1\le\nu\le-G
\end{equation}
hold in the region $\{r\ge r_0\}\cap\{v\ge v'\}$, where
$v'$ is as defined before $(\ref{3A})$.
As $r_0\to\infty$ then $G$ and $c$ can be chosen arbitrarily close to $1$.
\end{proposition}

\noindent\emph{Proof.} 
The right inequality of $(\ref{1-mPK})$ arises from
\begin{eqnarray*}
1-\mu	& = &	1-\frac{2m}r\\
	& = &	1-\frac{e^2}{r^2}+\frac2r\left(\frac{e^2}r-\varpi\right)\\
	&\le&	1-\frac{e^2}{r^2}-\frac{2\tilde{c}'}r\\
	& < &	1,
\end{eqnarray*}
where for the third inequality we use $(\ref{3A})$.

To show the left inequality of $(\ref{1-mPK})$, as well as $(\ref{upo1})$ and 
$(\ref{upo2})$,
we fix an $\epsilon>0$, and choose
$r_+<r''<r_0$ sufficiently close to $r_+$ so that
\begin{equation}
\label{idiotnta}
1-\frac{2\varpi_*}r+\frac{e^2}{r_*}\le\frac\epsilon{2\hat{C}^2},
\end{equation}
\begin{equation}
\label{idiotnta2}
r_*-\left(\varpi_*-\sqrt{\varpi_*^2-e^2}\right)>\epsilon
\end{equation}
for any
\[
r_*\in[r_+-(r''-r_+),r''], \varpi_*\in[\varpi_+-(r''-r_+)4^{-1}\hat{C}^2,\varpi_+-(r''-r_+)4^{-1}\hat{C}^2].
\]
Let $\tilde{v}$ be such that $r(\infty,\tilde{v})\ge r_+-(r''-r_+)$,
$\varpi(\infty,\tilde{v})\ge \varpi_+-(r''-r_+)4^{-1}\hat{C}^2$,
and consider the region
\[
\mathcal{R}=\{v\ge\tilde{v}\}\cap\{r(u,v)\le r''\}.
\]
Integrating $(\ref{puqu})$ from $(\infty,v)$,
in view of the right inequality of $(\ref{1-mPK})$, and Proposition \ref{decaystor}, 
it follows
that for $(u,v)\in\mathcal{R}$, 
\[
\varpi_+-(r''-r_+)4^{-1}\hat{C}^2\le\varpi(u,v)\le\varpi_+-(r''-r_+)4^{-1}\hat{C}^2
\]
and thus $(\ref{idiotnta})$ holds.
Integrating $(\ref{puqu})$ again, but using $(\ref{idiotnta})$ we obtain 
\begin{equation}
\label{giaparakatw}
|\varpi(u,v)-\varpi_+|\le\frac\epsilon2(r''-r_+).
\end{equation}

Consider the quantity $\varpi+\sqrt{\varpi^2-e^2}-r$
at $(u,v)$ on the curve $r=r''$.
We have
\begin{eqnarray}
\nonumber
\label{ammeta}
\varpi+\sqrt{\varpi^2-e^2}-r	& = &	\varpi-\varpi_++\sqrt{\varpi^2-e^2}-\sqrt{\varpi_+^2-e^2}-r+r_+\\
				&\le&	K|\varpi-\varpi_+|-(r''-r_+)\\
				&\le&	\left(\frac{K\epsilon}2-1\right)(r''-r_+),
\end{eqnarray}
for a positive constant $K$ easily computed,
where for the last inequality we have used $(\ref{giaparakatw})$.
Thus for $r(u,v)=r''$, we have
\begin{eqnarray*}
1-\mu(u,v)	& = &	\frac{1}{r^2}\left(r-\left(\varpi+\sqrt{\varpi^2-e^2}\right)\right)
					\left(r-\left(\varpi-\sqrt{\varpi^2-e^2}\right)\right)\\
		&\ge&	r''^{-2}\left(1-\frac{K\epsilon}2\right)(r''-r_+)\epsilon
\end{eqnarray*}
where for the last inequality we have used $(\ref{idiotnta2})$ and $(\ref{ammeta})$.
Choosing $\epsilon>0$ sufficiently small, the above gives a bound
\begin{equation}
\label{kalo1-m}
1-\mu(u,v)\ge\epsilon'>0
\end{equation}
for all $(u,v)$ on $r=r''$.

From the bound $(\ref{kalo1-m})$ together 
with the lower bound on $\kappa$ from Proposition
\ref{decaystor},
we obtain, on the curve $r=r''$, a bound
\[
\lambda=\kappa(1-\mu)\ge\hat{c}\epsilon'.
\]
But from the equation $(\ref{lqu})$ and $(\ref{3A})$, it follows that $\partial_u\lambda\le0$.
Thus, we have
$(\ref{upo1})$ in $r\ge r''$ as long as $G\le\hat{c}\epsilon'$.
From this, we immediately derive
\[
1-\mu=\kappa^{-1}\lambda\ge\hat{c}\epsilon'
\]
for $r\ge r''$,
which yields the left inequality of $(\ref{1-mPK})$.
Finally, $(\ref{upo2})$ follows after integration of
$(\ref{nqu})$ from $v=\infty$, in view of $(\ref{1-mPK})$. 

For $r_0$ sufficiently large, we have the alternative bound
\[
1-\mu\ge\lambda=\kappa(1-\mu)\ge \hat{c}c,
\]
where $c$ is the constant of $(\ref{pekcokmuhim})$,
and thus
\[
-\nu\ge\exp(2r_0^{-1}\hat{c}M).
\]
It follows that we can take $G\to1$ as $r_0\to\infty$.
$\Box$

\section{A red-shift estimate on long rectangles}
\label{rssection}
We have seen in Section \ref{UB} the importance of the ``red-shift''
effect in obtaining bounds for $\zn$.
In this section, we seek to take advantage of
the exponential factor in the second
term of $(\ref{prwtnmatia})$. 
For this, we shall restrict our 
estimates to a characteristic
rectangle $\mathcal{W}$, 
and seek to formulate assumptions on the boundary of $\mathcal{W}$
that optimize the bounds we can obtain for the two terms
on the right hand side of $(\ref{prwtnmatia})$, and thus, upon integration,
for $\zn$.
These assumptions will involve, on the one hand, the $v$-dimensions of the rectangle
(see $(\ref{mnkos})$),
which will allow us to control the second term of $(\ref{prwtnmatia})$ using
the exponential factor, and on the other hand, an \emph{a priori} bound
on the mass difference (see $(\ref{stageg})$) 
through one of the outgoing null edges of $\partial\mathcal{W}$,
which will allow us to control the first term through an energy estimate argument. 
Both assumptions are formulated with
respect to a constant $D$, and the final bounds in $\mathcal{W}$
depend on this constant
and the value of $r$. In particular, the power law dependence on $r$
in the bound $(\ref{|||})$ is fundamental. 

The rectangle-estimate of this section will be used in the induction
argument of Section \ref{partI}. 
The reader impatient to understand the structure of the argument
can turn there now.


We introduce the notation $\mathcal{D}'=\mathcal{D}\cap\{v\ge v'\}$,
where $v'$ is chosen so that $(\ref{3A})$ holds on $\mathcal{D}'$.
When we say in what follows that a constant $C_3$ depends ``only'' on 
given constants $C_1$ and $C_2$,
this is meant \emph{in addition} to the constants $C$ of bounds $(\ref{byil})$ 
and $(\ref{pioduvato})$,
$\hat{C}$ and $\hat{c}$ of Proposition~\ref{decaystor}, $\tilde{c}'$ of $(\ref{3A})$, and
$r_3$, $M$, and $\varpi_3>0$.

\begin{proposition}
\label{aptovoriz}
Let $r_0$ be a constant $r_0> r_1$, where $r_1$ is the constant of
$(\ref{r_1def})$,
and let $D$ be any constant $D>1$. 
Consider a characteristic rectangle $\mathcal{W}\subset\mathcal{D}'$ with Penrose
diagram depicted below\footnote{Translation: Let $P_1=(u_1,v_1)$, $P_2=(u_1,v_2)$,
$P_3=(u_1,v_3)$, $v_3>v_2>v_1$, let
$u_2$ be such that $r(u_2,v_2)=r_0$, define $\mathcal{W}=[u_2,u_1]
\times[v_1,v_3]$, $\mathcal{A}=[u_2,u_1]\times[v_1,v_2]$, $\mathcal{B}
=[u_2,u_1]\times [v_2,v_3]\cap \{r\le r_0\}$. Note that we allow $u_1=\infty$.}:
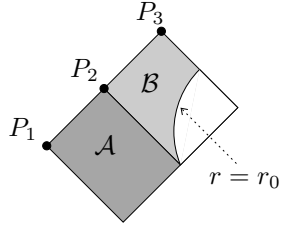
\[
\input{longrect.pstex_t}
\]
Assume 
\begin{equation}
\label{mnkos}
v(P_2)-v(P_1)\ge \frac{r_0^2}{4\tilde{c}'\hat{c}}\log D,
\end{equation}
\begin{equation}
\label{stageg}
\varpi(P_3)-\varpi(P_1)\le D^{-1},
\end{equation} 
and
\begin{equation}
\label{metoe}
\triangle_{\mathcal{A}\cup\mathcal{B}} r\le\epsilon
\end{equation}
for some $\epsilon>0$ satisfying
\begin{equation}
\label{epsepil}
\sqrt{\epsilon}<r_0^{-1}r_1\sqrt{\tilde{c}'\hat{c}},
\end{equation}
and
where for $\mathcal{C}\subset\mathcal{D}$,
we define 
\[
\triangle_{\mathcal{C}} f=\sup_{p_i\in\mathcal{C}}{|f(p_1)-f(p_2)|}.
\]
Then, there exists a positive constant
$\tilde{C}$ depending only on $r_0$ and $\epsilon$ such that
\begin{equation}
\label{apot1}
\left|\zn\right|\le \tilde{C}D^{-\frac12}
\end{equation}
in $\mathcal{B}$,
and
\begin{equation}
\label{apot2}
\triangle_{\mathcal{B}} \varpi\le \tilde{C}D^{-1}.
\end{equation}
\end{proposition}
\noindent\emph{Proof.} 
Consider $(u,v)\in\mathcal{B}$.
We will derive a pointwise bound for $\left|\zn(u,v)\right|$.
Let $P_1$ be given by $(u_1,v_1)$:
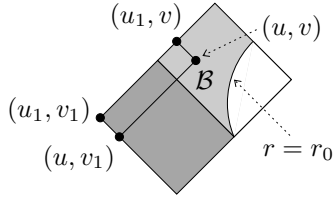
\[
\input{ektim.pstex_t}
\]
Integrating the equation $(\ref{zex})$
we obtain
\[
\zn(u,v)=
\int_{v_1}^v{-\frac{\theta}r(u,\tilde{v})e^{\int_{\tilde{v}}^v
-\frac{2\kappa}{r^2}\left(
\varpi-\frac{e^2}r\right)}d\tilde{v}}
+\zn(u,v_1)e^{\int_{v_1}^v
-\frac{2\kappa}{r^2}\left(
\varpi-\frac{e^2}r\right)(u,\tilde{v})d\tilde{v}}.
\]
Thus,
\begin{eqnarray*}
\left|\zn\right|(u,v)	&\le&	r_1^{-1}\sqrt{\int_{v_1}^v\theta^2(u,\tilde{v})d\tilde{v}}
						\sqrt{\int_{v_1}^v{e^{2\int_{\tilde{v}}^v
						-\frac{2\kappa}{r^2}\left(
						\varpi-\frac{e^2}r\right)}}}
						+\hat{C}e^{-{2\tilde{c}'
						\hat{c}r_0^{-2}}(v-v_1)}\\
			&\le&	r_1^{-1}\sqrt{\int_{v_1}^v\theta^2(u,\tilde{v})d\tilde{v}}
						\sqrt{\int_{v_1}^v
						{e^{-{4\tilde{c}'\hat{c}r_0^{-2}}
						(v-\tilde{v})}d\tilde{v}}}+
					\hat{C}e^{-\frac12\log D}\\
			&\le&	r_1^{-1}\sqrt{\int_{v_1}^v\theta^2(u,\tilde{v})d\tilde{v}}
					\sqrt{({4\tilde{c}'\hat{c}r_0^{-2}})^{-1}
						\left(1-e^{-\sqrt{D}}\right)}\\
			&   &	\hbox{}
						+\hat{C}e^{-
						\log \sqrt{D}}\\
			&\le&	r_0\left(2r_1\sqrt{\tilde{c}'\hat{c}}\right)^{-1}	
						\sqrt{\int_{v_1}^v\theta^2d\tilde{v}}+
					\hat{C}e^{-\sqrt{D}}.
\end{eqnarray*}
Here we have used $(\ref{mnkos})$, $(\ref{3A})$,
and the bound $r\le r_0$ on $\mathcal{B}$,
manifest from the Penrose diagram.
On the other hand, by $(\ref{stageg})$ and $(\ref{kappa9etiko})$, we have that
\begin{eqnarray*}
\frac12\int_{v_1}^v\theta^2(u,\tilde{v})d\tilde{v}	&\le&	
						\varpi(u,v)-\varpi(u,v_1)\\
				&\le&	(\varpi(u,v)-\varpi(u_1,v))+
					(\varpi(u_1,v)-\varpi(u_1,v_1))\\
				&\le&	\int_{u_1}^u{\left(\zn\right)^2(1-\mu)(-\nu) 
						(\tilde{u},v)d\tilde{u}}
							+D^{-1}\\
				&\le&	\epsilon\left(\sup_{u_*\in[u,u_1]}
						\left(\zn\right)^2(u_*,v)\right)
							+D^{-1}.
\end{eqnarray*}
Putting these two estimates together yields
\begin{eqnarray*}
\left|\zn\right|(u,v)	&\le&	r_0\left(2r_1\sqrt{\tilde{c}'\hat{c}}\right)^{-1}\sqrt{\epsilon\sup_{u_*\in[u,u_1]}
						\left(\zn\right)^2(u_*,v)
						+D^{-1}}\\
			&   &	\hbox{\ \ }+\hat{C}
						e^{-\sqrt{D}}\\
			&\le&	r_0\left(2r_1\sqrt{\tilde{c}'\hat{c}}
						\right)^{-1}\left(\sqrt{\epsilon}
						\sup_{u_*\in[u,u_1]}\left|\zn\right|
						(u_*,v)
						+D^{-\frac12}\right)\\
			&   & 	\hbox{\ \ }+\hat{C}e^{-\sqrt{D}}\\
			&\le&	\tilde{\epsilon}\sup_{u_*\in[u,u_1]}\left|\zn\right|
						(u_*,v)
						+C'D^{-\frac12},
\end{eqnarray*}
where
\begin{equation}
\label{tildeorism}
\tilde{\epsilon}=\frac{\sqrt{\epsilon}r_0}{2r_1\sqrt{\tilde{c}'\hat{c}}}.
\end{equation}
Now clearly, the above estimate also applies to
$\sup_{u_*\in[u,u_1]}\left|\zn\right|(u_*,v)$, i.e., we have
in addition
\[
\sup_{u_*\in[u,u_1]}\left|\zn\right|(u_*,v)
\le
\tilde{\epsilon}\sup_{u_*\in[u,u_1]}\left|\zn\right|(u_*,v)
+C'D^{-\frac12},
\]
and thus, in view of $(\ref{epsepil})$,
\begin{equation}
\label{divei}
\sup_{u_*\in[u,u_1]}\left|\zn\right|(u_*,v)
\le\frac{C'D^{-\frac12}}{1-\tilde{\epsilon}}.
\end{equation}
This gives in particular $(\ref{apot1})$.
The bound $(\ref{apot2})$ follows immediately by integrating $(\ref{puqu})$ using
$(\ref{divei})$, together with the bound $(\ref{stageg})$.
$\Box$

In the proof of the next proposition, fix $r_0>r_+$
and recall the constant $G_{r_0}$,
defined in Proposition \ref{taG}.
\begin{proposition}
\label{metobeta}
Let $R\ge r_0>r_+$, consider a characteristic
rectangle $\mathcal{W}\subset\mathcal{D}$
and suppose that on $\mathcal{W}\cap\{r=R\}$ we
have the bound
\[
|\zeta|\le \bar{A},
\]
as well as the bound
\[
\triangle_{\mathcal{W}\cap\{r\le R\}} \varpi\le \bar{A}^2R,
\]
for some positive constant $\bar{A}$.
Then, there is a constant $\tilde{\beta}>1$, depending
only on $r_0$, such that in 
\[
\{R\le r\le\tilde\beta R\}\cap{\mathcal{W}}
\]
we have the bound
\[
|\zeta|\le 3\bar{A},
\]
and
\[
\triangle_{\mathcal{W}\cap\{r\le \tilde{\beta}R\}} \varpi\le 9 \bar{A}^2\tilde{\beta}R.
\]
\end{proposition}

\noindent\emph{Proof.} Let $(u,v)$ be any point in $\mathcal{W}\cap\{R\le r\le
 \tilde\beta R\}$,
and let $(u_1,v_1)$ be as in the Penrose diagram below:
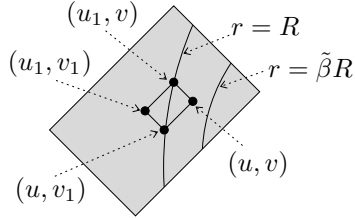
\[
\input{piopera2.pstex_t}
\]
Integrating the equation
\[
\partial_v\zeta=-\frac{\theta}r\nu
\]
on $\{u\}\times[v_1,v]$
 we obtain
\begin{eqnarray*}    
|\zeta|(u,v)	&\le&  \int_{v_1}^v{\left|\frac{\theta\nu}{r}(u,\tilde{v})\right|
					d\tilde{v}}+
					|\zeta(u,v_1)|\\
		&\le&	\sqrt{G^{-1}}\sqrt{\int_{v_1}^v{\theta^2(u,\tilde{v})d\tilde{v}}}
					\sqrt{\int_{v_1}^v{\frac{\lambda}{r^2}(u,
					\tilde{v})d\tilde{v}}}
					+\bar{A}\\
		&\le&	\sqrt{G^{-1}}\sqrt{\int_{v_1}^v{\theta^2(u,\tilde{v})
					d\tilde{v}}}\sqrt{
					\frac{\tilde{\beta}-1}{\tilde{\beta}R}}
					+\bar{A}.
\end{eqnarray*}
On the other hand,
\begin{eqnarray*}
\frac12\int_{v_1}^v{\theta^2(u,\tilde{v})d\tilde{v}}
					&\le&	\varpi(u,v)-\varpi(u_1,v_1)\\
					& = &	(\varpi(u,v)-\varpi(u_1,v))+
								(\varpi(u_1,v)-
								\varpi(u_1,v_1))\\
					&\le& 	\int_u^{u_1}{\left(\zn\right)^2
								(1-\mu)(-\nu)
								(\tilde{u},v)d\tilde{u}}
								+R \bar{A}^2\\
					&\le&	\left(\sup_{u_*\in[u,u_1]}
								{\left(\zn\right)^2}
								(u_*,v)\right)
								\int_u^{u_1}
								{|\nu|(\tilde{u},v)	
								d\tilde{u}}+R\bar{A}^2\\
					&\le&	(\tilde{\beta}-1)R
							\left(\sup_{u_*\in[u,u_1]}
							{\left|\zn\right|}(u_*,v)
								\right)^2+
								R\bar{A}^2.
\end{eqnarray*}
Combining this with the previous, we obtain
\begin{eqnarray}
\nonumber
\label{pavekt}
|\zeta|(u,v)	&\le&	\sqrt{G^{-1}}\sqrt{2(\tilde{\beta}-1)
					R\left(\sup_{u_*\in[u,u_1]}
					{\left|\zn\right|}(u_*,v)\right)^2+2R\bar{C}^2}
					\sqrt{\frac{\tilde{\beta}-1}{\tilde{\beta}R}}
					+\bar{A}\\
		&\le&	\frac12\sup_{u_*\in[u,u_1]}{|\zeta|(u_*,v)}+
					\frac{\bar{A}}2
					+\bar{A},
\end{eqnarray}
where to obtain the first two terms on the right in the latter inequality,
we have chosen $\tilde{\beta}>1$ such that
\begin{equation}
\label{evab}
\tilde{\beta}-1\le\frac{G}{8}.
\end{equation}
Since the estimate $(\ref{pavekt})$ is clearly also valid when one
replaces $|\zeta|$ with 
\[
\sup_{u_*\in[u,u_1]}{|\zeta|},
\]
we obtain:
\[
\sup_{u_*\in[u,u_1]}{|\zeta|(u_*,v)}\le
\frac12\sup_{u_*\in[u,u_1]}{|\zeta|(u_*,v)}
+\frac{\bar{A}}{2}
+\bar{A},
\]
and
thus
\[
\sup_{u_*\in[u,u_1]}{|\zeta|(u_*,v)}
\le
3\bar{A}.
\]

Finally, we have by
the above
\begin{eqnarray*}
\frac12\int_{u}^{u_1}{\left(\zn\right)^2(-\nu)(1-\mu)(\tilde{u},v)d\tilde{u}}&\le&
		\frac12G_1^{-2}9 {\bar{A}}^2R(\tilde{\beta}-1)\\
	&\le&	8\tilde{\beta}\bar{A}^2R,
\end{eqnarray*}
where the second estimate follows provided that $\tilde{\beta}$ is chosen to satisfy,
in addition to $(\ref{evab})$, the inequality
\[
\tilde{\beta}-1\le\frac{16G^2}{9}.
\]
Thus we have
\begin{eqnarray*}
\triangle_{\mathcal{W}\cap\{r\le \tilde{\beta}R\}}\varpi
					&\le&	\triangle_{\mathcal{W}\cap\{r\le 
					  R\}}\varpi+8\tilde{\beta}\bar{A}^2R\\
					&\le&	\tilde{C}^2R+
								8\tilde{\beta}\bar{A}^2R\\
							&\le&	9\bar{A}^2\tilde{\beta}R,
\end{eqnarray*}
and this completes the proof of the Proposition. $\Box$

\begin{proposition}
\label{sautatelos}
Let $r_0>r_+$, and assume we have a characteristic rectangle
$\mathcal{W}$ satisfying the assumptions 
of Proposition \ref{aptovoriz}.
Then, it follows that we have the estimate
\begin{equation}
\label{|||}
\left|\zn\right|\le Hr^\alpha D^{-\frac 12}
\end{equation}
in $\mathcal{W}\setminus{\mathcal{A}}$,
for some constant $H>0$ depending only on $r_0$ and $\epsilon$,
where
\[
\alpha=\frac{\log3}{\log{\tilde{\beta}}}.
\]
and $\tilde{\beta}$ is as in Proposition \ref{metobeta}.
\end{proposition}

\noindent\emph{Proof.} From the result of Proposition \ref{aptovoriz},
the assumptions of Proposition \ref{metobeta} hold on $r=r_0$,
where 
\[
\bar{A} =\max\left\{\tilde{C}D^{-\frac12},\sqrt{\tilde{C}r_0^{-1}}
D^{-\frac12}\right\}.
\] 
Denote $r_k=\tilde{\beta}^kr_0$.
By iterating the result of Proposition \ref{metobeta}, one obtains
for points with $r_{k-1}\le r(u,v)\le r_k$, a bound 
\[
|\zeta (u,v)|\le3^k\max\left\{\tilde{C},\sqrt{\tilde{C}r_0^{-1}}\right\}
D^{-\frac 12}.
\]
From
\[
k=\frac{\log r_k-\log r_0}{\log \tilde\beta},
\]
we have
\begin{eqnarray*}
\left|\zn(u,v)\right|	&\le&	G^{-1}3^{-\log r_0(\log\tilde{\beta})^{-1}}
					\max\left\{\tilde{C},\sqrt{\tilde{C}r_0^{-1}}\right\}3^{(\log\tilde\beta)^{-1}
					\log r_k}D^{-\frac 12}\\
			&\le&	G^{-1}3^{-\log r_0(\log\tilde{\beta})^{-1}}
					\max\left\{\tilde{C},\sqrt{\tilde{C}r_0^{-1}}\right\}
					e^{\log r_k^{(\log 3)(\log\tilde\beta)^{-1}}}
					D^{-\frac 12}\\
			&\le&	G^{-1}3^{-\log r_0(\log\tilde{\beta})^{-1}}
					\max\left\{\tilde{C},\sqrt{\tilde{C}r_0^{-1}}\right\}r_k^{\frac{\log 3}
					{\log\tilde\beta}}D^{-\frac 12}\\
			&\le&	Hr^{\frac{\log 3}
					{\log\tilde\beta}}D^{-\frac 12}.
\end{eqnarray*}
This completes the proof.
$\Box$

\section{A maximum principle}
\label{maxsec}
In this section, we shall return to the region where $r$ is
sufficiently large, and with the help of equation $(\ref{toallo})$, 
prove an estimate for $\theta$, $\phi$, and $\phi\lambda+\theta$
in the interior of a characteristic rectangle $\mathcal{X}$, in terms
of certain bounds on $\partial{\mathcal{X}}$.
This estimate, as we shall see, depends on
what is essentially a \emph{maximum principle}.

As with the previous section, the relevance of the assumptions
of the proposition below will become clear in the context of our
induction argument of Section~\ref{partI}.
\begin{proposition}
\label{sprpro}
Let $R>r_+$ be such that 
\begin{equation}
\label{Rblak1}
G_R^{-1}<2,
\end{equation}
where $G_R$ is the constant of 
Proposition \ref{taG} applied to $r_0=R$, assume
\begin{equation}
\label{Rblak2}
R\ge 8(e^2r_1^{-1}+M),
\end{equation}
and let $A$ and $B$ be two nonnegative constants.
Consider a characteristic
rectangle $\mathcal{X}=[u_1,u_2]\times[v_1,v_2]$,
and assume
\[
r(u_2,v_1)=R,
\]
\begin{equation}
\label{3ast}
|\theta|(u,v_1)\le A
\end{equation}
for all $u\in[u_1,u_2]$, 
\begin{equation}
\label{kaiv1}
|\phi\lambda+\theta|(u_1,v)\le B,
\end{equation}
\begin{equation}
\label{kaiv2}
|\phi|(u_1,v)\le A
\end{equation}
for $v_1\le v\le v_2$,
and 
\begin{equation}
\label{kaiv3}
|\phi|(u,v_2)\le A,
\end{equation}
for all $u_1\le u\le u_2$.
It follows that there exist constants $\sigma_1,\sigma_2>0$, depending
only on $R$,
such that
\begin{equation}
\label{phifragma}
|\phi|(u,v)\le\sigma_1\max\{A,B\},
\end{equation}
\begin{equation}
\label{plfragma}
|\phi\lambda+\theta|(u,v)\le B+\sigma_1r^{-1}(u,v)\max\{A,B\},
\end{equation}
\begin{equation}
\label{thetafragma}
|\theta|(u,v)\le\sigma_1\max\{A,B\}
\end{equation}
for all points $(u,v)\in\mathcal{X}$,
and 
\begin{equation}
\label{TOP-flux}
\varpi(u_2,v_2)-\varpi(u_2,v_1)
\le\sigma_2\left(\max\{A,B\}^2+B^2r(u_2,v_2)\right).
\end{equation}
\end{proposition}
\noindent\emph{Proof.}
We first show that the bound
\begin{equation}
\label{prwtaauto}
|\phi|\le\sigma'\max\{A,B\}
\end{equation}
holds throughout $\mathcal{X}$, for some sufficiently large
constant $\sigma'>1$. 
For this, define the number $u_*\ge u_{1}$ given by
\[
u_*=\sup_{u_1\le u\le u_2}
\left\{|\phi(u',v)|<\sigma'\max\{A,B\}
{\rm\ for\ all\ }v\in[v_1,v_2],u'\in[u_1,u]
\right\}.
\]
(Note that if $\sigma'>1$ then the set referred to on the right is non-empty,
and $u_*$ is in fact necessarily strictly greater that $u_{1}$ by continuity of $\phi$.)

If $u_*$=$u_2$ then clearly $(\ref{prwtaauto})$ holds throughout $\mathcal{X}$,
and we are done.
Otherwise, $u_*<u_2$, and there exists $v_*\in[v_1,v_{2}]$ such that
\begin{equation}
\label{devgivetai}
|\phi(u_*,v_*)|=\sigma'\max\{A,B\}.
\end{equation}

Consider first the possibility that $v_*\in(v_1,v_2)$:
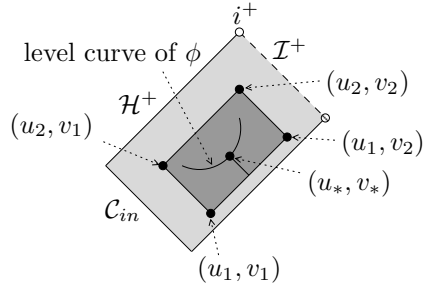
\[
\input{megist.pstex_t}
\]
Since $|\phi(u_*,v)|\le\sigma'\max\{A,B\}$ for all $v\in[v_1,v_2]$,
it follows that
\[
\partial_v\phi(u_*,v_*)=0.
\]
Integrating now $(\ref{toallo})$
on $[u_1,u_*]\times\{v_*\}$,
we obtain that
\begin{eqnarray}
\nonumber
\label{palitob}
|(\phi\lambda+\theta)(u_*,v_*)|	&\le&	B+\frac{2\sigma'(e^2r_1^{-1}+M)}{r(u_*,v_*)}
					\max\{A,B\}\\
				&\le&	\frac12 \sigma'\max\{A,B\}
\end{eqnarray}
where here we are using the fact that $r\ge R$, the condition
$(\ref{Rblak2})$, and, in addition, we are assuming that $\sigma'$ has
been chosen sufficiently
large. 		 	
Now we note that $\theta(u_*,v_*)=r^{-1}\partial_v\phi(u_*,v_*)=0$, so
\begin{eqnarray*}
|\phi|(u_*,v_*)	& = &	\lambda^{-1}|\phi\lambda+\theta|\\
		&\le&	\frac12G^{-1}\sigma'\max\{A,B\}\\
		& < &	 \sigma'\max\{A,B\},
\end{eqnarray*}
where, for the last inequality, we use that $R$ satisfies $(\ref{Rblak1})$.
But this contradicts $(\ref{devgivetai})$.

Since $v_*=v_2$ would contradict $(\ref{kaiv3})$ for $\sigma'>1$,
we clearly must have $v^*=v_1$.
Integrating $(\ref{toallo})$
as above, 
we again obtain the bound $(\ref{palitob})$ for $|\phi\lambda+\theta|(u_*,v_i)$. Thus,
we have
\begin{eqnarray*}
|\phi(u_*,v_1)|	&\le&	\lambda^{-1}|\phi\lambda+\theta|(u_*,v_1)+
					\lambda^{-1}|\theta|(u_*,v_1)\\
		&\le&	G^{-1}\frac12 \sigma'\max\{A,B\}
					+G^{-1}A\\
		& < &	 \sigma'\max\{A,B\},
\end{eqnarray*}
which again contradicts $(\ref{devgivetai})$. 
(Note that to derive the second inequality above we use
$(\ref{3ast})$ and to derive the third we require $\sigma'$ to be sufficiently large.)
Thus we have established $(\ref{prwtaauto})$.

Integrating $(\ref{toallo})$ from $u=u_1$, we obtain
\begin{equation}
\label{varimiz}
|\phi\lambda+\theta|(u,v)	\le	B+2r^{-1}(e^2r_1^{-1}+M)\sigma'\max\{A,B\}.
\end{equation}
From $(\ref{varimiz})$, $(\ref{plfragma})$ follows for sufficiently large $\sigma_1$,
as does $(\ref{thetafragma})$, applying the triangle inequality,
$(\ref{prwtaauto})$ and the bound
$\lambda\le G^{-1}\le 2$. 

Integrating the equation
\begin{equation}
\label{e3is}
\partial_v(r\phi)=\phi\lambda+\theta
\end{equation}
along $\{u_2\}\times[v_1,v_2]$, we obtain
\begin{eqnarray}
\label{prwtoprob}
\nonumber
|r\phi(u_2,v)|	&\le&	R\sigma_1\max\{A,B\}+BG_1^{-1}(r-R)\\
		&   &	\hbox{\ \ }+
					2M\log(r/R)G_1^{-1}\sigma'\max\{A,B\}\\
		&\le&	 \sigma''\max\{A,B\}(\log{r})+Br
\end{eqnarray}
and
thus
\begin{equation}
\label{privtosuv}
|\phi(u_2,v)|\le  \sigma''\left(\max\{A,B\}r^{-1}\log r+B\right).
\end{equation}
In view now of $(\ref{varimiz})$ and $(\ref{privtosuv})$ it follows
by the triangle inequality and $\lambda\le G^{-1}$ that
\begin{equation}
\label{suvepagetai}
|\theta(u_2,v)|\le \sigma'''\left(\max\{A,B\}r^{-1}\log r+B\right).
\end{equation}
From $(\ref{suvepagetai})$, $(\ref{pvqu})$, and $(\ref{makrua3})$,
 we obtain finally the energy 
estimate
\begin{eqnarray*}
\varpi(u_2,v)-\varpi(u_2,v_1)	&\le&	\frac12\hat{c}^{-1}\int_{v_i}^v{\theta^2(u_2,v^*)dv^*}\\
				&\le&	\frac12\hat{c}^{-1}G_1^{-1}\sigma'''
						\left(\max\{A,B\}^2\right.\\
				&   &	\hbox{\ \ }\left.\cdot
						\int_{v_1}^v{r^{-2}\log^2 r(u_1,v^*)\lambda dv^*}
						+B^2r(u_2,v)\right)\\
				&\le&	\sigma_2\left(\max\{A,B\}^2+
						B^2r(u_2,v)\right),
\end{eqnarray*}
for $\sigma_2$ suitably large. This yields $(\ref{TOP-flux})$.
$\Box$

\section{Two more estimates}
\label{akoma2sec}
\begin{proposition}
\label{piokovtaekei}
Let $r_*>r_0$ be given and consider a 
characteristic rectangle $\mathcal{Y}=[u_1,u_2]\times[v_1,v_2]\subset\mathcal{D}'$.
Assume
that for all $u\in[u_1,u_2]\cap\{r(u,v_2)\ge r_*\}$,
\begin{equation}
\label{alavaf}
|\phi|(u,v_2)\le A,
\end{equation}
and
\begin{equation}
\label{ekt-ev}
\varpi(u_1,v_2)-\varpi(u_2,v_1)\le A^2.
\end{equation}
It follows that there exists a constant $\sigma$ depending only on $r_*$,
such that for $(u,v)\in\mathcal{Y}\cap\{r\ge r_*\}$
\begin{equation}
\label{allottet}
|\phi|(u,v)\le\sigma A.
\end{equation}
Assuming in addition that 
\begin{equation}
\label{prosupo}
\left|\zn\right|(u,v_1)\le A,
\end{equation}
for all $u\in[u_1,u_2]\cap\{r(u,v_1)\le r_*\}$,
it follows that
\begin{equation}
\label{pros9apo1}
\left|\zn\right|(u,v)\le\sigma A,
\end{equation}
\begin{equation}
\label{pros9apo2}
|\phi|(u,v)\le\sigma A,
\end{equation}
in $\mathcal{Y}\cap\{r\le r_*\}$.
\end{proposition}

\noindent\emph{Proof.}
It follows from $(\ref{pumov})$, $(\ref{pvmov})$, and $(\ref{alavaf})$
that the energy estimates
\begin{equation}
\label{EnEs}
\frac12\int_{v_1}^{v_2}{\theta^2(u,v)\kappa^{-1} dv}\le A^2,
\end{equation}
for $u_1\le u\le u_2$, and
\[
\frac12\int_{u_1}^{u_2}{\left(\zn\right)^2(1-\mu)(-\nu)(u,v) du}\le A^2,
\]
for $v_1\le v\le v_2$,
hold.

Integrating the equation
\[
\partial_v\phi=\frac\theta{r}
\]
from $v=v_{2}$, i.e.~from right to left,
using the Schwarz inequality and the bound $(\ref{alavaf})$,
we obtain that for $(u,v)\in\{r\ge r_*\}$
\begin{eqnarray*}
|\phi(u,v)|	&\le&	|\phi(u,v_2)|+\sqrt{\int_v^{v_2}\theta^2
						\frac{1-\mu}{\lambda}(u,\tilde{v})
							d\tilde{v}}
					\sqrt{\int_v^{v_2}\frac{\lambda}{1-\mu}
						\frac{1}{r^2}(u,\tilde{v})d\tilde{v}}\\
		&\le&	A+
					\sqrt{2}Ac^{-1}
					\sqrt{\int_v^{v_2}\frac{\lambda}{r^2}(u,
						\tilde{v})d\tilde{v}}\\
		&\le&	\sigma A,
\end{eqnarray*}
and this yields $(\ref{allottet})$. 

Now consider the additional assumption $(\ref{prosupo})$. For $r\le r_*$,
we will use the ``red-shift'' technique. Integrating $(\ref{zex})$, we obtain
\begin{eqnarray*}
\left|\zn\right|(u,v)	&\le&	\sqrt{\int_{v_1}^{v}\theta^2\kappa^{-1}(u,\tilde{v})
						d\tilde{v}}\sqrt
					{\int_{v_1}^v{
					\frac{\kappa}{r^2}e^{-2\hat{c}\tilde{c}'r_*^{-2}
					(v-\tilde{v})}d\tilde{v}}}\\
			&   &	\hbox{\ \ }+\left|\zn\right|(u,v_1)\\
			&\le&	\sigma'\sqrt{A^2}+A\\
			&\le&	\sigma A
\end{eqnarray*}	
for sufficiently large $\sigma>1$. 
Thus we have $(\ref{pros9apo1})$. Integrating the equation
$\partial_u\phi=\zn\frac{\nu}r$ from $r=r_*$,
we immediately obtain $(\ref{pros9apo2})$.
$\Box$

\begin{proposition}
\label{arketamegaln}
Suppose $R$ is as in Proposition \ref{sprpro}, 
and let $r(u_1,v_1)=R$.
Consider a characteristic rectangle $\mathcal{Z}=[u_1,u_2]\times[v_1,v_2]\subset
\mathcal{D}$.
Suppose that 
\[
|\phi|\le A
\]
on $\mathcal{Z}\cap\{r=R\}$, and 
\[
|\phi|(u,v_2)\le A,
\]
\[
|\phi\lambda+\theta|(u_1,v)\le B,
\]
for $u_1\le u\le u_2$ and $v_1\le v\le v_2$, respectively. 
Then there exists a constant $\sigma>1$ depending only on
$R$ such that
\begin{equation}
\label{evakap}
|\phi|(u,v)\le\sigma\left(\max\{A,B\}r^{-1}R+B\right),
\end{equation}
\begin{equation}
\label{evakap2}
|\phi\lambda+\theta|(u,v)\le\sigma\left(B+r^{-2}\log r\max\{A,B\}\right),
\end{equation}
for all $(u,v)\in\mathcal{Z}\cap\{r\ge R\}$.
\end{proposition}

\noindent\emph{Proof.}
We first note that by integrating the equation $\partial_v(r\phi)=\phi\lambda+\theta$
along $\{u_1\}\times[v_1,v]$, we obtain
\[
|r\phi(u_1,v)|\le RA+G^{-1}(r(u_1,v)-R)B,
\]
and thus
\[
|\phi(u_1,v)|\le \sigma'B+A
\]
for $\sigma'$ sufficiently large.
Applying the argument of Proposition \ref{sprpro}
we immediately obtain
\begin{equation}
\label{GT}
|\phi(u,v)|\le \sigma''\max\{A,B\}
\end{equation}
for all $(u,v)\in\mathcal{Z}\cap\{r\ge R\}$.

Integrating now $(\ref{toallo})$ with the bound $(\ref{GT})$, 
we obtain
\[
|\phi\lambda+\theta|\le B+2\sigma''M\max\{A,B\}r^{-1}.
\]
Integrating  $(\ref{e3is})$ with the above bound, from $r=R$, we obtain
\[
|r\phi|\le RA+B(r-R)+2MG^{-1}\sigma''\max\{A,B\}\log (r/R),
\]
so 
\[
|\phi|\le B+\sigma'''\max\{A,B\}r^{-1}\log r.
\]
Once again integrating $(\ref{toallo})$, we obtain
\[
|\phi\lambda+\theta|\le B+Br^{-1}+2MG^{-1}\sigma'''r^{-2}\log (r/R)\max\{A,B\}.
\]
This yields $(\ref{evakap2})$ for sufficiently large $\sigma$.
Integrating yet again $(\ref{e3is})$ yields
\[
|r\phi|\le AR+G^{-1}B(r-R)+G^{-1}B\log r+\sigma''''\max\{A,B\}.
\]
Dividing by $r$, we finally obtain
\[
|\phi|\le \sigma(B+r^{-1}R\max\{A,B\}),
\]
for large enough $\sigma$.
This completes the proof. $\Box$

\section{Decay}
\label{partI}
In view of the estimates of the previous 3 sections,
all the tools necessary to extract decay for $\phi$ and $\theta$ are in 
place. 
The glue binding these estimates together
is the global geometry of $\mathcal{D}$: the relation between the $v$-length
of suitably positioned characteristic rectangles and the $r$-difference between their
vertices. 

Before continuing, a brief word about our argument is in order.
We will obtain decay by induction, each step improving the rate 
achieved in the previous.
The $0$'th step corresponds to the uniform boundedness that we have already proven.
Note that at this $0$'th step,
we also have a bound for the total energy flux on $\mathcal{H}^+$, namely
$\varpi_+-\varpi_1$,
and this is a (much) better bound than the bound obtained for 
the energy flux
by plugging in our pointwise bound for $\theta$, namely 
$\int_1^\infty{\hat{c}^{-1}\hat{r_0}^{-2}\hat{C}^2}=\infty$!
The catalyst that allows the improvement at each step
is the pigeon-hole principle applied to the energy flux on $\mathcal{H}^+$.
The principle is quite simple: If the flux is small on a long interval,
then every so often it is even smaller. 
We shall quantify this in the course of the proof by considering the event horizon
decomposed essentially dyadically\footnote{i.e.~such that the length of the segment
is on the order of the $v$-coordinate of the endpoint} 
into segments, and applying the pigeon hole principle in
\emph{each} dyadic interval to find a subinterval, of length $\sim v^p$ where
$0<p<1$ and $v$ is the $v$-coordinate of the endpoint, 
such that the flux in this subinterval is smaller than the flux
in the original by a factor
of $v^{-1+p}$. Starting from these ``good'' subintervals, we will construct
rectangles $\mathcal{W}$, $\mathcal{X}$, $\mathcal{Y}$, $\mathcal{Z}$, and,
using the estimates proven previously,
we will ``spread'' the better factor to all of $\mathcal{D}$,
leaving us in particular with a pointwise bound for $\theta$ on $\mathcal{H}^+$,
and a \emph{better} bound on the flux than that which arises from integrating
this pointwise bound on $\mathcal{H}^+$. We apply the pigeon-hole principle
again and continue the induction.

We proceed now with the argument. Because it is complicated to 
keep track of different regions at the same time, we have split the
argument into two main parts.

\subsection{The induction: Part I}
\label{PartIa}
Define
\[
\omega_*=\min\{\omega,2\}
\]
The result of this section is the following
\begin{theorem}
\label{3/2}
There exist constants $C_{\omega_*-\frac12}>0$, 
$\delta_{\omega_*-\frac12}>0$, such that
in the region $r\le\delta_{\omega_*-\frac12} v$, we have
\begin{equation}
\label{+}
|\phi|\le C_{\omega_*-\frac12}\left(v^{-\left(\omega_*-\frac12\right)}r^{-1}+
v^{-\omega_*}\right),
\end{equation}
\begin{equation}
\label{+++}
|\theta|\le C_{\omega_*-\frac12}\left(v^{-\left(\omega_*-\frac12\right)}r^{-1}
+v^{-\omega_*}\right),
\end{equation}
and in addition: 
\[
\varpi_+-\varpi(\infty,v)\le C_{\omega_*-\frac12}v^{-2\left(\omega_*-\frac12\right)}.
\]
\end{theorem}

\noindent\emph{Proof.}
A series of real numbers $w_k$ will be defined inductively.
Given $w_k$, define the statement $S_k$ as follows:
\begin{enumerate}
\item[($S_k$)]
There exist
constants $\bar{C}_k, \delta_k>0$, such that
for $r\le\delta_k v$,
\begin{equation}
\label{22y}
|\phi|\le \bar{C}_k\left(v^{-w_k}r^{-1}+ v^{-\omega_*}\right),
\end{equation}
\begin{equation}
\label{33y}
|\theta|\le \bar{C}_k\left(v^{-w_k}r^{-1}+v^{-\omega_*}\right),
\end{equation}
\begin{equation}
\label{11y}
\varpi_+-\varpi(\infty,v)\le\bar{C}_kv^{-2w_k}.
\end{equation}
\end{enumerate}
Choose a real number $p$ satisfying
\begin{equation}
\label{pepilogn}
0<p<\frac{1}{2(\alpha+1)+1}
\end{equation}
where $\alpha$ is the constant of Proposition~\ref{sautatelos}.
The 
inductive step we shall prove is
\begin{proposition}
\label{inductstep}
$S_k$ implies $S_{k+1}$, where $w_{k+1}$
is defined by
\begin{equation}
\label{indef}
w_{k+1}=\min\left\{w_k+p,\omega_*-\frac12\right\}.
\end{equation}
\end{proposition}
Assume for now Proposition \ref{inductstep} is true.
Setting $w_0=0$, Proposition \ref{decaystor} implies that
$S_0$ is true. It then
follows by induction
that $S_k$ is true for all $k\ge0$ where $w_{k}$ is defined inductively
by $(\ref{indef})$. 
Since $w_k=\omega_*$ for $k=[\frac{3}{2p}]+1$,
the theorem follows. $\Box$
\vskip1pc
\noindent\emph{Proof of Proposition \ref{inductstep}.}
Fix a sufficiently large $R$ satisfying Proposition \ref{taG}
with constant $G_R$. Set $h_{k+1}>1$ such that
\begin{equation}
\label{hepi}
4(h_{k+1}^2-1)(1+G_R^{-1})h_{k+1}\le \delta_k.
\end{equation}
We have
\begin{lemma}
Assume $S_k$ holds for some $w_k>0$.
There exist a sequence of
points $(\infty,v_{i,k+1})\in\mathcal{H}^+$ and
$(\infty,\tilde{v}_{i,k+1})\in\mathcal{H}^+$,
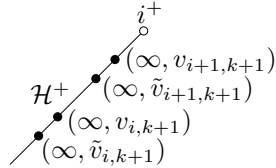
\[
\input{snmeia2.pstex_t}
\]
such that
\begin{equation}
\label{polla}
\frac1{4h_{k+1}^p}(h_{k+1}-1)v_{i,k+1}^p\le v_{i,k+1}-\tilde{v}_{i,k+1}
\le (h_{k+1}-1)v_{i,k+1}^p,
\end{equation}
\begin{equation}
\label{*?}
\frac{1+h_{k+1}}2v_{i,k+1}\le v_{i+1,k+1}\le h_{k+1}^2v_{i,k+1},
\end{equation}
\begin{equation}
\label{polla2}
\varpi(\infty,v_{i,k+1})-\varpi(\infty,\tilde{v}_{i,k+1})
\le 2h_{k+1}^{2w_k+1-p}\bar{C}_kv_{i,k+1}^{-2w_k-1+p},
\end{equation}
and 
\begin{equation}
\label{kalogia9}
|\theta(\infty, v_{i,k+1})|\le 
8(h_{k+1}-1)^{-1}r_+(r_1^{-1}+1)h_{k+1}^{p+w_k}\bar{C}_k
v_{i,k+1}^{-w_k-p}.
\end{equation}
\end{lemma}

\noindent\emph{Proof.} 
Set $\hat{v}_{i,k+1}=h_{k+1}^i$.
Consider the points $(\infty,\hat{v}_{i,k+1})$.
Bisect the segment $\{\infty\}\times[\hat{v}_{i,k+1},\hat{v}_{i+1,k+1}]$, 
and partition
the ``second'' half 
$\{\infty\}\times[\frac12(\hat{v}_{i,k+1}+\hat{v}_{i+1,k+1}),\hat{v}_{i+1,k+1}]$
into $[\hat{v}_{i,k+1}^{1-p}]$ subsegments of 
equal length $L_{i,k+1}$. We have\footnote{Note that $v^{1-p}_{i,k+1}$
is not an integer.}
\begin{equation}
\label{posotntaedw}
[\hat{v}^{1-p}_{i,k+1}]\ge\frac12\hat{v}^{1-p}_{i,k+1},
\end{equation}
\begin{equation}
\label{tomnkostou}
(h_{k+1}-1)\hat{v}_{i,k+1}^p\ge L_{i,k+1}\ge\frac12(h_{k+1}-1)\hat{v}_{i,k+1}^p.
\end{equation} 
Since--on account of $(\ref{11y})$--we have
\[
\varpi(\infty,\hat{v}_{i+1,k+1})-\varpi(\infty,\hat{v}_{i,k+1})\le 
\bar{C}_k\hat{v}_{i,k+1}^{-2w_k},
\]
it follows by $(\ref{posotntaedw})$ and the pigeon-hole
principle that for each $i$, on one of the aforementioned 
subsegments of length $L_{i,k+1}$,
call it $\{\infty\}\times[\tilde{v}_{i,k+1},v''_{i,k+1}]$,
we have
\begin{equation}
\label{peristeri}
\varpi(\infty,v''_{i,k+1})-\varpi(\infty,\tilde{v}_{i,k+1})
\le 2\bar{C}_k\hat{v}_{i,k+1}^{-2w_k-1+p}.
\end{equation}
Now bisect $\{\infty\}\times[\tilde{v}_{i,k+1},v''_{i,k+1}]$,
and consider again the ``second half'', i.e., the segment
\[
\{\infty\}\times[v'_{i,k+1},v''_{i,k+1}],
\]
where 
\[
v'_{i,k+1}=\tilde{v}_{i,k+1}+\frac12(v''_{i,k+1}-\tilde{v}_{i,k+1}).
\]
From $(\ref{22y})$ and the identity
$\partial_v\phi=r^{-1}\theta$,
we obtain
\[
2\bar{C}_k(r_1^{-1}+1)
\hat{v}_{i,k+1}^{-w_k}\ge\left|\int_{v'_{i,k+1}}
^{v''_{i,k+1}}
{\frac{\theta}r(\infty,v)dv}\right|.
\]
On the other hand, by $(\ref{tomnkostou})$ and the definition
of $v'_{i,k+1}$, we have
\[
v''_{i,k+1}-v'_{i,k+1}\ge\frac14(h_{k+1}-1)\hat{v}_{i,k+1}^p,
\]
so there must be a point in 
$\{\infty\}\times[v'_{i,k+1},v''_{i,k+1}]$, 
call it $(\infty,v_{i,k+1})$:
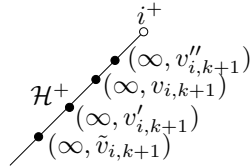
\[
\input{terata2a.pstex_t}
\]
%
where 
\begin{equation}
\label{kalo9nta}
|\theta|(\infty,v_{i,k+1})	\le 8(h_{k+1}-1)^{-1}r_+(r_1^{-1}+1)\bar{C}_k
\hat{v}_{i,k+1}^{-w_k}\hat{v}_{i,k+1}^{-p}.
\end{equation}

By their construction,
the points $v_{i,k+1}$ and $\tilde{v}_{i,k+1}$ satisfy $(\ref{polla})$ and
$(\ref{*?})$. Inequality $(\ref{kalo9nta})$ together with 
the inequlity
\begin{equation}
\label{sxesn}
v_{i,k+1}\le h_{k+1}^2\hat{v}_{k+1}
\end{equation}
yields $(\ref{kalogia9})$. Similarly, we have by $(\ref{peristeri})$
\begin{eqnarray*}
\varpi(\infty,v_{i,k+1})-\varpi(\infty,\tilde{v}_{i,k+1})&\le&
\varpi(\infty,v''_{i,k+1})-\varpi(\infty,\tilde{v}_{i,k+1})\\
&\le&2\bar{C}_k\hat{v}_{i,k+1}^{-2w_k-1+p},
\end{eqnarray*}
which immediately yields $(\ref{polla2})$, in view of $(\ref{sxesn})$.
$\Box$

Fix now $r_0>r_+$ and $\epsilon$ so that $(\ref{epsepil})$ of 
Proposition \ref{aptovoriz} holds, with $r_0-r_+<\frac\epsilon2$.
For each $i$, bisect the segment
$\{\infty\}\times[\tilde{v}_{i,k+1},v_{i,k+1}]$, 
and consider the constant-$v$ null segment passing through the midpoint.
Denoting by $q_{i,k+1}$ the point on this ingoing segment where $r=r_0$,
form the characteristic rectangle $\mathcal{W}_i=[u(q_{i,k+1}),\infty]\times
[\tilde{v}_{i,k+1},v_{i,k+1}]$:
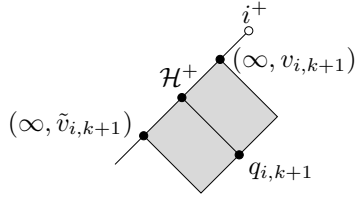
\[
\input{terata3a.pstex_t}
\]
Now let $I_{k+1}$ be a positive integer so that
\begin{equation}
\label{sta9eres}
\tilde{v}_{I_{k+1},k+1}\ge v',
\end{equation}
\begin{equation}
\label{sta9eres2}
r_0-r(\infty,\tilde{v}_{I_{k+1},k+1})\le\epsilon,
\end{equation}
\begin{equation}
\label{Dorismos}
D\equiv(2\bar{C}_k)^{-1}h_{k+1}^{-2w_k-1+p}v_{I_{k+1},k+1}^{2w_k+1-p}>1,
\end{equation}
\begin{eqnarray}
\label{sta9eres3}
\nonumber
\log v_{I_{k+1},k+1}&\le&\frac1{(2w_k+1-p)}
\left(\frac{(h_{k+1}-1)v_{I_{k+1},k+1}^p}{8h^{p}_{k+1}}
\right.\\
&&\hbox{}-\left.\frac{r_0^2}{4\tilde{c}'\hat{c}}\log(\bar{C}_k2h_{k+1}(2w_k+1-p))\right).
\end{eqnarray}
Assume always in what follows that $i\ge I_{k+1}$. By 
$(\ref{sta9eres})$, it follows that
$\mathcal{W}_i\subset \mathcal{D}'$, by
$(\ref{sta9eres2})$, it follows that $(\ref{metoe})$ holds,  
by $(\ref{sta9eres3})$ and the
inequality 
\begin{equation}
\label{togegov}
v(q_i)-\tilde{v}_{i,k+1}\ge \frac1{4(h_{k+1}-1)}\hat{v}^p_{i,k+1}
\ge \frac1{4(h_{k+1}-1)h_{k+1}^{2+p}}v_{i,k+1}^p,
\end{equation}
it follows that $(\ref{mnkos})$ holds with $D$ defined by $(\ref{Dorismos})$,
and finally, by $(\ref{polla2})$, it follows that $(\ref{stageg})$ holds.
Thus,
$\mathcal{W}_i$ satisfies the assumptions of
Proposition \ref{sautatelos}.
For $u\ge u(q_i)$, this gives the bound
\begin{equation}
\label{privtovbela}
\left|\zn\right|(u,v_{i,k+1})\le \sqrt{2h_{k+1}^{2w_k+1-p}\bar{C}_k}
Hr^\alpha v_{i,k+1}^{-w_k-\frac12+\frac p2},
\end{equation}
where $H$ is the constant of $(\ref{|||})$. 
Integrating the equation
\[
\partial_u\theta=-\zn\frac{\nu}r\lambda
\]
along $[u,\infty]\times\{v_{i,k+1}\}$,
we obtain from $(\ref{kalogia9})$ and $(\ref{privtovbela})$ that
\begin{eqnarray}
\label{belas}
\nonumber
|\theta|(u,v_{i,k+1})&\le&\alpha^{-1}
\sqrt{2h_{k+1}^{2w_k+1-p}\bar{C}_k}Hr^\alpha v_{i,k+1}^{-w_k-\frac12+\frac p2}
\\
&&\hbox{}+8(h_{k+1}-1)^{-1}r_+(r_1^{-1}+1)h_{k+1}^{p+w_k}\bar{C}_k
v_{i,k+1}^{-w_k-p},
\end{eqnarray}
for all $u\ge u(q_i)$.

Now we note a fundamental fact:
By our lower bound $|\lambda|\ge G_{r_0}$ of $(\ref{upo1})$ 
in $r\ge r_0$, the inequality $(\ref{togegov})$
and the equation
\[
v_{i,k+1}-v(q_i)=v(q_i)-\tilde{v}_{i,k+1},
\]
 it follows that 
for 
\[
\tilde{\delta}_{k+1}=\frac14(h_{k+1}-1)h_{k+1}^{2+p}G_{r_0}
\]
we have
\[
\{u:r(u,v_{i,k+1})\le\tilde{\delta}_{k+1}v_{i,k+1}^p\}\subset \{u\ge u(q_i)\}.
\]
In other words, our rectangles $\mathcal{W}_i$ ``penetrate'' 
into the region where $r$
is on the order of $v$ to the positive power $p$.
In particular, we have now that our bound $(\ref{belas})$ holds for
all $(u,v_{i,k+1})$ with $r(u,v_{i,k+1})\le\tilde{\delta}_{k+1}v_{i,k+1}^p$.

To lighten the notation, let us denote in what follows $v_{i, k+1}$ by $v_i$.
(We shall retain the $k$ subscripts on all other constants, however.)
Recall the constant $R$ we have fixed at the beginning of the proof
of this proposition.
Let $\bar{I}_{k+1}\ge I_{k+1}$ be such that there exists
a $u_{\bar{I}_{k+1}}\ge 1$ with $r(u_{\bar{I}_{k+1}},v_{\bar{I}_{k+1}})=R$.
For $i\ge\bar{I}_{k+1}$, again there exists
a unique $u_i$ such that $r(u_i,v_i)=R$. In what follows,
let us restrict to
$i\ge\bar{I}_{k+1}+1$.

Form the sequence of characteristic
rectangles
\[
\mathcal{X}_i=[u_{i-1},u_{i}]\times[v_{i},v_{i+1}]
\]
as depicted in the Penrose diagram below:
\[
\input{Xor9og.pstex_t}
\]
It is clear that $r\ge R$ on $\bigcup_i\mathcal{X}_i$. 
Since 
\[
r(u_{i-1},v_i)-r(u_i,v_i)
=r(u_{i-1},v_i)-r(u_{i-1},v_{i-1}),
\]
it follows
from 
Proposition \ref{taG}
that 
\[
u_i-u_{i-1}\le G_R^{-1}(v_i-v_{i-1})\le G_R^{-2}(u_i-u_{i-1}).
\]
Appealing to $(\ref{*?})$, we obtain  that 
\begin{equation}
\label{rsavv}
\frac{G_R^2(h_{k+1}-1)}2v_i\le  r\le (h_{k+1}^2-1)(1+G_R^{-1})v_i+R
\le 2(h_{k+1}^2-1)(1+G_R^{-1})v_i
\end{equation}
along $\{u_{i-1}\}\times[v_i,v_{i+1}]$ and
$[u_{i-1},u_i]\times\{v_{i+1}\}$, where to show the final inequality
we require that $i\ge\hat{I}_{k+1}\ge\bar{I}_{k+1}+1$, 
for some constant $\hat{I}_{k+1}$. In what follows we shall 
assume now $i\ge\hat{I}_{k+1}$. 

By $(\ref{rsavv})$ and $(\ref{hepi})$, we have
\begin{equation}
\label{eivaimesa}
\mathcal{X}_i\subset\{r\le\delta_k v\}.
\end{equation}
Thus, $(\ref{22y})$ and $(\ref{33y})$ apply in $\mathcal{X}_i$. 
Our first task is to obtain a bound for $\theta$ on $[u_{i-1},u_i]\times\{v_i\}$
by interpolating $(\ref{33y})$ with $(\ref{belas})$.\footnote{Remember
that $(\ref{belas})$ does \emph{not} apply on the whole segment.}
For this we will apply the following
\begin{lemma}
\label{deutlemm}
Let $\tilde{C},w,x,y,z,\tilde{\delta}, p$ be positive real numbers, and let $v_0\ge1$.
Suppose for all $u$ such that $r(u,v_0)\le\tilde{\delta}v^p$, we have
\begin{equation}
\label{111}
|\theta|(u,v_0)\le \tilde{C}(r^{\alpha}v_0^{-y}+v_0^{-z}),
\end{equation}
and for some $(u_0,v_0)$ we have
\begin{equation}
\label{222}
|\theta|(u_0,v_0)\le \tilde{C}(r^{-1}v_0^{-w}+v_0^{-x}).
\end{equation}
Then it follows that
\begin{equation}
\label{333}
|\theta|(u_0,v_0)\le \tilde{C}(3+\tilde{\delta}^{-p})v_0^{-\tilde{w}},
\end{equation}
where
\[
\tilde{w}=\min\left\{\frac{y}{\alpha+1}+
w\left(1-\frac{1}{\alpha+1}\right),w+p,z,x\right\}.
\]
\end{lemma}

\noindent\emph{Proof.}
If 
\[
r^\alpha v_0^{-y}\ge v_0^{-w}r^{-1}
\]
then 
\[
r\ge v_0^{\frac{1}{1+\alpha}\left(y-w\right)},
\]
and thus, by
$(\ref{222})$, we have
\begin{eqnarray}
\label{al1}
\nonumber
|\theta|(u_0,v_0)	&\le&	\tilde{C}v_0^{-w}r^{-1}+\tilde{C}v_0^{-x}\\
\nonumber	
		&\le&	\tilde{C}v_0^{-w-\frac{1}{1+\alpha}
					\left(y-w\right)}+\tilde{C}v_0^{-x}\\
		& = &	2\tilde{C}v_0^{-\min\left\{\frac{y}{(\alpha+1)}+
					w\left)1-\frac1{\alpha+1}\right),x\right\}},
\end{eqnarray}
while
if 
\[
r^\alpha v_0^{-y}\le v_0^{-w}r^{-1},
\]
then
\[
r\le v_0^{\frac{1}{1+\alpha}\left(y-w\right)},
\]
and thus by
$(\ref{111})$ and $(\ref{222})$, we have
\begin{eqnarray}
\label{al2}
\nonumber
|\theta|(u_0,v_0)	&\le& 	
\tilde{C}r^\alpha v_0^{-y}+\tilde{C}\tilde{\delta}^{-p}v_0^{-w-p}+
					\tilde{C}v_0^{-z}+\tilde{C}v_0^{-x}\\
\nonumber
		&\le&	\tilde{C}v_0^{-y+\frac{\alpha}{\alpha+1}
					\left(y-w\right)}+\tilde{C}\tilde{\delta}^{-p}v_0^{-w-p}+
						\tilde{C}v_0^{-z}+\tilde{C}v_0^{-x}\\
		& = &	\tilde{C}(3+\tilde{\delta}^{-p})
				v_0^{-\min\left\{\left(\frac{y}{(1+\alpha)}
					+w\left(1-\frac{1}{1+\alpha}\right)\right),
					w+p,x,z\right\}}.
\end{eqnarray}
Inequalities
$(\ref{al1})$ and $(\ref{al2})$ together yield the result of the lemma. $\Box$

In view of $(\ref{pepilogn})$, 
applying Lemma \ref{deutlemm}, with $w=w_k$, $z=w_k+p$, $y=w_k+\frac12-\frac p2$,
$x=\omega_*$, and $\tilde{\delta}=\tilde{\delta}_{k+1}$,
it follows that on $[u_{i-1},u_i]\times \{v_i\}$, we have
\[
|\theta|\le E_{k+1}v_i^{-w_{k+1}},
\]
where
\[
E_{k+1}=(3+\tilde{\delta}_{k+1}^{-p})
\max\left\{\alpha^{-1}
\sqrt{2h_{k+1}^{2w_k+1-p}\bar{C}_k}H,
8(h_{k+1}-1)^{-1}r_+(r_1^{-1}+1)h_{k+1}^{p+w_k}\bar{C}_k
\right\}.
\]

We can also bound $|\phi|$ and $|\phi\lambda+\theta|$ on part
of $\partial\mathcal{X}$.
In view of $(\ref{eivaimesa})$,
it follows from
$S_k$ and $(\ref{rsavv})$ that
on $[u_{i-1},u_i]\times \{v_{i+1}\}$,
\begin{equation}
\label{As}
|\phi|\le 2G_R^{-2}(h_{k+1}-1)^{-1}\bar{C}_k(v_i^{-w_k-1}+v_i^{-\omega_*}),
\end{equation}
while on 
$\{u_{i-1}\}\times[v_i,v_{i+1}]$, 
\begin{equation}
\label{Bs}
|\phi|\le 2G_R^{-2}(h_{k+1}-1)^{-1}\bar{C}_k(v_i^{-w_k-1}+v_i^{-\omega_*}),
\end{equation}
\begin{equation}
\label{kiautomnto3exvame}
|\phi\lambda+\theta|\le 2^{\omega_*}G^{-2\omega_*}(h^2_{k+1}-1)^{-\omega_*}\hat{C}v_i^{-\omega_*},
\end{equation}
where to obtain the latter inequality we use inequality $(\ref{3A})$.
Since $w_{k+1}<\omega_*$ and $w_k+1>w_{k+1}$, 
it follows that we can apply 
Proposition \ref{sprpro} on $\mathcal{X}_i$ 
with 
\[
A=\max\{E_{k+1},2G^{-2}(h_{k+1}-1)^{-1}\bar{C}_k\} v_i^{-w_{k+1}},
\] 
\[
B=2^{\omega_*}G^{-2\omega_*}(h^2_{k+1}-1)^{-\omega_*}\hat{C}v_i^{-w_{k+1}},
\]
to obtain
\begin{equation}
\label{...}
|\phi|\le\tilde{E}_{k+1}v_i^{-w_{k+1}},
\end{equation}
\[
|\theta|\le{\tilde{E}}_{k+1}v_i^{-w_{k+1}},
\]
\begin{equation}
\label{Cs}
|\lambda\phi+\theta|\le
{\tilde{E}}_{k+1}\left(v_i^{-\omega_*}+r^{-1}v_i^{-w_{k+1}}\right)
\end{equation}
in $\mathcal{X}_i$,
and 
\begin{equation}
\label{Ds}
\varpi(u_i,v_{i+1})-\varpi(u_i,v_i)\le\tilde{E}_{k+1}^2
\left(v_i^{-2w_{k+1}}
+r(u_i,v_{i+1})v_i^{-2\omega_*}\right),
\end{equation}
where
\begin{eqnarray*}
{\tilde{E}}_{k+1}&=&\max\left\{\sigma_1,\sqrt{\sigma_2}\right\}
\cdot\max\left\{E_{k+1},2G^{-2}(h_{k+1}-1)^{-1}\bar{C}_k,\right.\\
&&\hbox{\ }\left.2^{\omega_*}G^{-2\omega_*}(h^2_{k+1}-1)^{-\omega_*}\hat{C}\right\}.
\end{eqnarray*}
Since $w_{k+1}\le\omega_*-\frac12$, $(\ref{Ds})$ and $(\ref{rsavv})$
give the bound
\begin{equation}
\label{kiedwflux}
\varpi(u_i,v_{i+1})-\varpi(u_i,v_i)\le(1+2(h_{k+1}^2-1)(1+G_R^{-1}))
\tilde{E}_{k+1}^2v_i^{-2w_{k+1}}.
\end{equation}

We can now easily obtain $(\ref{11y})_{k+1}$:
By $(\ref{privtovbela})$, it follows that we have a bound
\[
\varpi(u_i,v_i)-\varpi(\infty,v_i)\le 
2h_{k+1}^{2w_k+1-p}\bar{C}_kH^2(2\alpha+1)^{-1}
R^{2\alpha+1}v_{i,k+1}^{-2w_k-1+p},
\]
and thus, by $(\ref{kiedwflux})$,
a bound
\begin{equation}
\label{giatoY}
\varpi(u_i,v_{i+1})-\varpi(\infty,v_i)
\le \hat{E}_{k+1} v_i^{-2w_{k+1}},
\end{equation}
where 
\[
\hat{E}_{k+1}=
2h_{k+1}^{2w_k+1-p}\bar{C}_kH^2(2\alpha+1)^{-1}
R^{2\alpha+1}+(1+2(h_{k+1}^2-1)(1+G_R^{-1}))
\tilde{E}_{k+1}^2.
\]
In particular, 
\[
\varpi(\infty,v_{i+1})-\varpi(\infty,v_i)\le
\varpi(u_i,v_{i+1})-\varpi(\infty,v_i)
\le \hat{E}_{k+1} v_i^{-2w_{k+1}}, 
\]
so
\begin{eqnarray*}
\varpi_+-\varpi(\infty,v_i)	&\le&	\sum_{j\ge i}
					 (\varpi(\infty,v_{j+1})-\varpi(\infty,v_j))\\
				&\le&	\hat{E}_{k+1}\sum_{j\ge i} v_j^{-2w_{k+1}}\\
				&\le&	\hat{E}_{k+1}\sum_{j\ge i} 
						\hat{v}_{j,k+1}^{-2w_{k+1}}\\
				&\le&	\hat{E}_{k+1}(1-h_{k+1}^{-i2w_{k+1}})^{-1}
						\hat{v}_{i,k+1}^{-2w_{k+1}}.
\end{eqnarray*}
For $v\ge v_{\hat{I}_{k+1}}$, let $i$ be such that $v_i\le v\le v_{i+1}$.
We have
\begin{eqnarray*}
\varpi_+-\varpi(\infty,v)	&\le& 	\varpi_+-\varpi(\infty, v_i)\\
				&\le&	\hat{E}_{k+1}\left(1-h_{k+1}^{-i2w_{k+1}}\right)
						^{-1}
						\hat{v}_{i,k+1}^{-2w_{k+1}}\\
				&\le&	\hat{E}_{k+1}h^{4w_{k+1}}_{k+1}
						\left(1-h_{k+1}^{-\hat{I}_{k+1}2w_{k+1}
						}\right)^{-1}
						v^{-2w_{k+1}}.
\end{eqnarray*}
Thus, for 
\[
\bar{C}_{k+1}\ge \max\left\{h^{4w_{k+1}}_{k+1}
\left(1-h_{k+1}^{-\hat{I}_{k+1}2w_{k+1}}\right)^{-1}\hat{E}_{k+1},
(\varpi_+-\varpi_1)h_{k+1}^{\hat{I}_{k+1}+1}\right\}
\]
we have
$(\ref{11y})_{k+1}$.

We still need to
obtain our pointwise bounds $(\ref{22y})_{k+1}$ and $(\ref{33y})_{k+1}$.
Form the rectangles $\mathcal{Y}_i=[u_i,\infty]\times [v_i,v_{i+1}]$:
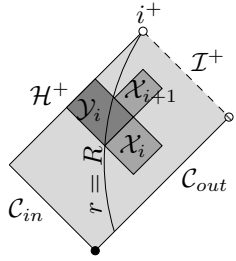
\[
\input{Yor9og.pstex_t}
\]
From $(\ref{giatoY})$ and
the pointwise bounds $(\ref{privtovbela})$ and $(\ref{...})$, it follows that the
assumptions of Proposition \ref{piokovtaekei} hold for $\mathcal{Y}_i$,
with $r_*=R$ and
\[
A=\sqrt{2h_{k+1}^{2w_k+1-p}\bar{C}_kH^2
R^{2\alpha+1}+(1+2(h_{k+1}^2-1)(1+G_R^{-1}))
\tilde{E}_{k+1}^2}v_{i,k+1}^{-w_{k+1}}.
\]
Thus
\begin{equation}
\label{kaistoRetsi}
|\phi|\le \bar{E}_{k+1}v_i^{-w_{k+1}},
\end{equation}
\begin{equation}
\label{gialigopiokatw}
\left|\zn\right|\le \bar{E}_{k+1}v_i^{-w_{k+1}},
\end{equation}
in $\mathcal{Y}_i$,
where 
\[
\bar{E}_{k+1}=\sqrt{2h_{k+1}^{2w_k+1-p}\bar{C}_kH^2
R^{2\alpha+1}+(1+2(h_{k+1}^2-1)(1+G_R^{-1}))
\tilde{E}_{k+1}^2}.
\]
Form now the
sequence of rectangles $\mathcal{Z}_i=[u_i,u_{i+2}]\times[v_i,v_{i+1}]$. 
By $(\ref{kiautomnto3exvame})$ and $(\ref{kaistoRetsi})$,
it follows that the $\mathcal{Z}_i$ satisfy the assumptions of 
Proposition \ref{arketamegaln},
with $A=\bar{E}_{k+1}v_i^{-w_{k+1}}$, and 
$B=2^{\omega_*}G^{-2\omega_*}(h_{k+1}^2-1)^{-\omega_*}\hat{C}v_i^{-\omega_*}$.
Thus, we have
\begin{equation}
\label{pragmkovte1}
|\phi|\le{\tilde{C}}_{k+1}\left(r^{-1} v_i^{-w_{k+1}}+v_i^{-\omega_*}\right),
\end{equation}
\begin{equation}
\label{pote;}
|\lambda\phi+\theta|\le{\tilde{C}}_{k+1}\left(r^{-2}(\log r) v_i^{-w_{k+1}}
+v_i^{-\omega_*}\right)
\end{equation}
in $\mathcal{Z}_i\cap\{r\ge R\}$,
where
\[
{\tilde{C}}_{k+1}=\sigma\max\left\{\bar{E}_{k+1},
2^{\omega_*}G_R^{-2\omega_*}(h_{k+1}^2-1)^{-\omega_*}\hat{C}\right\}.
\]

The inequalities $(\ref{pragmkovte1})$ and $(\ref{pote;})$,
together with the triangle inequality
and the bound
$1\le\lambda^{-1}\le G$,
 immediately yield
\begin{equation}
\label{pragmkovte2}
|\theta|\le 2G_R^{-1}{\tilde{C}}_{k+1}
\left(r^{-1}v_i^{-w_{k+1}}+v_i^{-\omega_*}\right)
\end{equation}
in $\mathcal{Z}_i\cap\{r\ge R\}$.
Finally, integrating the equation
\[
\partial_u\theta=-\zn\frac\nu{r}\lambda
\]
in $\mathcal{Y}_i$, from $r=R$, using the bound $(\ref{gialigopiokatw})$
and the initial bound provided by $(\ref{pragmkovte2})$,
yields the inequality
\begin{equation}
\label{thkaistoY}
|\theta|\le (2G_R^{-1}{\tilde{C}}_{k+1}(1+R^{-1})+
\bar{E}_{k+1}\log (R/r_1))v_i^{-w_{k+1}}
\end{equation}
in $\mathcal{Y}_i$.

Define the region
\[
\mathcal{R}=\bigcup_{i\ge \hat{I}_k}
\left(\mathcal{X}_i\cup\mathcal{Y}_i\right).
\]
Given a $(u,v)\in\mathcal{R}$, 
setting $i$ such that $v_i\le v\le v_{i+1}$,
we have by $(\ref{kaistoRetsi})$ and $(\ref{pragmkovte1})$
\begin{eqnarray*}
|\phi(u,v)|	&\le& 	\max\{R\bar{E}_{k+1}, \tilde{C}_{k+1}\}
				\left(r^{-1}v_i^{-w_{k+1}}+v_i^{-\omega_*}\right)\\
		&\le& 	\max\{R\bar{E}_{k+1}, \tilde{C}_{k+1}\}
				h_{k+1}^{2\omega_*}\left(r^{-1}v^{-w_{k+1}}+v^{-\omega_*}
				\right).
\end{eqnarray*}
On the other hand, by $(\ref{pragmkovte2})$ and $(\ref{thkaistoY})$
we have the  bound
\begin{eqnarray*}
|\theta(u,v)|	&\le& 	(2G_R^{-1}{\tilde{C}}_{k+1}(1+R^{-1})+
				\bar{E}_{k+1}R\log (R/r_1))
				\left(r^{-1}v_i^{-w_{k+1}}+v_i^{-\omega_*}\right)\\
		&\le& 	(2G_R^{-1}{\tilde{C}}_{k+1}(1+R^{-1})+
				\bar{E}_{k+1}R\log (R/r_1))
				h_{k+1}^{2\omega_*}\left(r^{-1}v^{-w_{k+1}}+v^{-\omega_*}
				\right).
\end{eqnarray*}
For 
\[
\delta_{k+1}\le 2(h^2_{k+1}-1)(1+G^{-1}_R)h^{-2}_{k+1}
\]
it follows by $(\ref{rsavv})$ that 
\[
\{r\le\delta_{k+1}v\}\cap\{v\ge v_{\hat{I}_k}\}\subset\mathcal{R}.
\]
The bound $(\ref{22y})_{k+1}$ 
then follows for
\[
\bar{C}_{k+1}\ge 
\max\left\{\left(\sup_{v\le v_{\hat{I}_k}}|\phi|\right)\left(v_{\hat{I}_k}^{w_k}
r(1,\hat{v}_{\tilde{I}_k})+v^{\omega_*}_{\tilde{I}_k}\right),
h_{k+1}^{2\omega_*}R\bar{E}_{k+1}, h_{k+1}^{2\omega_*}\tilde{C}_{k+1}\right\}
\]
and
$(\ref{33y})_{k+1}$ follows for
\begin{eqnarray*}
\bar{C}_{k+1}&\ge& 
\max\left\{\left(\sup_{v\le v_{\hat{I}_k}}|\theta|\right)\left(v_{\hat{I}_k}^{w_k}
r(1,\hat{v}_{\tilde{I}_k})+v^{\omega_*}_{\tilde{I}_k}\right),\right.\\
&&\hbox{\ }\left.
(2G_R^{-1}{\tilde{C}}_{k+1}(1+R^{-1})+
				\bar{E}_{k+1}R\log (R/r_1))
				h_{k+1}^{2\omega_*}
\right\}.
\end{eqnarray*}
This completes the proof of the proposition. $\Box$

\subsection{Decay on $\mathcal{I}^+$}
So far we have only obtained decay in the region
$r\le \delta_{\omega_*-\frac12}v$.
In this section we will prove decay in $r\ge\delta_{\omega_*-\frac12}v$,
in particular, decay along $\mathcal{I}^+$.  This decay will now be measured
in the $u$ direction.

\begin{theorem}
\label{nulinf1}
There exists a constant $\tilde{C}>0$ such that
in $r\ge \delta_{\omega_*-\frac12}v$, we have the bounds
\begin{equation}
\label{**}
|\theta|\le \tilde{C}v^{-1}u^{-(\omega_*-1)},
\end{equation}
\[
|\zeta|\le \tilde{C}u^{-\left(\omega_*-\frac12\right)},
\]
\begin{equation}
\label{ekei;}
|r\phi|\le \tilde{C}u^{-(\omega_*-1)}.
\end{equation}
\end{theorem}

\noindent\emph{Proof.}
Fix $0<\delta'<1$ and consider the curve $\Gamma$ defined by $u=\delta' v$. 
For $\delta'$ sufficiently close
to $1$, $\Gamma$ is clearly contained in $r\le\delta_{3/2}v$. This 
follows immediately by noting, on the one hand, that
$r(1,v)v^{-1}\to1$ as $v\to\infty$,
while, on the other hand, if $r(u,v)\ge R$, then
$r(1,v)-r(u,v)\ge G_R^{-1}u$, and $G_R^{-1}\to 1$ as $R\to\infty$.
Choosing such a $\delta'$, we have 
that on $\Gamma$, $r\ge h(1-\delta')v$, for some $h>0$.
Thus, it follows from Theorem \ref{3/2} that $(\ref{ekei;})$ holds
on $\Gamma$ where $\tilde{C}$ is replaced by a constant we shall call 
$\tilde{C}'$.\footnote{In the remainder of this paper, we leave the explicit
computation of constants to the reader.}

If $(u,v)$ is such that $u\le \delta' v$, then we can
integrate $(\ref{e3is})$ along $\{u\}\times[\delta'^{-1}u,v]$ to obtain
\begin{eqnarray}
\nonumber
\label{a*}
|r\phi|(u,v)		&\le& 	\tilde{C}'u^{-(\omega_*-1)}+
						\int_{\delta'^{-1}u}^v
						{\hat{C}r^{-\omega_*}(u,\bar{v})d\bar{v}}\\
			&\le&	\tilde{C}'u^{-(\omega_*-1)}+G^{-1}\hat{C}r^{-(\omega_*-1)}
						(u,\delta'^{-1}u),
\end{eqnarray}
which yields $(\ref{ekei;})$
for $\tilde{C}$ sufficiently large. 
We then obtain
\begin{eqnarray}
\label{gtt}
\nonumber
|\theta|(u,v)	&\le&	G^{-1}|\phi\lambda+\theta|(u,v)+r^{-1}|r\phi|(u,v)\\
		&\le&	\tilde{C}'\left(v^{-\omega_*}+v^{-1}u^{-(\omega_*-1)}\right),
\end{eqnarray}
and this yields $(\ref{**})$,
again for sufficiently large $\tilde{C}$.
Finally, we note that in the proof of Theorem \ref{3/2}, we have in fact 
shown that
for $r(u,v_*)=r_0$, 
\[
|\zeta|(u,v_*)\le\tilde{C}'v_*^{-\left(\omega_*-\frac12\right)},
\]
for some $\tilde{C}'$.
Integrating $(\ref{sign2})$ using $(\ref{+++})$, we obtain the bound
\begin{eqnarray*}
|\zeta|(u,\delta'^{-1}u)	&\le&	\tilde{C}'v_*^{-\left(\omega_*-\frac12\right)}
						+\int_{v_*}^
								{\delta'^{-1}u}
						\frac{|\theta\nu|}r(u,\bar{v})d\bar{v}\\
				&\le&	\tilde{C}'v_*^{-\left(\omega_*-\frac12\right)}\\
				&   &	\hbox{\ \ }
					+C_{\omega_*-\frac12}G^{-1}
								\left(r_0^{-1}
								v_*^{-\left
						(\omega_*-\frac12\right)}+
						\log (r(u,\delta'^{-1}u)/r_0)
							v_*^{-\omega_*}\right)\\
				&\le&	\tilde{C}''v_*^{-\left(\omega_*-\frac12\right)}.
\end{eqnarray*}
Now let $v\ge\delta'^{-1}u$.  Integrating from $\Gamma$ using
our bound $(\ref{gtt})$ for $\theta$ we obtain
\begin{eqnarray*}
|\zeta|(u,v)		&\le&	\tilde{C}''u^{-\left(\omega_*-\frac12\right)}+
						\int_{\delta'^{-1}u}^{v}
						\frac{|\theta\nu|}r(u,\bar{v})d\bar{v}\\
			&\le&	\tilde{C}''u^{-\left(\omega_*-\frac12\right)}+
						\tilde{C}'''\left(
						r^{-\omega_*}+
						u^{-\omega_*-1}r^{-1}(u,\delta'^{-1}u)
						\right)\\
			&\le&	\tilde{C}u^{-\left(\omega_*-\frac12\right)}
\end{eqnarray*}
for $\tilde{C}$ sufficiently large.
This completes the proof.
$\Box$

We can use the above theorem
to obtain a better decay estimate in $r$ for the quantity $\phi\lambda+\theta$:
\begin{proposition}
There exists 
a constant $\hat{C}'$ such that
\begin{equation}
\label{prwtnkalu}
|\phi\lambda+\theta|\le \hat{C}'(r^{-3}\log r+r^{-\omega}).
\end{equation}
\end{proposition}

\noindent\emph{Proof.}
The inequality $(\ref{prwtnkalu})$ holds on the initial outgoing null curve.
Integrating $(\ref{toallo})$ using $(\ref{ekei;})$,
we obtain immediately $(\ref{prwtnkalu})$ in the region
$r\ge\delta_{\omega_*-\frac12} v$. Using $(\ref{+})$, 
the estimate extends to $r\le\delta_{\omega_*-\frac12}v$. $\Box$

We will remove the $\log$ term in $(\ref{prwtnkalu})$ in the
context of the proof of the following theorem. The $r^{-3}$ term
cannot be improved, however, because it arises independently of
the rate of the $u$-decay of $|r\phi|$. This is ultimately the source of the
exponent $3$ in Price's law, and is the reason why we have restricted
consideration to $\omega\le 3$ in Assumption ${\bf \Delta}'$.

We noted immediately before the statement of Theorem \ref{3/2} in
Section \ref{partI} that the obstruction at $\omega_*$ arose 
from the bound $(\ref{3A})$ applied in $(\ref{Cs})$.
As this bound has now been improved, we can improve Theorem \ref{3/2}
with the following

\begin{theorem}
\label{5/2}
There exist constants $C_{\omega-\frac12}>0$, $\delta_{\omega-\frac12}>0$ 
such that
in the region $r\le\delta_{\omega-\frac12} v$, we have
\[
|\phi|\le C_{\omega-\frac12}\left(v^{-\left(\omega-\frac12\right)}r^{-1}+
v^{-\omega}\right),
\]
\[
|\theta|\le C_{\omega-\frac12}\left(v^{-\left(\omega-\frac12\right)}r^{-1}+
v^{-\omega}\right),
\]
and in addition
\[
\varpi_+-\varpi(\infty,v)
\le C_{\omega-\frac12}v^{-(2\omega-1)}.
\]
\end{theorem}

\noindent\emph{Proof.}
We first show the above where $\omega<3$.
For this, one proceeds exactly as in the proof of Theorem \ref{3/2}, except that we now
define 
\[
w_{k+1}=\min\left\{w_k+p,s\right\},
\]
for $p>0$ defined as before.
Inequality $(\ref{prwtnkalu})$ implies 
$|\phi\lambda+\theta|\le C(3-\omega)r^{-\omega}$,
and thus the $v_i^{-\omega_*}$ in $(\ref{As})$, $(\ref{Bs})$, and
$(\ref{Cs})$ can be replaced by $v_i^{-s-\omega}$, and, in $(\ref{Ds})$,
the $v_i^{-2\omega_*}$ can be replaced by $v_i^{-2\omega}$. 
Finally, since $w_k\le \frac52$,
$(\ref{kiedwflux})$ holds as well,
and the rest follows as before.

If $\omega=3$, then first carry through the argument
of the above paragraph with $\omega$ replaced with some $\omega'$ satisfying
$2<\omega'<\omega$. One improves Proposition \ref{nulinf1} to obtain in particular
$|r\phi|\le\tilde{C}\left(u^{-(\omega'-\frac12)}+u^{-2}\right)$ 
for some constant $\tilde{C}$. 
>From this one improves $(\ref{prwtnkalu})$ to obtain
the uniform decay in $r$
\begin{equation}
\label{deutkalu}
|\phi\lambda+\theta|\le \hat{C}'r^{-3}.
\end{equation}
for sufficiently large $\hat{C}'$. But now one can carry through the argument
of the previous paragraph with $\omega$. 
$\Box$

\subsection{The induction: Part II}
\label{finalsec}

Theorem \ref{5/2} removes the obstruction of $(\ref{3A})$
in $(\ref{Cs})$. The obstruction
at $v^{-\left(\omega-\frac12\right)}$, on the other hand, arises
the \emph{second} term of $(\ref{TOP-flux})$ from Proposition \ref{sprpro},
i.e.~in the second term of $(\ref{Ds})$. 
To circumvent this obstruction, we must apply the propositions
of Section \ref{maxsec} and \ref{akoma2sec} 
to rectangles $\mathcal{X}$, $\mathcal{Y}$, $\mathcal{Z}$ whose $v$-dimensions are smaller, 
i.e.~so that $r(u_i,v_{i+1})\sim v_{i+1}^s$
with $0<s<1$. We proceed to do this in the present section.

\begin{theorem}
\label{final1}
Let $\epsilon>0$.
Then there exist constants $C_{\omega-\epsilon}>0$, $\delta_{\omega-\epsilon}>0$ such that
for $r\le\delta_{w-\epsilon}v$, we have
\[
|\phi|\le C_{\omega-\epsilon}\left(v^{-(\omega-\epsilon)}r^{-1}+v^{-\omega}\right),
\]
\[
|\theta|\le C_{\omega-\epsilon}\left(v^{-(\omega-\epsilon)}r^{-1}+v^{-\omega}\right),
\]
\[
|\phi\lambda+\theta|\le C_{\omega-\epsilon}\left(v^{-(\omega-\epsilon)}r^{-1}+v^{-\omega}\right).
\]
\end{theorem}
 
\noindent\emph{Proof.}
Consider the statement  $S'_\tau$
\begin{enumerate}
\item[$(S'_\tau)$]
There exist constants
constants $\bar{C}_\tau, \delta_\tau>0$, such that
for $r\le\delta_\tau v$,
\begin{equation}
\label{22y''}
|\phi|\le \bar{C}_\tau\left(v^{-(\omega-\tau)}r^{-1}+v^{-\omega}\right),
\end{equation}
\begin{equation}
\label{33y''}
|\theta|\le \bar{C}_\tau\left(v^{-(\omega-\tau)}r^{-1}+v^{-\omega}\right),
\end{equation}
\[
|\lambda\phi+\theta|\le\bar{C}_\tau\left(v^{-(\omega-\tau)}r^{-1}+v^{-\omega}\right),
\]
\[
\varpi(v+v^\tau)-\varpi(v)
\le\bar{C}_\tau v^{-2(\omega-\tau)}.
\]
\end{enumerate}

We will prove the following
\begin{proposition}
\label{meto'}
$S'_\tau$ implies $S'_{\frac\tau2}$.
\end{proposition}

Assuming for the moment the above Proposition, and given any $\epsilon>0$, there
exists an $n$ such that $\omega-\frac1{2^n}>\omega-\epsilon$. Since $S'_\frac12$ is true by
Theorems \ref{3/2} and \ref{5/2}, then by 
induction, $S'_{\frac1{2^n}}$ is true. But $S'_{\frac1{2^n}}$ immediately implies
the theorem. $\Box$.
\vskip1pc
\noindent\emph{Proof of Proposition \ref{meto'}.}
This we shall also prove by induction!
Consider the statement $S''_k$ defined by
\begin{enumerate}
\item[$(S''_k)$]
There exist
constants $\bar{C}_k, \delta_k>0$, such that
for $r\le\delta_k v$,
\begin{equation}
\label{22y'''}
|\phi|\le \bar{C}_k\left(v^{-w_k}r^{-1}+v^{-\omega}\right),
\end{equation}
\begin{equation}
\label{33y'''}
|\theta|\le \bar{C}_k\left(v^{-w_k}r^{-1}+v^{-\omega}\right),
\end{equation}
\begin{equation}
\label{44y'''}
|\phi\lambda+\theta|\le \bar{C}_k\left(v^{-w_k}r^{-1}+v^{-\omega}\right),
\end{equation}
\begin{equation}
\label{11y'''}
\varpi(v+v^\tau)-\varpi(v)\le\bar{C}_kv^{-2w_k}.
\end{equation}
\end{enumerate}
Choose a $p$
such that
\begin{equation}
\label{pepilogn2}
0<p<\frac{\tau}{2\alpha+3}.
\end{equation}
We will prove
\begin{proposition}
\label{telprop;}
$S''_k$ implies $S''_{k+1}$
where 
\[
w_{k+1}=\min\left\{w_k+p, \omega-\frac\tau2\right\}.
\]
\end{proposition}
Assuming the result of Proposition \ref{telprop;}, Proposition \ref{meto'} follows
immediately by induction. For setting $w_0=\omega - \tau$, $S''_0$ is true, as it is implied by $S'_\tau$,
and thus by induction $S''_k$ is true for all $k\ge 0$. There clearly exists a $k$
such that $w_k=\omega-\frac\tau2$, and for such a $k$, $S''_k$ immediately gives $S'_{\frac\tau2}$.
$\Box$
\vskip1pc
\noindent\emph{Proof of Proposition \ref{telprop;}.}
We fix $h_{k+1}>0$ sufficiently small,
set $\hat{v}_{0,k+1}=1$, and define inductively
\[
\hat{v}_{i+1,k+1}=\hat{v}_{i,k+1}+h_{k+1}\hat{v}_{i,k+1}^\tau.
\]
In analogy with the proof of Theorem \ref{3/2}, we obtain 
a sequence of segments  $\{\infty\}\times[\tilde{v}_{i,k+1},v_{i,k+1}]$,
such that $v_{i,k+1}\sim \hat{v}_{i,k+1}$ and
\[
\varpi(v_{i,k+1})-\varpi(\tilde{v}_{i,k+1})\le\tilde{C}v_{i,k+1}^{-2w_k-\tau+p},
\]
\[
|\theta|(\infty, v_{i,k+1})\le \tilde{C}v_{i,k+1}^{-w_k+\frac p2-\frac \tau2}.
\]
Constructing as before characteristic rectangles $\mathcal{W}_i$, and
applying Proposition \ref{sautatelos}, we obtain that there exist
constants $\tilde{C}'$ and $\delta>0$ such that,
for $r\le \delta v_i^p$, we have
\[
|\theta(u,v_i)|\le \tilde{C}'\left(v_i^{-w_k-p}+r^\alpha v_i^{-w_k+\frac p2-\frac\tau2}\right).
\]
Again, we have here dropped the $k+1$ subscript from $v_{i,k+1}$.
On the other hand, 
from our assumption $S''_k$, we have that for $r(u,v_i)\le \delta_k v$,
\[
|\theta(u,v_i)|\le \bar{C}_k\left(v_i^{-w_k}r^{-1}+v_i^{-\omega}\right).
\]
Applying Lemma \ref{deutlemm} 
with $w=w_k$, $x=\omega$, $y=w_k+\frac\tau2-\frac p2$, $z=w_k+p$,
in view of our choice $(\ref{pepilogn2})$, it follows that for
$r\le \delta v_i$, if $\delta$ is chosen sufficiently small, we have
\[
|\theta(u,v_i)|\le \tilde{C}'v_i^{-w_{k+1}},
\]
for some $\tilde{C}'$.
We then construct rectangles $\mathcal{X}_i$, $\mathcal{Y}_i$ 
from the sequence $v_i$ as before, 
arranging by appropriate choice
of $h_{k+1}$ for $\bigcup_{i\ge I}\mathcal{X}_i$ to
be contained in the region $r\le \delta_{k}v$ for sufficiently large $I$. 
On the other hand, $\bigcup_{i\ge I}\left(\mathcal{X}_i\cup\mathcal{Y}_i\right)$ covers 
the region
$r\le \delta_{k+1}v^\tau$ for small enough $\delta_{k+1}<\delta_k$. Since 
$w_k\ge \omega -\tau$, in the region
$\delta_{k+1}v^\tau\le r\le \delta_{k+1}v$, $(\ref{22y'''})_k$, $(\ref{33y'''})_k$, 
$(\ref{44y'''})_k$
immediately yield $(\ref{22y'''})_{k+1}$, $(\ref{33y'''})_{k+1}$, and $(\ref{44y'''})_{k+1}$,
since in that region, the second term of these inequalities always dominates.
Thus, to show the proposition, it suffices to restrict to 
$\bigcup_{i\ge I}\left(\mathcal{X}_i\cup\mathcal{Y}_i\right)$.
The assumptions
of Proposition \ref{aptovoriz} clearly apply to $\mathcal{X}_i$ 
with $A=\tilde{C}'v_i^{-w_{k+1}}$, 
$B=\tilde{C}'v_i^{-\omega}$, after redefinition of the constant
$\tilde{C}'$. 
Since $2\omega-\tau\ge 2w_{k+1}$, it follows that the assumptions
of Proposition \ref{metobeta} then apply to the rectangles $\mathcal{Y}_i$ 
with $A=\tilde{C}'v_i^{-w_{k+1}}$.
In particular this yields $(\ref{11y'''})_{k+1}$. Finally, 
the assumptions of Proposition \ref{sautatelos} apply to the rectangles 
$\mathcal{Z}_i$ with $A=\tilde{C}'v_i^{-w_{k+1}}$
and $B=\tilde{C}'v_i^{-\omega}$, again after redefinition
of $\tilde{C}'$. All together, the bounds
obtained yield 
$(\ref{22y'''})_{k+1}$, $(\ref{33y'''})_{k+1}$, and $(\ref{44y'''})_{k+1}$ in
$\bigcup_{i\ge I}\left(\mathcal{X}_i\cup\mathcal{Y}_i\right)$ and the proof of the
proposition is complete.
$\Box$

From the above theorem, we can deduce also decay in a ``neighborhood'' of
$\mathcal{I}^+$.
\begin{theorem}
\label{final2}
Let $\epsilon>0$ and define $\omega$ as in the previous theorem. Then
there exists a constant $\tilde{C}_\epsilon>0$ such that
in $r\ge \delta_{\omega-\epsilon}v$, we have the bounds
\[
|\theta|\le \tilde{C}_\epsilon v^{-1}u^{-\omega+1+\epsilon}
\]
\[
|\zeta|\le \tilde{C}_\epsilon u^{-\omega+\epsilon}
\]
\[
|r\phi|\le \tilde{C}_\epsilon u^{-\omega+1}.
\]
\end{theorem}

\noindent\emph{Proof.} The proof is similar to Theorem \ref{nulinf1}. $\Box$

\section{Unanswered questions}
\label{sharpsec}
In view of Theorem 2 of \cite{md:cbh}, we have, in addition to
Theorem \ref{int-the} of the Introduction, the following complementary  
result:
\begin{theorem} Under the assumptions of Corollary \ref{cor-1}, if $(\ref{kritiknp})$ does
not hold, and if there exists a $V>0$ such that, for $v\ge V$, 
\begin{equation}
\label{po9oume}
|\theta|(\infty,v)\ge cv^{-p}, 
\end{equation}
for some $p<9$, $c>0$,
then the Cauchy development of initial data is \emph{inextendible}
as a $C^1$ metric. 
\end{theorem}

Heuristic analysis~\cite{gpp:de1, rpr:ns, lb:sbh, bo:lt} and 
numerical studies~\cite{gpp:de, bo:lte, mc:bhsf}
seem to indicate that the inequality $(\ref{po9oume})$ holds
with $p=3$ for generic compactly supported data. This remains, however,
an open problem.

On the other hand, it seems that the red-shift techniques of this paper 
could
yield the following
\begin{conjecture}
\label{conj2} 
Given any $\epsilon>0$, and $\omega>1$, there exist characteristic asymptotically
flat initial data, on
which $|\phi|(0,v)\le Cr^{-1}$, $|\theta+\phi\lambda|(0,v)\le Cr^{-\omega}$, and for 
which 
\[
|\phi(\infty,v)|\le Cv^{-\omega-\epsilon},
\]
for some constant $C>0$.
\end{conjecture}

A proof of the above conjecture might proceed by prescribing initial data
on the event horizon and a conjugate constant-$v$ ray, and solving this
characteristic initial value problem. Note that this problem is well-posed
in spherical symmetry, as the spacelike and timelike directions can be
interchanged.

Conjecture \ref{conj2} would say that 
the decay rate of Theorem \ref{5/2} is sharp (modulo an
$\epsilon$) in the context of the characteristic initial value problem
formulated in Section 3. At the same time, it would
indicate that any argument guaranteeing $(\ref{po9oume})$ will have
to come to terms with whatever is special about ``generic''
compactly supported data. 

\section{The uncoupled case}
\label{uncoupledsec}
In this section, we will adapt the approach of this paper
to the problem of the wave equation on a fixed
Schwarzschild or--more generally--a non-extreme
Reissner-Nordstr\"om background.\footnote{The reader can formulate 
for himself the most general assumptions
on the background metric for which the the argument applies.}
As we shall see, this uncoupled problem is basically described by
replacing the right-hand side of $(\ref{puqu})$, $(\ref{pvqu})$, and 
$(\ref{fdb1})$
with $0$. While at first glance, 
it might seem that this only makes things simpler,
in fact, 
we must be careful! Whereas the right-hand side
of $(\ref{fdb1})$ is always handled as an error term,
and thus, indeed, things are easier when it vanishes,
the right-hand side of equations $(\ref{puqu})$ and $(\ref{pvqu})$ 
provide the energy estimate for $\phi$, fundamental to our
argument! It turns out, however,
that since we now have a \emph{static} background, we can retrieve the
very same energy estimate as before, not
from the Einstein equations, but from Noether's theorem.
To faciliate the comparison, we will actually
redefine $\varpi$ in such a way so that
the equations $(\ref{puqu})$ and $(\ref{pvqu})$ still hold precisely as before.
The applicability of our argument to the uncoupled case will then be clear.

Now for the details.
Let $(\mathcal{M},g)$ be the 
maximally analytic non-extreme Reissner-Nordstr\"om
spacetime of~\cite{he:lssst}, with parameters $0\le |e|< M$. Note that this 
metric arises as a spherical
symmetric solution of 
the system $(\ref{Einstein-xs})$--$(\ref{emtensor-xs})$, 
if $F_{\mu\nu}$ is appropriately
defined, and if 
$\phi$ is defined to vanish identically. Let $\mathcal{Q}$ denote
the Lorentzian quotient of $(\mathcal{M},g)$,
let $\pi_1$ denote the natural projection $\pi_1:\mathcal{M}\to\mathcal{Q}$, 
let $r:\mathcal{Q}\to{\bf R}$ be the area radius function,
and let $\mathcal{D}\subset\mathcal{Q}$
be a subset with Penrose diagram as depicted below:
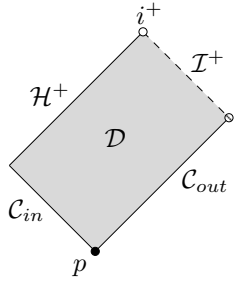
\[
\input{arxiko.pstex_t}
\]

Let $m$ be the Hawking mass, as defined in Section~\ref{assumsec},
let $(u,v)$ be a regular coordinate system covering $\mathcal{D}$,
defined as in Section~\ref{ecsec}, and let $\Omega$, $\nu$, $\lambda$, $\kappa$
be as defined in Section~\ref{assumsec}. 
In particular, we have $\nu<0$, $\kappa>0$.
The equations $(\ref{puqu})$--$(\ref{fdb1})$ 
are satisfied with $\zeta$ and $\theta$
vanishing. In particular, the renormalized Hawking mass is a constant
that can easily be seen to equal $M$, since, at infinity, the
renormalized Hawking mass tends to the Bondi mass. 
In what follows, we shall always denote 
the renormalized Hawking mass by $M$,
as we will reserve the symbol $\varpi$ for a new quantity to be defined
shortly. 
We have thus that
\begin{equation}
\label{KL?}
\partial_u\kappa=0,
\end{equation}
\begin{equation}
\label{lqu-u}
\partial_u\lambda=\frac{2\nu\kappa}{r^2}\left(M-\frac{e^2}r\right),
\end{equation}
and similarly
\begin{equation}
\label{simly}
\partial_v\frac{1-\mu}{\nu}=0,
\end{equation}
\begin{equation}
\label{nqu-u}
\partial_v\nu=\frac{2\nu\kappa}{r^2}\left(M-\frac{e^2}r\right).
\end{equation}
From $(\ref{KL?})$ and $(\ref{simly})$, 
it is easy to see that
\[
K=-\frac12\frac{1-\mu}\nu\frac\partial{\partial u}
+\frac12\kappa^{-1}\frac\partial{\partial v}
\]
defines a Killing vector on $\pi_1^{-1}(\mathcal{D})$.

Let $\phi$ now be a $C^2$ solution to the wave equation $\Box_g\phi=0$
on $\pi_1^{-1}(\mathcal{D})$. Define the zeroth spherical harmonic
$\phi_0:\mathcal{D}\to{\bf R}$ by 
\[
\phi_0(p)=\frac{1}{4\pi r^2}\int_{\pi_1^{-1}(p)}\phi.
\]
It is clear that
$\pi_1^*\phi_0$ satisfies $\Box_g\pi_1^*\phi_0=0$.

We proceed to compute the energy identity for $\phi_0$
arising from Noether's theorem applied to the Killing field $K$.
First, we compute the components of the energy-momentum tensor
$T_{\mu\nu}$ of $\pi_1^*\phi_0$. This can be written
\[
T_{\mu\nu}dx^\mu\otimes dx^\nu=T_{vv}dv\otimes dv+T_{uv}(du\otimes dv+dv\otimes du)
+T_{uu}du\otimes du+T_{AB}dx^A\otimes dx^B
\]
where $x^A$ denote local coordinates on the sphere.
Defining
$\theta=r\partial_v\phi_0$, $\zeta=r\partial_u\phi_0$,
we compute
\[
T_{uv}=0,
\]
\[
T_{uu}=r^{-2}\zeta^2,
\]
\[
T_{vv}=r^{-2}\theta^2,
\]
\[
T_{AB}=-g_{AB}2g^{uv}\partial_u\phi\partial_v\phi=
-g_{AB}\kappa^{-1}\nu^{-1}\frac{\theta\zeta}{r^2}.
\]
Defining now
$P^\mu=g^{\mu\nu}T_{\nu\gamma}K^\gamma$, and $\bar{P}=\pi_{1*} P$,
we have
\[
g\left(P,\frac\partial{\partial u}\right)=
\bar{g}\left(\bar{P},\frac\partial{\partial u}\right)=
-\frac12\frac{1-\mu}\nu T_{uu}=-\frac12\frac{1-\mu}{\nu}r^{-2}\zeta^2,
\]
\[
g\left(P,\frac\partial{\partial v}\right)=
\bar{g}\left(\bar{P},\frac\partial{\partial v}\right)=\frac12\kappa^{-1} 
T_{vv}=\frac12\kappa^{-1}r^{-2}\theta^2.
\]
Given an arbitrary characteristic rectangle $\mathcal{R}=[u_1,u_2]\times[v_1,v_2]
\subset\mathcal{D}$,
integrating the equation
\[
\nabla\cdot P=0
\]
in $\pi_1^{-1}(\mathcal{R})$,
using the divergence theorem,
we obtain 
\begin{eqnarray*}
\int_{u_1}^{u_2}{4\pi r^2\bar{g}\left(\bar{P},\frac\partial{\partial u}\right)(u,v_1)du}+
\int_{v_1}^{v_2}{4\pi r^2\bar{g}\left(\bar{P},\frac\partial{\partial v}\right)(u_1,v)dv}
=\\
\int_{u_1}^{u_2}{4\pi r^2\bar{g}\left(\bar{P},\frac\partial{\partial u}\right)(u,v_2)du}+
\int_{v_1}^{v_2}{4\pi r^2\bar{g}\left(\bar{P},\frac\partial{\partial v}\right)(u_2,v)dv},
\end{eqnarray*}
and consequently
\begin{eqnarray}
\nonumber
\label{UNC}
\int_{u_1}^{u_2}{\frac12\zeta^2\frac{1-\mu}{(-\nu)}(u,v_1)du}+
\int_{v_1}^{v_2}{\frac12\theta^2\kappa^{-1}(u_1,v)dv}
=\\
\int_{u_1}^{u_2}{\frac12\zeta^2\frac{1-\mu}{(-\nu)}(u,v_2)du}+
\int_{v_1}^{v_2}{\frac12\theta^2\kappa^{-1}(u_2,v)dv}.
\end{eqnarray}
We have thus obtained precisely the same energy identity which was
so crucial in the coupled case! 

So as to facilitate comparison with the notation of this paper,
we first recast $(\ref{UNC})$ in differential from:
Defining the $1$-form $\eta$ by
\[
\eta=\frac12(1-\mu)\left(\zn\right)^2\nu du+
\frac12(1-\mu)\left(\tl\right)^2\lambda dv,
\]
$(\ref{UNC})$ is equivalent to the statement
that $\eta$ is closed:
\[
d\eta=0.
\]
Thus, since $\mathcal{D}$ is simply connected,
it follows that, if $\phi_0$ has finite energy initially,
there exists
a unique $0$-form $\varpi$, such that
\begin{equation}
\label{vatairiazei}
\varpi(1,\infty)=M,
\end{equation}
in the obvious limiting sense,
and 
\begin{equation}
\label{akribns}
d\varpi=\eta,
\end{equation}
i.e.~such that $\varpi$ satisfies
equations $(\ref{puqu})$ and $(\ref{pvqu})$.

Thus, we can consider the ``uncoupled problem'' 
as described by the system $(\ref{puqu})$, $(\ref{pvqu})$,
$(\ref{KL?})$, 
$(\ref{sign1})$, $(\ref{sign2})$,
$(\ref{lqu})$\footnote{We include this equation in our system
since we can no longer recover $\lambda$ from $\kappa$, $\varpi$ and $r$.}.
Note that from the above equations, one
derives as before the equations
\begin{equation}
\label{zex-u}
\partial_v\left(\zn\right)=
-\frac{\theta}r-\zn\left(\frac{2\kappa}{r^2}\left(
M-\frac{e^2}r\right)\right),
\end{equation}
\begin{equation}
\label{avafora-u}
\partial_u(\phi\lambda+\theta)=\frac{2\kappa\nu\phi}{r^2}
				\left(M-\frac{e^2}r\right),
\end{equation}
taking the place of $(\ref{zex})$ and $(\ref{toallo})$.

Examining the propositions of this paper, one
immediately sees that the estimates employed for the expression 
\begin{equation}
\label{suzposot}
\varpi-
\frac{e^2}{r},
\end{equation}
when integrating $(\ref{nqu})$ or $(\ref{zex})$, 
also hold for
\begin{equation}
\label{posot1}
M-\frac{e^2}{r},
\end{equation}
while 
the estimates for
\begin{equation}
\label{suzposot2}
\frac1r\left(\zn\right)^2\nu,
\end{equation}
when integrating $(\ref{fdb1})$,
also hold (trivially) for $0$.\footnote{We have chosen the normalization
$(\ref{vatairiazei})$ precisely so that $(\ref{MDEF})$ holds.}
 Thus, this paper can be read
\emph{mutatis mutandis} for the uncoupled case, replacing 
the expression $(\ref{suzposot})$ by $(\ref{suzposot2})$
when integrating $(\ref{nqu})$ or $(\ref{zex})$, and $(\ref{posot1})$ by $0$,
when integrating $(\ref{fdb1})$. 
There is really only one point in the paper where the above leads to
an essential
simplification for the
uncoupled case: In the proof of 
Proposition \ref{ZBD}, 
one can immediately derive equation $(\ref{3A})$
and continue from there.

The authors hope that besides aiding the 
reader primarily interested in the linearized uncoupled problem,
this section at the same time convinces him that the
non-linear coupled problem is not much more 
difficult once the linear problem
is looked at in a more robust setting.

\appendix

\section{The initial value problem in the large}
\label{ivpinl}
\subsection{Basic Lorentzian geometry}
\label{lorapp}
We review here certain basic Lorentzian geometric concepts and notations,
that will be used in the sequel without comment.
The reader unfamiliar with these notions should consult a 
basic textbook, for instance, Hawking and Ellis~\cite{he:lssst}.

An $n$-dimensional \emph{spacetime} $(\mathcal{M}, g)$ is an
$n$-dimensional $C^2$ Hausdorff
manifold $\mathcal{M}$ together with a 
time-oriented $C^1$ Lorentzian\footnote{Our convention will be
signature $(-1,\stackrel{n-1}{1,\ldots,1})$.} metric $g$. 
Tangent vectors are called \emph{timelike}, 
\emph{spacelike}, or \emph{null}, according to whether their
length is negative, positive, or zero, respectively. Timelike and null vectors
collectively are known as \emph{causal}. A time-orientation is defined
by a continuous timelike vector field $T$ defined throughout $\mathcal{M}$.
Causal vectors are called \emph{future directed} if their inner product with 
$T$ is negative, and \emph{past directed} if their inner 
product is positive. A parametrized $C^1$-curve $\gamma$ in 
$\mathcal{M}$ is said to be timelike, spacelike, or null, 
according to whether its tangent vector has negative, positive, 
or zero length, everywhere.
If $\dot\gamma$ is everywhere causal, $\gamma$ is called
causal, and if $\dot\gamma$ is future directed, $\gamma$
is called future directed.
An affine parametrization on a parametrized geodesic $\gamma$ is one such
that $\frac{\partial}{\partial s}$ is parallel along $\gamma$. 

 For a subset $S\subset\mathcal{M}$, we define the 
\emph{causal future} of $S$, denoted by $J^+(S)$, to be
the set of points which can be reached from $S$ by a future pointing
causal curve. We define 
the \emph{causal past of $J^-(S)$} 
to be the set of points which can be reached from $S$ by a past
pointing causal curve.
The chronological future $I^+(S)$ is defined similarly, where
``causal'' is replaced by ``timelike''.
A subset $S\subset\mathcal{M}$ 
is said to be \emph{achronal} if $S\cap I^+(S)=\emptyset$.

A submanifold of $\mathcal{M}$ is called \emph{spacelike}
if its induced metric is Riemannian, \emph{null} if its induced metric
is degenerate, and timelike if its induced metric is \emph{Lorentzian}.
The reader can check that for embedded curves, this definition coincides
with the previous. Also, one can easily see that a hypersurface
is spacelike, timelike, or null if and only if its normal is 
everywhere timelike, spacelike, or null, respectively.

A curve is called \emph{inextendible} if it is not a proper subset of
another curve. A spacetime such that all inextendible future directed
causal geodesics
extend for arbitrary large positive values of an affine parameter
are called \emph{future causally geodesically complete}, otherwise 
\emph{future causally geodesically incomplete}.
A spacelike hypersurface $\mathcal{S}$
is said to be \emph{Cauchy}
if every inextendible parametrized
causal curve in $\mathcal{M}$ intersects $\mathcal{S}$ once and only once.
A spacetime possessing a Cauchy surface is said to be
\emph{globally hyperbolic}. 

\subsection{The maximal Cauchy development}
\label{cccon}
For convenience, in the rest of this section
all functions, tensors, maps and actions referred
to are by assumption $C^\infty$.

We say that a $4$-dimensional spacetime $(\mathcal{M}, g)$,
together with an anti-symmetric $2$-tensor $F_{\mu\nu}$
and a function $\phi$ satisfy the Einstein-Maxwell-scalar
field equations if
\[
R_{\mu\nu}-\frac12Rg_{\mu\nu}=2T_{\mu\nu},
\]
\[
F^{,\mu}_{\mu\nu}=0, F_{[\mu\nu,\lambda]}=0,
\]
\[
g^{\mu\nu}\phi_{;\mu\nu}=0,
\]
\[
T_{\mu\nu}=\phi_{;\mu}\phi_{;\nu}
-\frac12g_{\mu\nu}\phi^{;\alpha}\phi_{;\alpha}
+F_{\mu\alpha}F^{\alpha}_{\ \nu}-\frac14g_{\mu\nu}F^{\alpha}_{\ \lambda}
F^{\lambda}_{\ \alpha}.
\]
Here $R_{\mu\nu}$ denotes the Ricci curvature of $g_{\mu\nu}$, and
$R$ the scalar curvature. 
We call a connected
Riemannian $3$-manifold $(\Sigma, \stackrel\circ{g})$, together
with a $2$-symmetric tensor $K_{ij}$, two one forms $E_i$, $B_i$, 
and functions $\stackrel{\circ}\phi$,
$\phi'$ initial data for the Einstein-Maxwell-scalar field,
equations if the following constraint equations are satisfied:
\[
\stackrel\circ{R} - K_{ij} K^{ij} + (K^i_i)^2 = (\phi')^2 + 
\stackrel\circ{{\phi}^{;i}}\stackrel\circ{\phi}_{;i} + 
E^iE_i  + 
B^iB_i,
\]
\[
K_{ij}^{;j} - K^j_{j;i} = \phi'\stackrel\circ{\phi}_i+
2\epsilon_{ijk}  E^j B^k, 
\]
where $\stackrel\circ{ R}$ is the scalar curvature of the metric 
$\stackrel\circ{g}$ on $\Sigma$, $\epsilon_{ijk}$ is the fully anti-symmetric tensor 
representing its volume form, and raised, covariant, etc.~latin indices are with respect
to $\stackrel\circ{g}$ and its Levi-Civita connection $\stackrel\circ\nabla$.

The theorem of Choquet-Bruhat and Geroch~\cite{chge:givp}, applied to the above
equations, yields
\begin{theorem}
\label{cbg}
Given an initial data set $\{\Sigma, \stackrel{\circ}{g}_{ij}, K_{ij}, 
E_i, B_i,
\stackrel{\circ}\phi, \phi'\}$, there exists a unique
solution $\{\mathcal{M},g,F_{\mu\nu},\phi\}$
to the Einstein-Maxwell-scalar field equations such that
\begin{enumerate}
\item
$(\mathcal{M},g)$ is a globally hyperbolic spacetime.
\item
$\Sigma$ is a submanifold of $(\mathcal{M}, g)$ with first fundamental form
$\stackrel{\circ}{g}_{ij}$ and second fundamental form $K_{ij}$, with respect
to a future-pointing unit normal $N$. Moreover, $\phi|_\mathcal{S}=\stackrel
\circ{\phi}$, $N^\alpha\phi_{;\alpha}=\phi'$, $N^\alpha F_{\alpha i}=E_i$, 
$-\frac12N^{\alpha}\epsilon_{\alpha i}^{\ \ \gamma\delta}F_{\gamma\delta}=B_i$.
\item 
Given a second solution $\{\tilde{\mathcal{M}}, \tilde{g}, \tilde{F}_{\mu\nu},
\tilde{\phi}\}$ such that the above two properties holds,
there exists a map
\[
\Phi:\tilde{\mathcal{M}}\to\mathcal{M}
\]
such that $\Phi^*g=\tilde{g}$, $\Phi^*F=\tilde{F}$, $\Phi^*\phi=\tilde{\phi}$,
where $\Phi^*$ denotes the induced pull back map.
\end{enumerate}
\end{theorem}
The above collection $\{\mathcal{M},g, F_{\mu\nu}, \phi\}$, is known as the
\emph{Cauchy development} or the \emph{maximal domain of development}
of $\{\Sigma,\stackrel{\circ}{g},K,E_i,B_i,
\stackrel{\circ}{\phi},\phi'\}$. The set
$\mathcal{M}\cap J^+(\Sigma)$,\footnote{By
our convention, this is a manifold
with (spacelike) boundary.} together with the
restriction of $F_{\mu\nu}$, $\phi$ to this set,  is
the \emph{future Cauchy development}. 

\subsection{Penrose's singularity theorem}
If $\mathcal{E}\subset\mathcal{M}$ is a closed spacelike $2$-surface, then
the intersection of a tubular neighborhood of $\mathcal{E}$ with 
$J^+(\mathcal{E})\setminus (I^+(\mathcal{E})\cup \mathcal{E})$ 
is a null $3$-surface with $2$ connected
components. $\mathcal{E}$ can be considered as a smooth boundary of either;
if with respect to future-pointing normals, the mean curvature
of $\mathcal{E}$ is everywhere negative for both connected components, 
then we say $\mathcal{E}$ is \emph{trapped}.
Penrose's celebrated singularity theorem is given by
\begin{theorem}
\label{celebrated}
Consider a globally hyperbolic spacetime $(\mathcal{M},g)$
with a non-compact Cauchy surface $\Sigma$. 
Suppose for all null vectors
$L$, the Ricci curvature of $g$ satisfies the inequality
$Ric(L,L)\ge0$. Then, it follows that 
if $\mathcal{M}$ contains a closed trapped
surface $\mathcal{E}$, 
then $(\mathcal{M},g)$ is 
future-causally geodesically incomplete.
\end{theorem}
It is easy to check that the assumptions of the theorem are
satisfied by the future Cauchy development of non-compact initial data
for the Einstein-Maxwell-scalar field system, if it is assumed
that the data contain a trapped surface $\mathcal{E}$.

\subsection{Asymptotically flat data}
The study of isolated self-gravitating systems leads to the idealization
of so-called asymptotically flat initial data. Let 
$\{\Sigma,\stackrel{\circ}{g},K,E_i,B_i,
\stackrel{\circ}{\phi},\phi'\}$ be
a data set, and let $\mathcal{U}\subset \Sigma$ be open. $\mathcal{U}$
is said to be an \emph{asymptotically flat end} if
$\mathcal{U}$
is diffeomorphic to ${\bf R}^3\setminus B_1(0)$, and if
with respect to the coordinate chart induced by this diffeomorphism,
the following are satisfied:
\[
\left|g_{ij}(x)-
\left(1+\frac M{|x|}\right)\delta_{ij}(x)\right|+ 
|x||K_{ij}(x)|\le C |x|^{-1-\alpha}
\]  
\[
|\phi'(x)|+ |\nabla\stackrel\circ{\phi}| \le C|x|^{-2-\alpha} 
\]

\begin{equation}
\label{nomag}
\left|E_{i}(x)- \frac {e}{4\pi} \frac{x_i}{|x|^3}\right| + |H_i(x)| \le C|x|^{-2-\alpha}
\end{equation}
where $|\cdot|$ denotes the Euclidean metric, constant $\alpha>0$ and $M$ and $e$
are the mass and the charge respectively.
The data set is said to 
be \emph{asymptotically flat with $k$ ends} if 
there exists a compact set $\mathcal{K}\subset\Sigma$ such that
\[
\Sigma\setminus \mathcal{K}=\bigcup_{i=1}^k \mathcal{U}_k
\]
where the above union is disjoint, and each $\mathcal{U}_k$ is
an asymptotically flat end.

\subsection{The cosmic censorship conjectures}
\label{eikasies}
With this, we can formulate \emph{strong cosmic censorship}\footnote{This
formulation is due to Christodoulou~\cite{chr:givp}.}:
\begin{conjecture}
\label{scc}
For a ``generic'' asymptotically flat complete
initial data set for the Einstein-Maxwell-scalar field system,
the Cauchy development is inextendible as a manifold with $C^0$ metric.
\end{conjecture}

Physically speaking, strong cosmic censorship, as formulated above,
ensures that observers
travelling to the end of the Cauchy development are ``destroyed''.
It is for this reason that our assumption on inextendibility be
with respect to continuous metrics, and not more regular metrics. 
See~\cite{chr:givp}.

In addition to the above conjecture, there is the so-called
\emph{weak cosmic censorship} conjecture. A definite formulation 
is given by
\begin{conjecture} 
\label{wcccon}
For a ``generic'' complete asymptotically flat
initial data set the Einstein-Maxwell-scalar field system,
the Cauchy development posseses a complete null infinity in the sense of
Christodoulou~\cite{chr:givp}.
\end{conjecture}
In the case where the Maxwell field identically vanishes, the 
above conjecture was
established by Christodoulou~\cite{chr:ins}, when one restricted to initial data
which are spherically symmetric. If the Maxwell field does not vanish,
then the completeness of null infinity for complete asymptotically flat
spherically symmetric initial data
follows immediately from~\cite{md:sssts}. It should be noted, however, 
that the fact that the initial data set has two asymptotically flat ends renders
weak cosmic censorship almost trivial
in this case.

\section{Spherical symmetry}
\label{sphsymsec}
A solution $\{\mathcal{M},g,F_{\mu\nu}, \phi\}$
of the Einstein-Maxwell-scalar 
field system $(\ref{Einstein-xs})$--$(\ref{emtensor-xs})$ is said to be
\emph{spherically symmetric} if the group $SO(3)$ acts nontrivially
by isometry on $(\mathcal{M}, g)$, and preserves the matter fields $\phi$
and $F_{\mu\nu}$,
in the sense that $\phi$ is constant on group orbits and
$F_{\mu\nu}$ is invariant with respect to the induced action of $SO(3)$
in $\Lambda^2(T\mathcal{M})$.

We will call an initial data set $\{\Sigma,\stackrel{\circ}g,
K_{ij}, E_i, B_i, \stackrel{\circ}\phi, \phi'\}$
spherically symmetric if $SO(3)$
acts by isometry on $(\Sigma, \stackrel\circ{g})$, 
if this action preserves $K$, $\phi$,
$E_i$, and $B_i$ and if, moreover,
$\mathcal{S}=\Sigma/SO(3)$ inherits 
the structure of a connected $1$-dimensional Riemannian manifold,
possibly with boundary.\footnote{We require, in other words,
$\Sigma$ to be a warped product of a connected
subset of $\mathcal{R}$ with $S^2$, where $SO(3)$ acts
transitively on the $S^2$ factor. It is this that ensures
that $\mathcal{Q}$ described in what follows is indeed a manifold (with boundary).}
(In particular, it is clear that $\Sigma$ can have at most $2$
ends.)

We have the following:
\begin{theorem}
Let $\{\Sigma,\stackrel\circ{g}, B_i, E_i,
\stackrel\circ\phi, \phi'\}$
be an asymptotically flat
spherically symmetric initial data set. Then its
Cauchy development $\{\mathcal{M},g,
F_{\mu\nu},\phi\}$ is also spherically symmetric. 
\end{theorem}
We omit the proof. The above theorem says that our symmetry assumptions
are compatible with the problem of evolution. 

From the point of view of p.d.e.~theory, the fundamental fact
is that the problem of evolution now reduces to a system
of equations on a $2$-dimensional domain. The setting
for this system is the quotient space $\mathcal{Q}=\mathcal{M}/SO(3)$.
In fact, $\mathcal{Q}$ inherits the structure
of a $2$-dimensional Lorentzian manifold, possibly with boundary,
with metric $\bar{g}$, and a natural
function $r:\mathcal{Q}\to{\bf R}$, called the \emph{area-radius}, 
defined by
\begin{equation}
\label{r-orismos}
r(p)=\sqrt{\frac{Area(p)}{4\pi}}.
\end{equation}
(In particular, the $\infty>Area(p)\ge 0$ for all group orbits $p$;
these group orbits will be spacelike spheres or points.)
The metric $g$ on $\mathcal{M}$ can then be written as
\[
g=\bar{g}+ r^2\gamma,
\] 
where $\gamma$ is the standard metric on the unit $2$ sphere.
The function $\phi$ descends to a function on $\mathcal{Q}$, which,
without fear of confusion, we shall also denote by $\phi$.
It was shown in~\cite{md:si} that the
energy momentum tensor $T_{\mu\nu}$
can be written as 
\begin{eqnarray}
\label{edw}
\nonumber
T_{\mu\nu}dx^\mu\otimes dx^\nu&=&\frac{e^2}{2r^4}\bar{g}+\frac{e^2}{r^2}\gamma
\\
&&\hbox{}+\phi_\mu\phi_\nu dx^\mu\otimes dx^\nu-\frac12g_{\mu\nu}g(\nabla\phi,\nabla\phi)
dx^\mu\otimes dx^\nu
\end{eqnarray}
where $e$ is the constant of $(\ref{nomag})$. 
Thus $(\ref{Maxwell-xs})$ decouples and the evolution 
of $g$ and $\phi$ can be determined by a second order
system of equations for $\bar{g}$, $r$, $\phi$ on $\mathcal{Q}$.
Introducing null coordinates, and defining quantities as
in Assumptions ${\bf \Gamma}'$--${\bf E}'$,
then it was shown in~\cite{md:si} that this system can be 
written as $(\ref{puqu})$--$(\ref{sign2})$.

Let $\Gamma$ denote the subset of $\mathcal{Q}$ such that $r=0$. If nonempty,
$\Gamma$ is a connected timelike boundary for
$\mathcal{Q}$, intersecting
$\Sigma/SO(3)$. $\Gamma$ is called the \emph{axis of symmetry} and corresponds
precisely to the fixed points of the $SO(3)$ action. By $(\ref{edw})$, it is clear
that  regularity requires that if $\Gamma$ is non-empty,
then $e=0$. Now, if $(\Sigma,\stackrel\circ{g})$ is assumed in addition 
to be \emph{complete}, then $\Sigma$ has one end if $\Gamma$ is non-empty,
and two ends if $\Gamma$ is empty. Thus, if $e\neq0$, in view of the above
remarks, it follows that a complete $\Sigma$ 
necessarily has \emph{two} asymptotically flat ends.

\section{Penrose diagrams}
\label{pdveo}
In the introduction to this paper, we have used so-called \emph{Penrose diagram} notation
to conveniently convey certain global geometric information about spherically symmetric 
spacetimes. Briefly put, the domain of a Penrose diagram is the 
image of a map $\Phi:(\mathcal{Q},\bar{g})\to({\bf R}^{1+1},-dudv)$ 
into $2$-dimensional Minkowski space, such that $\Phi$ preserves the causal structure,
and is bounded. (Alternatively, one can think of the diagram as the domain of a bounded global
null coordinate system in the $(u,v)$ plane.) Since one can determine the causal 
relations of two points in a bounded subset of Minkowski space
by inspection, the causal geometry of $\mathcal{Q}$ can be easily perceived
by displaying its image under $\Phi$.

There is more, however, to Penrose diagrams than 
this purely causal aspect. The map $\Phi$
defines a boundary of $\mathcal{Q}$, namely 
$\partial(\Phi(\mathcal{Q}))\subset{\bf R}^2$. 
The natural function $r:\mathcal{Q}\to{\bf R}$ defined in Appendix~\ref{sphsymsec}
allows one to distinguish special subsets of this boundary, in particular, a set to
be called \emph{null infinity}, denoted $\mathcal{I}^+$. Causal 
concepts extend to $\overline{\mathcal{Q}}$.
With this, one will be able to define the black hole region  
$\mathcal{Q}\setminus J^-(\mathcal{I}^+)$ and the event horizon
$(\partial J^-(\mathcal{I}^+))\cap\mathcal{Q}$.

We proceed in the following sections to discuss these issues in more detail.
We begin in Section~\ref{2mink} with a review of the geometry of $2$-dimensional
 Minkowski space.
In Section~\ref{pdiag}, we define the concept of a Penrose diagram, and the
induced concepts of black hole and event horizon. 
In Section~\ref{conv}, we illustrate certain conventions on incidence
relations when  displaying domains of ${\bf R}^2$, and finally, we give the Penrose
diagrams of some celebrated special solutions of the Einstein equations, in
Section~\ref{paradeigmata}.

\subsection{$2$-dimensional Minkowski space}
\label{2mink}
We call the spacetime $({\bf R}^2,-dt^2+dx^2)$ \emph{$2$-dimensional Minkowski
space}. It is time-oriented by the vector $\frac\partial{\partial t}$.
To examine the causal properties, it is more convenient to
introduce null coordinates $u=t-x$, $v=t+x$. 
One easily sees that a vector $a\frac\partial{\partial u}+
b\frac\partial{\partial v}$ is causal if $ab\ge0$, in which
case it is future pointing if $a>0$, timelike if $ab>0$,
and null if $ab=0$. Thus
\[
J^+(u_0,v_0)=\{(u,v)|u\ge u_0, v\ge v_0\},
\]
\[
J^-(x_0,t_0)=\{(u,v)|u\le u_0, v\le v_0\},
\]
\[
I^+(u_0,v_0)=\{(u,v)|u> u_0, v> v_0\},
\]
\[
I^-(x_0,t_0)=\{(u,v)|u< u_0, v< v_0\}.
\]
If the original $x$ and $t$ coordinates are
mapped to the Euclidean plane as horizontal and vertical, then the constant-$u$ 
and constant-$v$ lines are at $45$ degree, and $135$ degree angles from
the horizontal, respectively. Thus $p\in J^+(q)$ if the angle $\alpha$ between 
$p$ and $q$, as measured from the horizontal:
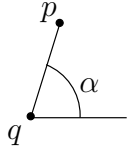
\[
\input{penrose0.pstex_t}
\]
satisfies $45\le\alpha\le135$.
Similarly $p\in J^-(q)$ if $-45\ge\alpha\ge -135$ degrees.
Null curves are precisely those of constant $u$ or $v$,
i.e.~line segments at $45$ and $135$ degrees, whereas
timelike curves have tangent with angle from the horizontal $\alpha$
satisfying $45<\alpha<135$ or $-45>\alpha>-135$.

As an exercise, the reader can read off from below
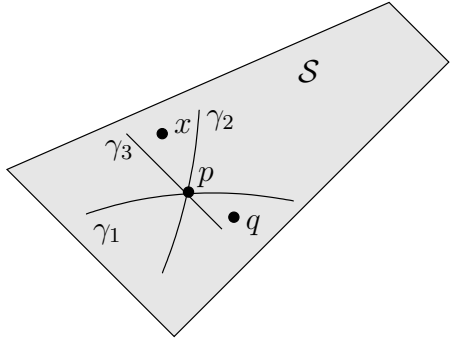
\[
\input{penrose.pstex_t}
\]
the following incidence and causal relations:
$p,q,x\in \mathcal{S}$, $\gamma_1,\gamma_2, \gamma_3\subset S$, 
$q\not\in J^-(p)\cup J^+(p)$, $x\in I^+(p)$,
$p=\gamma_1\cap\gamma_2\cap\gamma_3$, and that $\gamma_1$,
$\gamma_2$, and $\gamma_3$ are spacelike, timelike, and null, respectively.

\subsection{Penrose diagrams}
\label{pdiag}

We will define a \emph{Penrose diagram} to consist of the following information:
\begin{enumerate}
\label{prw}
\item $(\mathcal{Q},\bar{g})$ a $2$-dimensional $C^2$ manifold, possibly with
boundary, with a time-oriented $C^1$ Lorentzian metric.
\item A $C^2$ diffeomorphism preserving the causal structure:
\[
\Phi:\mathcal{Q}\to \mathcal{PD}
\]
where $\mathcal{PD}\subset {\bf R}^2$ is a bounded region.
\item
\label{tele}
A continuous nonnegative function $r:\mathcal{Q}\to{\bf R}$.
\end{enumerate}

The domain of the Penrose diagram is $\mathcal{PD}$; this set
and its closure inherit a topology and causal structure from ${\bf R}^{1+1}$.

For Cauchy developments of spherically symmetric
asymptotically flat initial data for a wide variety of Einstein-matter systems,
the Lorentzian quotient $(\mathcal{Q},g)$, together
with the area radius function $r$, will indeed admit a 
Penrose diagram, in view of the existence of a globally defined null
coordinate system $(u,v)$, which can be chosen in addition to be
bounded. The coordinate functions together determine the map $\Phi$.

In general, given a Penrose diagram $\{(\mathcal{Q},g),r,\Phi\}$,
we can define a subset
\[
\mathcal{I}^+\subset\partial\mathcal{PD}=\overline{\mathcal{PD}}\setminus
\mathcal{PD}
\]
as follows:
\[
\mathcal{I}^+=
\{p\in\partial\mathcal{PD}:\forall_{R,\epsilon>0}, \exists_{q\in (J^-(p)\setminus
I^-(p))\cap B_\epsilon(p)\cap\mathcal{PD}}:r(q)\ge R\}
\]

For $\mathcal{Q}$ which arises from the Cauchy development
of spherically symmetric asymptotically flat data as above, 
a lot can be said about $\mathcal{I}^+$.
In particular, if  $\mathcal{Q}$ has one end, then 
$\mathcal{I}^+$ will be a connected null curve,
with past limit point common with the limit point of the quotient
of the Cauchy surface. (This common limit point is often called
\emph{spacelike infinity}, and denoted $i^0$.) In this case we call
$\mathcal{I}^+$ \emph{future null infinity}.\footnote{If $\mathcal{Q}$
has $2$ ends, then $\mathcal{I}^+$ will have $2$-connected components
of this form.}
See~\cite{md:sssts}.

With this definition of null infinity, we can define the 
black hole as the region
\[
\mathcal{Q}\setminus\Phi^{-1}(J^-(\mathcal{I}^+))
\]
and the
\emph{future event horizon}, denoted $\mathcal{H}^+$,
by
\[
\mathcal{H}^+=\partial(\mathcal{Q}\setminus\Phi^{-1}(J^-(\mathcal{I}^+)))\cap
\mathcal{Q}.
\]
Since the black hole is a future subset of $\mathcal{Q}$, $\mathcal{H}^+$ is
in fact its past boundary. In the context of
the spacetimes considered here, it is easy to see that these 
definitions do not depend on $\Phi$. Similarly, we can define white holes and
$\mathcal{H}^-$, by interchanging future and past. 

Finally, we note that Penrose diagrams are intimately related to null coordinate
systems. Indeed, given a Penrose diagram, the pull back of the standard $u$ and
$v$ coordinates on ${\bf R}^2$ define null coordinates on $\mathcal{Q}$,
and similarly, given $\{(\mathcal{Q}, g), r\}$ and a null coordinate system
$(u,v)$ with bounded range satisfying 
$J^+(u_0,v_0)=\{(u,v)|u\ge u_0, v\ge v_0\}\cap\mathcal{Q}$,
$J^-(u_0,v_0)=\{(u,v)|u\le u_0, v\le v_0\}\cap\mathcal{Q}$,
these coordinates determine a map to ${\bf R}^2$ defining a Penrose diagram. 


\subsection{Conventions}
\label{conv}
In displaying subsets of Minkowski space,
in particular Penrose diagrams, it will be useful to 
have certain conventions on inclusion and incidence relations.

Let us adopt the following then: Curves depicted by
dashed lines are assumed not to be contained in shaded regions that they
bound, and points depicted by white circles are assumed not to be
contained in either the shaded regions or the lines to which they
are adjacent. Finally, circles filled in with lines are included
in the adjacent line segment to which
the filling is parallel.
Example: from
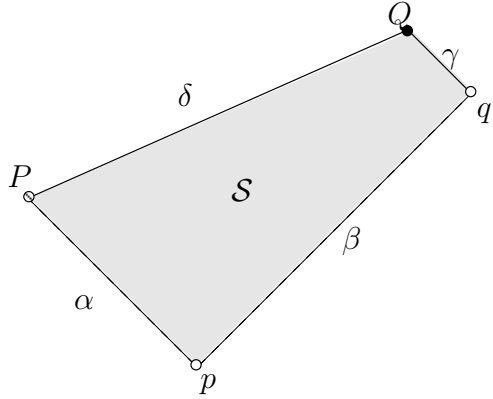
\[
\input{penrose2.pstex_t}
\]
we are to  understand all of the following: 
$p,P,q,Q,\in\overline{\mathcal{S}}$, 
$\alpha,\beta,\gamma,\delta\subset\overline{\mathcal{S}}$,
$\mathcal{S}\subset I^+(p)$, $\alpha, \beta\subset J^+(p)\setminus I^+(p)$,
$\gamma\subset J^+(q)\setminus I^+(q)$, $J^+(q)\cap \mathcal{S}=\gamma$,
$\overline\alpha\cap\overline\beta=\{p\}$, but
$p\not\in \alpha\cup\beta$, $\delta\subset \mathcal{S}$ is spacelike, 
$P\not\in\delta$, but $P\in\alpha$, $Q\in\delta$.
Finally, dotted (as opposed to dashed) lines with arrows are 
reserved to help label sets.

Also, it goes without saying that when displaying a Penrose diagram,
in view of the fact that $\Phi:\mathcal{Q}\to\mathcal{PD}$
is $1-1$, we can use the same symbol
for a subset of $\mathcal{Q}$ and its image in $\mathcal{PD}$.

\subsection{Examples}
\label{paradeigmata}
We provide for reference the Penrose diagrams of 
$3+1$ dimensional Minkowski space:
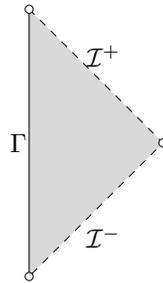
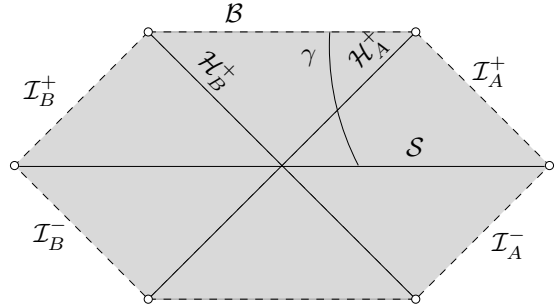
\[
\input{minkowski.pstex_t}
\]
the Schwarzschild solution: 
\[
\input{schwarzschild.pstex_t}
\]
and the 
Reissner-Nordstr\"om solution\footnote{We mean here
the Cauchy development of the complete time symmetric Cauchy surface $\Sigma$,
whose quotient $\mathcal{S}=\Sigma/SO(3)$ is depicted, \emph{not}
a so-called maximally analytic extension.}:
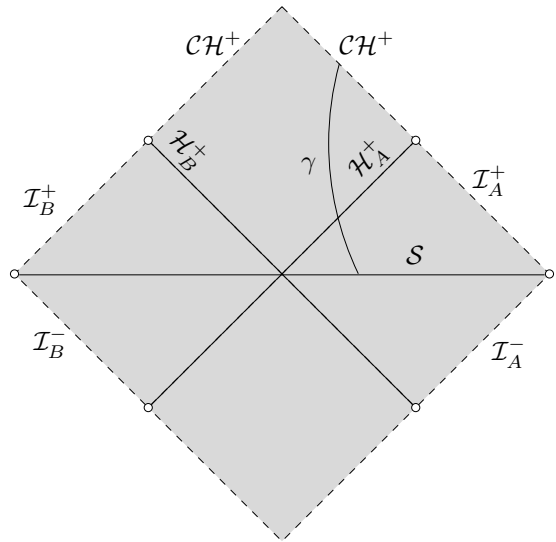
\[
\input{Re-No.pstex_t}
\]
Note that the future-inextendible timelike geodesic depicted by $\gamma$ has
finite length in the above two spacetimes. All three spacetimes
are spherically symmetric solutions of $(\ref{Einstein-xs})$--$(\ref{emtensor-xs})$.

\end{document}